\documentclass[trackchanges, twocolumn]{aastex7}
\usepackage{graphicx}
\usepackage{amsmath}
\usepackage{natbib}
\usepackage{subcaption}
\usepackage{hyperref}
\usepackage{booktabs}
\usepackage{array,tabularx,makecell}
\newcolumntype{Y}{>{\raggedright\arraybackslash}X}

\usepackage{xcolor}

\begin{document}

\title{ The GASKAP-HI Survey towards the Magellanic Clouds: Cold Atomic Gas Survival and Evolution in the Large Magellanic Cloud}

\author[0009-0005-1781-5665]{Hongxing Chen}\email{hchen792@wisc.edu}
\affiliation{University of Wisconsin–Madison, Department of Astronomy, 475 N Charter St, Madison, WI 53703, USA}

\author[0000-0002-3418-7817]{Sne{\v z}ana Stanimirovi{\'c}}
\affiliation{University of Wisconsin–Madison, Department of Astronomy, 475 N Charter St, Madison, WI 53703, USA}
\email{sstanimi@astro.wisc.edu}

\author[0000-0002-4899-4169]{James Dempsey}
\affiliation{Research School of Astronomy \& Astrophysics, The Australian National University, Canberra ACT 2611, Australia}
\email{james.dempsey@anu.edu.au}

\author[0000-0002-1272-3017]{Jacco Th. van Loon}
\affiliation{Lennard-Jones Laboratories, Keele University, ST5 5BG, UK}
\email{j.t.van.loon@keele.ac.uk}

\author[0000-0003-0742-2006]{Yik Ki Ma}
\affiliation{Max-Planck-Institut f\"ur Radioastronomie, Auf dem H\"ugel 69, 53121 Bonn, Germany}
\affiliation{Research School of Astronomy \& Astrophysics, The Australian National University, Canberra ACT 2611, Australia}
\email{yikki.ma@anu.edu.au}

\author[0000-0002-2712-4156]{Hiep Nguyen}
\affiliation{Research School of Astronomy \& Astrophysics, The Australian National University, Canberra ACT 2611, Australia}
\email{vanhiep.nguyen@anu.edu.au}

\author[0000-0001-9504-7386]{Nickolas~M.~Pingel}
\affiliation{Department of Astronomy, Indiana University, 727 East Third Street, Bloomington, IN 47405, USA}
\email{nmpingel@iu.edu}

\author[0000-0002-6300-7459]{John M. Dickey}
\affiliation{School of Natural Sciences, Private Bag 37, University of Tasmania, Hobart, TAS, 7001, Australia}
\email{John.dickey@utas.edu.au}

\author[0000-0002-9888-0784]{Min-Young Lee}
\affiliation{Korea Astronomy and Space Science Institute, 776 Daedeok-daero, Daejeon 34055, Republic of Korea}
\affiliation{Department of Astronomy and Space Science, University of Science and Technology, 217 Gajeong-ro, Daejeon 34113, Republic of Korea}
\email{mlee@kasi.re.kr}

\author[0000-0003-2730-957X]{N. M. McClure-Griffiths}
\affiliation{Research School of Astronomy \& Astrophysics, The Australian National University, Canberra ACT 2611, Australia}
\affiliation{SKA Observatory, Jodrell Bank, Lower Withington, Macclesfield, SK11 9FT, UK}
\email{Naomi.mcclure-griffiths@anu.edu.au}

\author[0000-0002-7743-8129]{Claire E. Murray}
\affiliation{Space Telescope Science Institute, 3700 San Martin Drive, Baltimore, MD 21218, USA}
\affiliation{Department of Physics \& Astronomy, Johns Hopkins University, 3400 N. Charles Street, Baltimore, MD 21218, USA}
\email{clairemurray56@gmail.com}

\author[0000-0002-8057-0294]{Lister Staveley-Smith}
\email{lister.staveley-smith@uwa.edu.au}
\affiliation{International Centre for Radio Astronomy Research, University of Western Australia, Crawley, WA 6009, Australia}

\author[0000-0002-4814-958X]{Denis Leahy}
\affiliation{Department of Physics \& Astronomy, University of Calgary, Calgary, AB T2N 1N4, Canada}
\email{leahy@ucalgary.ca}

\author[0000-0001-6846-5347]{Callum Lynn}
\affiliation{Research School of Astronomy \& Astrophysics, The Australian National University, Canberra ACT 2611, Australia}
\email{callum.lynn@anu.edu.au}

\author[0000-0002-5501-232X]{Antoine Marchal}
\affiliation{Research School of Astronomy \& Astrophysics, The Australian National University, Canberra ACT 2611, Australia}
\affiliation{Laboratoire de Physique de l’École Normale Supérieure, ENS, Université PSL, CNRS, Sorbonne Université, Université Paris Cité, F75005, Paris, France}
\affiliation{LUX, Observatoire de Paris, Université PSL, Sorbonne Université, 75014 Paris,France}
\email{antoine.marchal@anu.edu.au}

\author[0000-0003-3351-6831]{Daniel R. Rybarczyk}
\affiliation{University of Wisconsin–Madison, Department of Astronomy, 475 N Charter St, Madison, WI 53703, USA}
\email{rybarczyk@astro.wisc.edu}

\author[0000-0002-9214-8613]{Helga Dénes}
\affiliation{College of Sciences and Engineering, Universidad San Francisco de Quito, Quito, 170901, Ecuador}
\email{helgadenes@gmail.com}

\author[0000-0002-1495-760X]{Steven Gibson}
\affiliation{Department of Physics and Astronomy, Western Kentucky University, Bowling Green, KY 42101, USA}
\email{steven.gibson@wku.edu}

\author[0000-0001-7105-0994]{Katherine Jameson
}
\affiliation{Caltech Owens Valley Radio Observatory, Pasadena, CA 91125, USA}
\email{kjameson@caltech.edu}

\author[0000-0002-6637-9987]{Ian Kemp}
\affiliation{International Centre for Radio Astronomy Research (ICRAR), Curtin University, Bentley, WA 6102, Australia}
\email{ian.kemp@postgrad.curtin.edu.au}

\author[0000-0002-4110-8769]{Ioana~A.~Stelea}
\email{stelea@wisc.edu}
\affiliation{University of Wisconsin–Madison, Department of Astronomy, 475 N Charter St, Madison, WI 53703, USA}

\begin{abstract}
We use atomic hydrogen (HI) absorption detections from the GASKAP-HI survey to investigate the properties of cold atomic gas in the Large Magellanic Cloud (LMC). Using the radiative transfer method, we decompose 155 sightlines into 330 cold neutral medium (CNM), 2 thermally unstable neutral medium (UNM), and 310 warm neutral medium (WNM) components. We find that the CNM in the LMC exhibits higher optical depths (median 0.46), lower spin temperatures (median $\sim$37 K), broader linewidths (median $\sim$4.9 km s$^{-1}$), and slightly lower CNM fractions (median $\sim$23\%) than in the Milky Way. 
We examine the connection between the CNM, molecular gas, and star formation, finding that CNM correlates more closely with molecular gas than WNM, while molecular gas shows a tighter relation with star formation. Molecular hydrogen (H$_2$) formation begins near $N{_\mathrm{HI,CNM}}\sim10^{20}~\mathrm{cm^{-2}}$, and nearly all sightlines with $N_{\mathrm{HI,CNM}}>10^{21}~\mathrm{cm^{-2}}$ contain molecular gas. The CNM fraction increases with visual extinction ($A_V$), and 
the LMC maintains CNM fractions comparable to those in the Milky Way at substantially lower $A_V$, likely due to higher local densities and a longer line-of-sight depth.
Sightlines near expanding shells tend to show higher CNM fractions, although this is partly driven by higher total HI column densities. Finally, the CNM kinematics generally follow the HI disk rotation, with about 7\% of components showing velocity offsets greater than $25~\mathrm{km~s^{-1}}$, likely tracing inflows or outflows driven by stellar feedback or large-scale interactions within the Magellanic System.

\end{abstract}


\section{Introduction}

Neutral atomic hydrogen (HI), the most abundant gaseous component of the interstellar medium (ISM), predominantly exists in two stable phases: the cold neutral medium (CNM) and the warm neutral medium (WNM) \citep{McKee1977}. In the Milky Way, their typical kinetic temperatures and volume densities are $(T_k, n) \sim (25$–$250~\mathrm{K}, 10$–$100~\mathrm{cm^{-3}})$ for the CNM and $(T_k, n) \sim (4000$–$8000~\mathrm{K}, 0.1$–$1~\mathrm{cm^{-3}})$ for the WNM \citep{Wolfire2003}. Between these two stable phases lies the thermally unstable neutral medium (UNM), which occupies the intermediate temperature and arises from turbulent mixing, shocks, or transient heating and cooling processes \citep{Wolfire2003,Bialy2019}.

The coexistence of the WNM and CNM is governed by a balance between radiative heating and cooling processes within a dynamic pressure equilibrium. This balance depends on local environmental conditions, including the ambient interstellar radiation field, dust properties, metallicity, and interstellar turbulence \citep{Field1969, Wolfire1995, Wolfire2003, Liszt2002, Glover2012, Glover2014, Bialy2019}. 
A galaxy’s ability to sustain HI in the cold phase is closely tied to its capacity for star formation, since molecular hydrogen -- the direct fuel for star formation -- forms primarily out of the CNM \citep{Krumholz2009, Kennicutt2012, Bialy2019}.  The CNM is therefore expected to correlate positively with the star formation rate \citep{Smith2023}. Cold gas properties are strongly influenced by dust grains, which shield the CNM from ultraviolet (UV) radiation, helping to maintain the low temperatures required for its survival, while simultaneously catalyzing molecular gas formation \citep{Safranek-Shrader2017}. Both simulations and observations in the Milky Way suggest a close link between dust and CNM properties. For instance, \citet{McClure-Griffiths2023}, building on the model of \citet{Bialy2019}, showed that dust optical extinction ($A_V$) significantly shapes the predicted CNM temperature range, and reported a positive correlation between the CNM fraction ($f_\text{CNM}$) and $A_V$ in Galactic observations.


Observations of the HI 21-cm line with radio telescopes provide a powerful way of directly measuring a galaxy’s atomic gas content. While HI emission can be detected across the entire galaxy, it does not by itself allow the CNM and WNM to be disentangled. To isolate these phases, HI absorption measurements against background radio continuum sources are required, as absorption
(in moderately sensitive observations)
directly traces CNM and/or UNM \citep[e.g.,][]{Heiles2003, Murray2015, Dempsey2022}. By combining HI absorption measurement with adjacent HI emission, one can derive the excitation (or spin) temperature ($T_s$) and column density ($N_{\mathrm{HI}}$) of CNM and WNM components along the line of sight (LOS). However, while HI emission can be readily obtained and has been studied for decades \citep[e.g.][]{Staveley-Smith1997,Kim1998, Thilker2004, Kalberla2005, HI4PI2016, Pingel2022}, HI absorption observations are very limited by telescope sensitivity and the scarcity of bright background continuum sources \citep[e.g.][]{Heiles2003, Heiles2003b, Murray2015, Murray2018, Jameson2019, Dempsey2022, Chen2025}. This limitation is particularly severe for galaxies beyond the Milky Way, which cover only a small solid angle on the sky and therefore intersect far fewer background sources than the Milky Way. In addition, their greater distances lead to stronger beam dilution effects and lower signal-to-noise ratios for the absorption features. 

In this context, the Magellanic Clouds (MCs) offer a unique opportunity to study the CNM/WNM in an external galactic environment, as they are the nearest galaxies to the Milky Way. Their close distance ($\sim$50 kpc; \citealt{Freedman2001}) enables HI absorption studies on spatial scales approaching those achievable in the Milky Way. Previous surveys have successfully 
mapped HI emission \citep{Staveley-Smith1997,Kim1998,Stanimirovic1999,Staveley-Smith2003,Pingel2022} and
carried out HI absorption measurements in both the Small Magellanic Cloud (SMC; \citealt{Dickey2000,Jameson2019,Dempsey2022}) and the Large Magellanic Cloud (LMC; \citealt{Dickey1994, Marx-Zimmer2000, Liu2021}). 

In this paper, we focus on the LMC, an irregular, low-mass, gas-rich galaxy viewed nearly face-on, with a moderately sub-solar metallicity of $\sim$0.5 $Z_\odot$ \citep{Olszewski1991, deGrijs2014} and a stronger ambient UV radiation field  \citep[e.g.,][]{Tumlinson2002,Welty2006} compared to the Milky Way. The lower metallicity and higher UV radiation fields are actually not independent. Lower metallicity implies a lower dust-to-gas ratio and reduced dust shielding, allowing UV photons to penetrate more deeply into the ISM. For example, recent numerical simulations \citep[e.g.][]{Kim2024} show that reduced dust attenuation of FUV radiation at low metallicity can enhance photoelectric heating. 
\cite{Dickey1994} and  \cite{Marx-Zimmer2000} conducted HI absorption studies of the LMC using the Australia Telescope Compact Array (ATCA). They targeted 50 continuum sources, primarily concentrated around the 30 Doradus region (RA = 84.68$^\circ$, Dec = $-$69.10$^\circ$), the most active star-forming complex and the highest HI column density region in the LMC. \citet{Dickey1994} found that the cold gas in the LMC likely has lower temperatures than in the Milky Way, and that the abundance of cold gas decreases with increasing distance from 30 Doradus. \citet{Marx-Zimmer2000} reported cold gas temperatures of $\sim$30 K and suggested enhanced cooling rates in the vicinity of 30 Doradus and within a supergiant shell, likely driven by enhanced pressures in these regions.
A subsequent, more sensitive ATCA survey by \citet{Liu2021} extended HI absorption measurements across the entire LMC, targeting 92 continuum sources and providing a less biased view of the cold atomic gas properties. This study reported an average CNM spin temperature of $T_{\rm CNM} \sim 30$ K and an average CNM fraction of $\sim$14\%.

The recent Galactic-ASKAP HI (GASKAP-HI) survey, utilizing the Australian Square Kilometre Array Pathfinder (ASKAP) radio telescope \citep{Hotan2021}, brings unprecedented spectral and spatial resolution for HI in the Magellanic Clouds \citep{Dickey2013,Pingel2022,Dempsey2022, Chen2025,Dempsey2026}. 
It has also provided the largest sample to date of HI absorption detections in the Magellanic Clouds. 
In this study, we focus on the LMC and analyze HI absorption/emission spectra using the Gaussian decomposition method with radiative transfer \citep[developed by][]{Heiles2003} to decompose spectra into individual WNM and CNM components. This process derives key physical properties such as spin temperature and column density for each cold structure identified in absorption across different regions of the LMC. We then explore the spatial distribution of the cold HI properties. This large sample size offers a more complete and unbiased view of cold gas in a sub-solar metallicity and higher UV radiation environment. By comparing our results with similar studies of the Milky Way, we investigate how cold gas properties vary with different environments. In addition, we investigate how local environmental factors -- such as expanding shells, molecular gas, dust, and star formation -- affect the cold gas properties. Finally, we also investigate the CNM kinematics to explore possible cold gas inflows/outflows.
In this study we only focus on the LMC observations, the decomposition of HI absorption spectra for the Small Magellanic Cloud (SMC) will be presented in future studies.

This paper is organized as follows.
Section \ref{sec:obs} describes the observations and the extraction of HI absorption spectra toward background radio continuum sources, along with the corresponding HI emission spectra.
Section \ref{sec:decomp} details the radiative transfer method used to derive the physical properties of the CNM and WNM components.
In Section \ref{sec:comp_result}, we present the properties of individual components, such as optical depth, spin temperature, and linewidth. We also compare results for individual components with Galactic HI absorption studies. 
Section \ref{sec:los_result} examines line of sight quantities, including total column densities and the CNM fraction, in comparison to Milky Way results.
Section \ref{sec:environment} investigates how the cold gas properties correlate with local environmental effects, including molecular gas, dust, star formation, and the existence of expanding shells.
Section \ref{sec:kine} explores the kinematics of the CNM and identifies several inflow/outflow cold clouds.
Finally, Section \ref{sec:conclu} summarizes our main findings and conclusions.

\section{HI Observations} \label{sec:obs}
The Magellanic Clouds observations were conducted by the GASKAP-HI survey \citep{Dickey2013}. The LMC data were collected as part of the GASKAP-HI Pilot Phase II survey, which targeted ten fields (five towards the LMC, four toward the Bridge, and one towards the SMC) from June 2020 to March 2022. Each field received 10 hours of observation using the standard GASKAP-HI configuration \citep[see][]{Pingel2022}.
The spectral resolution is $0.24~ \text{km}~ \text{s}^{–1}$. 
Our analysis of HI spectra of the LMC focuses on the LSR velocity range from 170 $ \text{km}~ \text{s}^{–1}$ to 325 $ \text{km}~ \text{s}^{–1}$.
For each observation, the HI emission data were processed using the ASKAPSoft package \citep{Hotan2021} and WSClean software package \citep{Offringa2014}, which calibrated the data and generated a continuum image and a continuum source catalog. Detailed descriptions of these observations and the initial data processing are provided by \cite{Pingel2022}. 

\subsection{HI absorption spectra}
To measure the HI absorption against continuum sources, \citet{Dempsey2022} developed an absorption pipeline that produced a continuum-included spectral-line subcube from the GASKAP-HI data for each source, 
and extracted the HI absorption spectra. We briefly summarize the method here. 

The initial continuum source catalog was constructed from the ASKAPsoft continuum catalog, selecting sources with flux densities $S_{\rm cont} \geq 15~\mathrm{mJy}$. For each source, a spectral line subcube ($50'' \times 50''$ in size, with $1''$ pixel size and with 0.24  $ \text{km}~ \text{s}^{–1}$ velocity resolution) centered on the source position was generated. To maximize signal-to-noise and minimize contamination from extended emission (e.g., from the WNM), a minimum baseline cutoff of 1.5 k$\lambda$ (corresponding to $\sim$315 m) was applied during subcube imaging. This resulted in a typical synthesised beam size of $16'' \times 14''$. 
Within each subcube, the on-source spectrum was produced by combining all pixels within the source ellipse identified from the ASKAPsoft continuum catalog. Then, the mean continuum flux density was measured in line-free channels for each pixel. Following \citet{Dickey1992}, a weighted-mean on-source spectrum was created, with each pixel being weighted by the square of the pixel mean continuum flux density. Finally, the absorption spectrum ($e^{-\tau}$) was produced by dividing the combined spectrum by its mean continuum flux density. 
These absorption spectra, collected at a resolution of 0.24~km~s$^{-1}$, were smoothed to 0.98~km~s$^{-1}$ (4 times the native bin width) to improve signal-to-noise.

The noise in the absorption spectrum was measured by combining the noise in the off-line region and emission in the primary beam of the dish \citep{Jameson2019}. The base noise of the spectrum was given by the standard deviation of the spectrum in the line-free region. To model the increase in system temperature caused by emission received by the antenna at different frequencies, the emission data from the Parkes Galactic All-Sky Survey (GASS) were used. The GASS emission was averaged across a 7-pixel (33 arcmin) radius annulus centered on the source position, excluding the central 1 pixel. The $1\sigma$ noise profile for the GASKAP-HI absorption spectrum was estimated by
\begin{equation}
    \sigma_\tau(v)=\sigma_{\mathrm{cont}} \frac{T_{\mathrm{sys}}+\eta_{\mathrm{ant}} T_{\mathrm{em}}(v)}{T_{\mathrm{sys}}},
\end{equation}
where $T_{\mathrm{sys}}=50$ K is the system temperature and $\eta_{\mathrm{ant}}=0.67$ is the antenna efficiency \citep{Hotan2021}, $\sigma_{\mathrm{cont}}$ is the standard deviation of the line-free region of the spectrum, and $T_{\mathrm{em}}(v)$ is the mean brightness temperature as measured in GASS.

\begin{figure}
    \centering
    \includegraphics[width=0.5\textwidth]{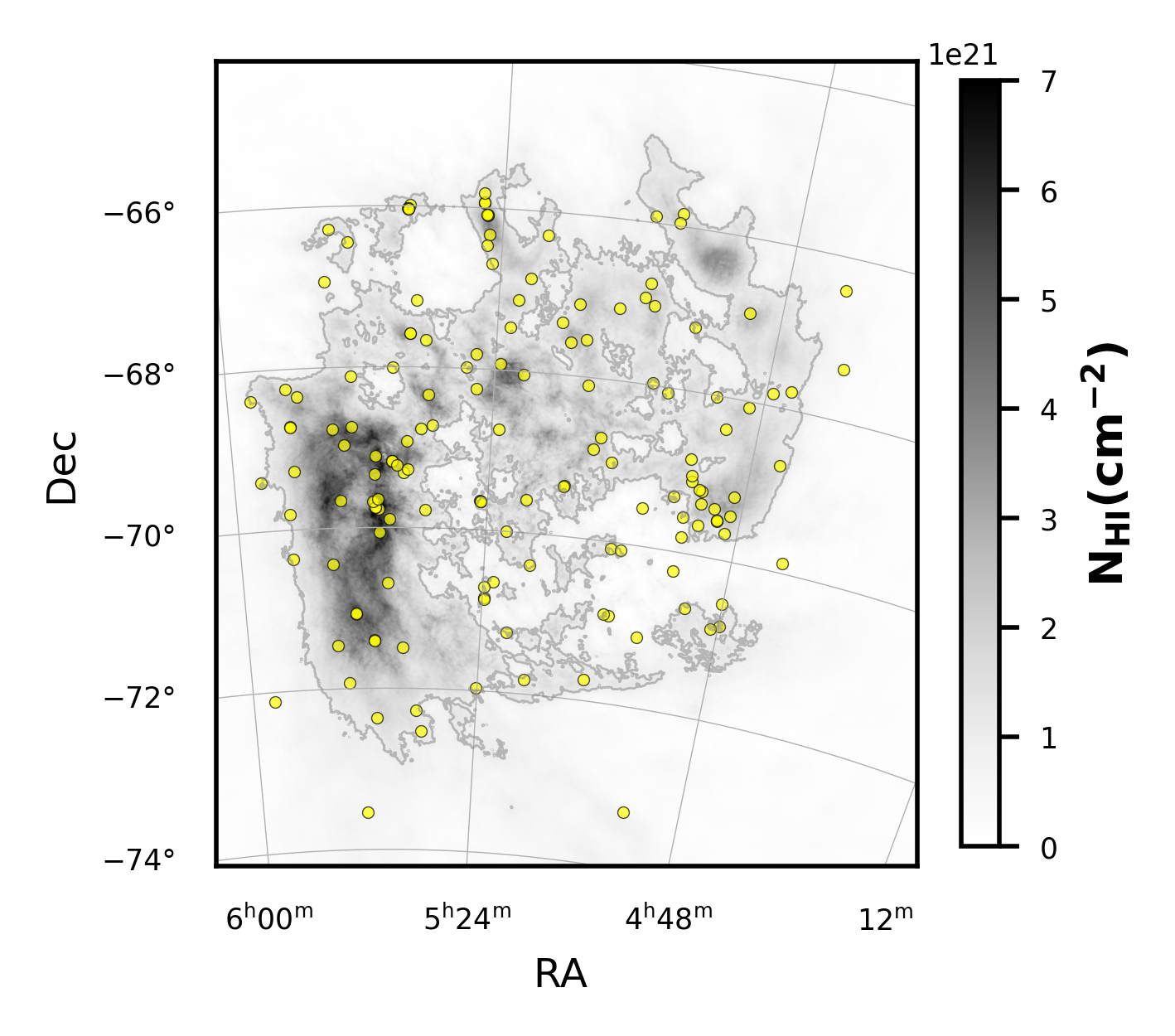}
    \caption{Selected background radio continuum sources for the LMC. The HI column density images are from the  GASKAP-HI phase II pilot survey with contours indicating $1\times10^{21} ~\text{cm}^{-2}$. Our sources are colored as yellow points. Note that some sources may overlap. The GASKAP-HI emission data are taken from archived data obtained through the CSIRO ASKAP Science Data Archive, CASDA (\url{https://research.csiro.au/casda}).}
    \label{fig:Map_source}
\end{figure}

Absorption features were identified in the absorption spectra by one channel of $\geq 3\sigma$ and an adjacent channel of $\geq 2.8\sigma$ noise level in optical depth. 
\cite{Dempsey2026} identified 1,365 continuum sources toward the five fields in the LMC, of which 205 show detected HI absorption within the LMC velocity range.
Restricting the sample to sightlines within the main LMC region (defined by $N_{\rm HI} > 5\times10^{20}~\mathrm{cm^{-2}}$) and with sufficient absorption sensitivity of $e^{-\tau}$ ($\sigma_{\tau} < 0.2$), yields 438 sources, of which 192 show detected HI absorption. This corresponds to a detection rate of 44\%. This rate is lower than those reported in previous HI absorption surveys of the LMC, including 63\% in \citet{Dickey1994}, 45\% in \citet{Marx-Zimmer2000}, and 79\% in \citet{Liu2021}. However, those earlier surveys covered smaller areas and were often biased toward dense regions or sightlines more likely to show absorption. In contrast, GASKAP provides the first large, relatively unbiased HI absorption survey across the LMC, including many sightlines through lower density or more diffuse environments. The lower detection rate is therefore expected. Despite this, GASKAP provides the largest number of HI absorption detections in the LMC to date.
For comparison, the HI absorption detection rate in the SMC is 73\% in the GASKAP pilot II survey \cite{Dempsey2026}, while Milky Way surveys can reach higher detection rates, such as the 88\% reported by \citet{Murray2018}.

After a manual inspection of each spectrum, we excluded sources with poor absorption data quality, strong saturation in absorption (only two sources; i.e., channels where the measured absorption spectra $e^{-\tau}$ becomes negative because of noise or calibration residuals, leading to non-physical values of $\tau$), or duplicate recordings (i.e., sources separated by less than one beam size). This left a final sample of 155 sources for analysis. Their locations are shown in Figure~\ref{fig:Map_source}, overlaid on the HI column density map. Among these, 25 sources have semi-major axes larger than the absorption beam and are classified as extended sources, while the remaining 130 are treated as point sources. 

The median optical depth (in $e^{-\tau}$) noise level for these 155 sources is $\sigma_{\tau}=0.035$ with a range of $0.002\leq \sigma_\tau \leq 0.175$. 
In comparison, earlier HI absorption studies in the LMC reported substantially higher noise levels. \citet{Dickey1994} found a median $\sigma_{\tau}=0.08$ with a range of $0.02 \leq \sigma_{\tau} \leq 0.23$, while \citet{Marx-Zimmer2000} reported $\sigma_{\tau}=0.16$ with a range of $0.10 \leq \sigma_{\tau} \leq 0.25$. More recently, \citet{Liu2021} obtained a median $\sigma_{\tau}=0.08$, spanning $0.006 \leq \sigma_{\tau} \leq 0.21$. The GASKAP survey provides a substantial improvement in optical depth sensitivity. The velocity resolution for the previous three surveys are all $\sim$ 1.6 $\rm km ~ s^{-1}$, while the GASKAP survey is 1 $\rm km ~ s^{-1}$.

\subsection{HI emission spectra}\label{sec:HIemi}
As described in \cite{Pingel2022}, the GASKAP-HI emission data cube was combined with single-dish HI observations from the Parkes Galactic All-Sky Survey \citep[GASS;][]{McClure-Griffiths2009} to cover emission across all angular scales, ensuring the recovery of diffuse emission filtered out by the interferometer. The final emission data cube of the LMC provided a composite dataset with $30''$ angular resolution, velocity resolution of $1.0~ \text{km}~ \text{s}^{–1}$, and sensitivity of 1.1 K per channel. 


For radiative transfer calculations, we require an HI emission profile that would be observed in the direction of background sources if the continuum sources were absent, the so-called ``expected” HI emission profile based on \cite{Heiles2003}.
We therefore calculate the HI emission spectrum for each source as the mean spectrum within an annulus centered on the source position, with an outer radius of $56''$ (8 pixels, $\sim 2 $ beam widths) and an inner exclusion radius of $28''$ (4 pixels, $\sim 1 $ beam widths) to avoid contamination from the central absorption. 

We derive the 1$\sigma_{\rm T_B}$ noise level from the standard deviation of emission spectra within the annulus. The typical maximum $\sigma_{T_B}$ is $\sim5$ K near regions of peak emission, while 27 sources exhibit maximum $\sigma_{T_B} > 10$ K. These high-noise sightlines are predominantly located near active star-forming regions, where the HI spectra are often complex and structured. We note that the same $30''$ beamwidth corresponds to sub-pc scales in the Milky Way, but to $\sim$ 7 pc in the LMC because of its greater distance. As a result, each beam in the LMC encompasses a larger number of  structures. In dense environments with clumpy or filamentary CNM/WNM structures, the gas may not uniformly fill the beam, leading to increased spectral complexity and larger uncertainties. The impact of worse physical resolution was mentioned in \cite[][hereafter Paper~I]{Chen2025} and is further discussed in Section~\ref{sec:beam_dilu}.
\defcitealias{Chen2025}{Paper~I}


\section{Decomposing individual CNM/WNM structures}\label{sec:decomp}
We decompose the HI emission and absorption spectra into individual CNM and WNM components using the radiative transfer framework developed by \citet{Heiles2003} and implemented in \citetalias{Chen2025}. In this approach, the absorption spectrum constrains the CNM optical depth, while the emission spectrum contains contributions from both CNM and WNM, together with contribution from the diffuse radio continuum emission. The emission spectrum (expected brightness temperature) is modeled as a combination of CNM and WNM Gaussian components under radiative transfer, accounting for the relative ordering of CNM clouds and the fractional location of WNM components (F value) along the line of sight. The number of components and the optimal model are determined using the Bayesian Information Criterion (BIC), which balances goodness of fit and model complexity. 
We allow small velocity offsets between emission and absorption components to account for local kinematic variations and finite spectral resolution. We adopt a tolerance of maximum $\pm 4~\mathrm{km~s^{-1}}$, which is comparable to the typical linewidths observed in molecular clouds in the LMC \citep{Hughes2010,Wong2019,Green2024}. Such offsets can arise from small-scale turbulence, cloud-to-cloud motions within the beam, or slight differences in the effective volume sampled by emission and absorption measurements. This allowance ensures that physically associated CNM structures are not artificially separated due to minor centroid shifts.
Full mathematical expressions, fitting procedures, and parameter exploration are provided in Appendix~\ref{sec:method_app}.

\subsection{Fitting results}\label{sec:result}


\begin{figure}
    \centering
    \includegraphics[width=0.46\textwidth]{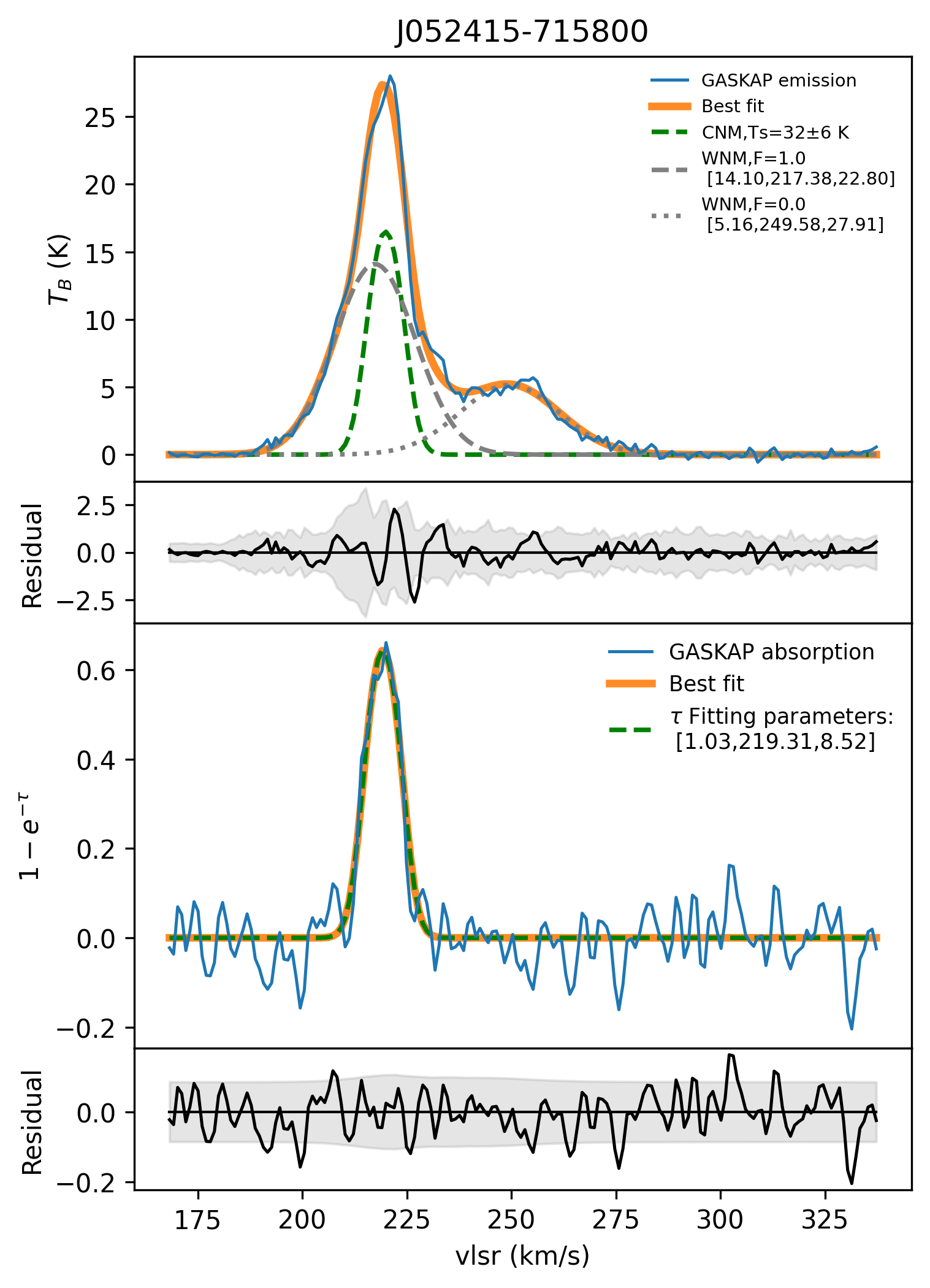}
    \caption{Gaussian decomposition for Source J052415-715800. The top panel shows the HI emission spectrum and its fitting residuals with shaded area presenting $\sigma_{\rm T_B}$ noise of the emission spectrum. The bottom panel presents the HI optical depth (absorption spectrum) and its fitting residuals with shaded area presenting $\sigma_{\tau}$ noise of the absorption spectrum. The total fits for both the emission and absorption spectra are shown in orange, while the individual CNM components are represented in green, and the individual WNM components are shown in gray. Following the 'CNM' label, the weighted mean spin temperature is presented. The 'F' value following each 'WNM' label quantifies the fraction of WNM in front of CNM, as stated in the Appendix~\ref{sec:method_app}. The three values in every square brackets in the top and bottom panels are, respectively, the amplitude, the central velocity, and the FWHM of each fitted Gaussian component.  All spectral fits are presented in \url{https://retarchen.github.io/fitting_results_LMC/spectra.html}.}
    \label{fig:spectra_single}
\end{figure}

Figure~\ref{fig:spectra_single} presents an example of HI absorption and emission spectra along with their corresponding Gaussian decomposition into CNM and WNM components. Across the entire sample, the HI absorption spectra contain between one and eight CNM Gaussian components, while the HI emission spectra include between one and five WNM components.

Additionally, we calculate the column density for each CNM and WNM component. For the CNM we use:
\begin{equation}
N_{\text{HI, CNM, }comp}=1.823\times 10^{18} \ T_s \int \tau(v) \ \mathrm{d} v \quad\mathrm{cm}^{-2},
\end{equation}
where $T_s$ and $\tau(v)$ are the spin temperature and optical depth of each component. We assume that WNM components are optically thin and calculate the column density using:
\begin{equation}
N_{\text{HI, WNM, }comp}=1.823\times 10^{18} \int T_{B,\text{WNM}}(v) \ \mathrm{d} v \quad\mathrm{cm}^{-2},
\end{equation}
where $T_{B,\text{WNM}}(v)$ is the brightness temperature. The total CNM/WNM column density for each LOS is the sum of all CNM/WNM components, referred to as $N_{\text{HI, CNM, }all}$ and $N_{\text{HI, WNM, }all}$. 

We estimate the CNM fraction along each line-of-sight by:
\begin{equation}
    f_{\text{CNM}}=\frac{N_{\text{HI, CNM, }all}}{N_{\text{HI, WNM, }all}+N_{\text{HI, CNM, }all}}.
\end{equation}

We also calculate the maximum kinetic temperature for each component via:
\begin{equation}\label{eq:Tk}
T_{k, \max }=\frac{m_{\mathrm{H}}}{8 k_{\mathrm{B}} \ln 2} \Delta v_0^2=21.866 \cdot \Delta v_0^2,
\end{equation}
where  $\Delta v_0$ is the linewidth (FWHM) for each component (see Appendix~\ref{sec:method_app} for more details), $m_{\mathrm{H}}$ is the hydrogen mass and  $k_{\mathrm{B}}$ is the Boltzmann constant \citep{Draine2011}.

We include three data products in this paper: the fitting results for CNM/UNM components (Table~\ref{tab:CNM}), WNM components (Table~\ref{tab:WNM}), and the combined line-of-sight results (Table~\ref{tab:full})\footnote{These tables can be downloaded at \url{https://retarchen.github.io/fitting_results_LMC/index.html}}. 
Table~\ref{tab:CNM} provides the properties for each CNM component.
Table~\ref{tab:WNM} lists 
properties for each WNM component.
Table~\ref{tab:full} summarizes the line-of-sight results.


\begin{table*}[htbp]
\centering
\begin{tabular}{lllllllll}
\toprule
Name & $\tau$ & $v_\tau$ ($\rm km ~ s^{-1}$) & $\Delta v_0$ ($\rm km ~ s^{-1}$) & $ T_s$ (K) & $T_{k,\max}$ (K) & $N_{\rm HI,c}$ ($10^{20}$ cm$^{-2}$) & Order & $v_{\rm shift}$ \\
\midrule
J043855-672153 & $0.08 \pm 0.00$ & $246.97 \pm 0.12$ & $3.70 \pm 0.29$ & $23 \pm 4$ & 300 & $0.14 \pm 0.03$ & 1 & 3.57 \\
J043855-672153 & $0.11 \pm 0.00$ & $241.01 \pm 0.11$ & $4.87 \pm 0.28$ & $31 \pm 4$ & 519 & $0.32 \pm 0.04$ & 0 & 4.00 \\
J044047-695217 & $0.05 \pm 0.01$ & $239.54 \pm 0.32$ & $4.38 \pm 0.75$ & $84 \pm 3$ & 420 & $0.38 \pm 0.09$ & 0 & 2.70 \\
J044056-662423 & $0.31 \pm 0.00$ & $231.12 \pm 0.03$ & $3.94 \pm 0.06$ & $3 \pm 0$ & 340 & $0.07 \pm 0.01$ & 0 & -1.50 \\
J044415-684211 & $0.69 \pm 0.02$ & $241.33 \pm 0.08$ & $6.59 \pm 0.19$ & $56 \pm 9$ & 950 & $4.97 \pm 0.86$ & 0 & 0.88 \\
J044451-674636 & $0.23 \pm 0.02$ & $256.02 \pm 0.11$ & $2.75 \pm 0.27$ & $34 \pm 9$ & 165 & $0.41 \pm 0.12$ & 0 & 0.11 \\
J044710-675049 & $0.91 \pm 0.04$ & $261.85 \pm 0.18$ & $7.68 \pm 0.43$ & $28 \pm 4$ & 1291 & $3.79 \pm 0.60$ & 0 & 2.61 \\
J044746-704843 & $0.23 \pm 0.02$ & $245.48 \pm 0.23$ & $4.43 \pm 0.55$ & $27 \pm 4$ & 430 & $0.52 \pm 0.12$ & 0 & 2.39 \\
J044809-703144 & $0.29 \pm 0.05$ & $232.35 \pm 0.20$ & $2.50 \pm 0.47$ & $9 \pm 2$ & 136 & $0.12 \pm 0.04$ & 0 & -1.12 \\
J044902-705155 & $0.12 \pm 0.02$ & $234.58 \pm 0.24$ & $3.53 \pm 0.57$ & $56 \pm 7$ & 273 & $0.44 \pm 0.11$ & 0 & 0.23 \\
\bottomrule
\end{tabular}%
\caption{First 10 rows of properties for CNM or UNM after  fitting. From left to right, each column shows the source name, peak optical depth, central velocity, linewidth (FWHM), spin temperature, maximum kinetic temperature, HI column density, the component order (starting from 0) and the velocity shift when the absorption components fit to the emission.}
\label{tab:CNM}
\end{table*}

\begin{table*}[htbp]
\centering
\begin{tabular}{lllllll}
\toprule
Name & $T_{B,\rm WNM}$(K) & $v$ ($\rm km ~ s^{-1}$) & $\Delta v_0$ ($\rm km ~ s^{-1}$) & $T_{k,\max}$ (K) & $N_{\rm HI,w}$ ($10^{20}$ cm$^{-2}$) & F \\
\midrule
J043855-672153 & $9.33 \pm 0.19$ & $249.88 \pm 0.14$ & $23.50 \pm 0.36$ & 12080 & $4.26 \pm 0.11$ & 1.0 \\
J044056-662423 & $3.55 \pm 0.06$ & $249.78 \pm 0.34$ & $40.49 \pm 0.76$ & 35857 & $2.79 \pm 0.07$ & 0.0 \\
J044415-684211 & $18.52 \pm 0.32$ & $244.81 \pm 0.11$ & $29.35 \pm 0.26$ & 18838 & $10.55 \pm 0.21$ & 0.0 \\
J044415-684211 & $13.17 \pm 1.09$ & $240.64 \pm 0.25$ & $9.56 \pm 0.58$ & 1997 & $2.44 \pm 0.25$ & 0.0 \\
J044451-674636 & $4.45 \pm 0.40$ & $266.71 \pm 0.24$ & $6.81 \pm 0.76$ & 1014 & $0.59 \pm 0.08$ & 0.5 \\
J044451-674636 & $19.64 \pm 0.22$ & $254.62 \pm 0.19$ & $28.13 \pm 0.30$ & 17297 & $10.72 \pm 0.17$ & 0.0 \\
J044710-675049 & $11.44 \pm 0.12$ & $234.73 \pm 0.18$ & $19.53 \pm 0.41$ & 8336 & $4.34 \pm 0.10$ & 0.0 \\
J044710-675049 & $16.36 \pm 1.96$ & $254.41 \pm 0.49$ & $11.92 \pm 0.65$ & 3108 & $3.79 \pm 0.50$ & 0.5 \\
J044710-675049 & $7.99 \pm 1.56$ & $267.76 \pm 2.49$ & $17.24 \pm 2.45$ & 6499 & $2.67 \pm 0.65$ & 1.0 \\
J044746-704843 & $3.07 \pm 0.55$ & $252.08 \pm 4.51$ & $34.14 \pm 4.79$ & 25483 & $2.03 \pm 0.46$ & 1.0 \\
\bottomrule
\end{tabular}%
\caption{First 10 rows of WNM properties after fitting. From left to right, each column shows the source name, peak brightness temperature, central velocity, linewidth (FWHM), maximum kinetic temperature, HI column density, and the F value for each WNM component.} 
\label{tab:WNM}
\end{table*}

\begin{table*}[htbp]
\centering
\scriptsize
\setlength{\tabcolsep}{2pt}
\renewcommand{\arraystretch}{1.4}
\begin{tabular}{cccccccccc}
\toprule
Name & RA (deg) & Dec (deg) &
$N_{\rm HI,c}$ ($10^{20}$ cm$^{-2}$) &
$N_{\rm HI,w}$ ($10^{20}$ cm$^{-2}$) &
$N_{\rm HI,u}$ ($10^{20}$ cm$^{-2}$) &
$f_{\rm CNM}$ & $T_{\rm sky}$ (K) & Note & LMC/x \\
\midrule
J043855-672153 & 69.73 & $-$67.36 & $0.46 \pm 0.05$ & $4.26 \pm 0.11$ & $0.00 \pm 0.00$ & $0.10 \pm 0.01$ & 2.73 & galaxy & x \\
J044047-695217 & 70.20 & $-$69.87 & $0.38 \pm 0.09$ & $0.00 \pm 0.00$ & $0.00 \pm 0.00$ & $1.00 \pm 0.00$ & 2.73 & QSO & x \\
J044056-662423 & 70.24 & $-$66.41 & $0.07 \pm 0.01$ & $2.79 \pm 0.07$ & $0.00 \pm 0.00$ & $0.02 \pm 0.00$ & 2.73 & ? & (x) \\
J044415-684211 & 71.07 & $-$68.70 & $4.97 \pm 0.86$ & $12.99 \pm 0.33$ & $0.00 \pm 0.00$ & $0.28 \pm 0.03$ & 2.73 & \makecell[l]{QSO (z=2.54)\\at 4"} & x \\
J044451-674636 & 71.21 & $-$67.78 & $0.41 \pm 0.12$ & $11.31 \pm 0.19$ & $0.00 \pm 0.00$ & $0.04 \pm 0.01$ & 2.73 & ? & (x) \\
J044710-675049 & 71.79 & $-$67.85 & $3.79 \pm 0.60$ & $10.79 \pm 0.82$ & $0.00 \pm 0.00$ & $0.26 \pm 0.03$ & 2.73 & ? & (x) \\
J044746-704843 & 71.95 & $-$70.81 & $0.52 \pm 0.12$ & $9.50 \pm 0.63$ & $0.00 \pm 0.00$ & $0.05 \pm 0.01$ & 2.73 & ? & (x) \\
J044809-703144 & 72.04 & $-$70.53 & $0.12 \pm 0.04$ & $6.66 \pm 0.58$ & $0.00 \pm 0.00$ & $0.02 \pm 0.01$ & 3.39 & ? & (x) \\
J044902-705155 & 72.26 & $-$70.87 & $0.57 \pm 0.13$ & $9.93 \pm 0.24$ & $0.00 \pm 0.00$ & $0.05 \pm 0.01$ & 7.19 & ? & (x) \\
J044926-691203 & 72.36 & $-$69.20 & $14.94 \pm 5.27$ & $19.45 \pm 0.20$ & $0.00 \pm 0.00$ & $0.43 \pm 0.09$ & 3.09 & \makecell[l]{IRAS 04496-6917\\= YSO? at 5"} & LMC \\
\bottomrule
\end{tabular}
\caption{First 10 rows of full LOS dataset. From left to right, it shows the source name, RA and DEC coordinates, total CNM column density, total WNM column density, CNM fraction and background $T_{\text{sky}}$. The ``Note" column lists potential counterparts or associated structures identified by manual inspection using SIMBAD and Aladin, based on positional coincidence with known objects such as QSOs, stars, infrared sources, H II regions, or supernova remnants. Entries marked with``?" indicate cases where no secure identification could be established, although many of these sources are likely extragalactic. The final column ``LMC/x" indicates whether the background continuum source is inferred to lie within the LMC (``LMC") or is a background source (``x"). Parentheses denote cases where the classification is likely but not fully confirmed. Question marks in this column highlight sources projected onto extended LMC structures, such as supernova remnants, where it is difficult to distinguish between an embedded LMC source and a background object. 
}
\label{tab:full}
\end{table*}

\subsection{Fitting results limitations}\label{sec:beam_dilu}



\subsubsection{Resolution effects in the emission and absorption spectra} \label{sec:resolution}

As discussed in \citetalias{Chen2025}, the finite resolution of a telescope can affect the HI emission and absorption spectra. In the LMC, one GASKAP beam ($30''$) subtends $\sim$7 pc, whereas the same survey towards the Milky Way \citep{Nguyen2024} corresponds to sub-parsec scales. This could affect the fitting results in two ways.

First, unresolved spatial and velocity blending can bias the recovered component. If multiple CNM or WNM clouds lie within a single beam and have similar velocities, their spectral profile may overlap and be recovered as a single broader component rather than as several distinct ones. This can reduce the number of fitted components and broaden the inferred linewidth of the recovered component.

Second, there can exist a mismatch between the HI emission and absorption measurements. The absorption spectra toward background point sources probe a very narrow line of sight, whereas the derived HI emission spectra represent beam-averaged gas over a larger surrounding region. As a result, the emission and absorption spectra do not necessarily trace the same CNM clouds. In particular, compact CNM structures identified in absorption may have their emission brightness underestimated because their emission is diluted within the larger beam. This can make narrow CNM emission features harder to detect and more difficult to associate with the corresponding absorption components. As pointed out by \cite{Murray2018}, some CNM absorption components may therefore appear to have unphysically low spin temperatures ($T_s \lesssim 10$ K) because their associated emission is diluted or absent. In our sample, 9 out of 330 CNM components have $T_s \lesssim 10$ K, which has only a minor effect on the overall statistics. In addition, because the HI emission spectra are calculated by excluding the central beam, narrow emission components detected in the surrounding annulus may be misclassified as WNM, even though they likely trace CNM that is spatially offset from the source center. In our case, the minimum line width of our WNM component is $\sim 6 ~\rm km ~ s^{-1}$, corresponding to a maximum kinetic temperature of $\sim 775$ K.  Future higher resolution HI emission observations will be essential to mitigate these effects and obtain more reliable measurements.

\subsubsection{Source location and size}

We detect HI absorption in the direction of 155 radio continuum sources which are a mixture of sources behind (mostly quasars and radio galaxies) and inside (H II regions, star-forming complexes, or supernova remnants) the LMC.
As we show in Appendix~\ref{sec:extend_point}, 22 sources are identified as internal to the LMC, 115 sources are classified as background objects, and 18 sources remain unclassified.
For radio sources inside the LMC, the absorption traces only the cold gas in front of the continuum source, whereas the corresponding HI emission spectrum includes gas along the entire line of sight. This geometric mismatch can, in principle, lead to an underestimation of the number of cold components and hence a lower inferred CNM fraction. To assess this effect, we compare the cold gas properties in the direction of sources classified as internal to the LMC with those of background sources in Appendix~\ref{sec:extend_point}. Although internal sources tend to show slightly higher spin temperatures and broader linewidths, as well as higher total column densities and CNM fractions, the two groups overall span similar ranges. These differences are likely driven by the fact that internal sources preferentially probe denser, more active environments, rather than by source location itself.
   
In addition to source location, continuum sources can be classified by angular size into point sources, which are smaller than the telescope beam, and extended sources, which are larger than the beam. For point sources (which are mostly the background continuum sources), the absorption spectrum is effectively a pencil-beam measurement and is not affected by beam dilution or smearing. In contrast, absorption against extended sources (which are mostly internal to the LMC) averages over multiple sightlines across the source, which can reduce the apparent optical depth through partial covering and velocity blending. In our sample, 11 out of 25 extended sources are also classified as internal to the LMC. As shown in Appendix~\ref{sec:extend_point}, the extended and point-like samples also show slightly different distributions. As with the internal-source sample, these trends are most naturally explained by the environments preferentially probed by extended sources, many of which are internal objects or lie near dense regions such as shells, rather than by source size alone.
We therefore conclude that neither source location nor angular size has a strong impact on our overall results, and that the differences between subsamples mainly reflect environmental selection effects. Future higher-resolution studies \citep[e.g. ][]{Park2026} combining HI absorption and emission cubes around extended continuum sources will be able to spatially resolve the absorbing structures and directly map the CNM distribution, enabling a more detailed assessment of small-scale structure of the CNM distribution and line-of-sight blending effects.

\subsubsection{Fitting and radiative transfer uncertainties}

In Table~\ref{tab:CNM} we provide uncertainties on fitted absorption and emission Gaussian parameters, as well as uncertainties for parameters such as spin temperature which is a result of radiative transfer modeling as well as varying geometric line of sight constraints. An obvious question is how well does our methodology (fitting and radiative transfer considerations) recover underlying physical properties of HI in the LMC. In external galaxies like the LMC, due to larger physical scales probed, this question is even more relevant than in the Milky Way.

\cite{Murray2017} used synthetic HI absorption and emission spectra to quantify how well a radiative transfer methodology similar to what we use here recovers true physical parameters. This study showed that the recovery depends on line-of-sight complexity. Sightlines at high Galactic latitudes were shown to have excellent recovery, while sightlines through the Milky Way plane had only $\sim50$\% completeness in terms of matching absorption peaks with simulated 3D structures. By extension, we would expect lower completeness for more complex LMC regions such as the 30 Doradus. The blending of individual HI components is also likely to be more prominent in the LMC relative to the Milky Way due to worse angular resolution. Future studies similar to that of \cite{Murray2017} will be important to quantify completeness.

\citet{Murray2017} also compared the spin temperature derived from Gaussian decomposition and radiative transfer with intrinsic gas temperatures in high-resolution simulations and found that the inferred temperatures and column densities agree with the true values within a factor of two for most gas structures. However, when they compared the spin temperature
for aggregated distributions (e.g., histograms, medians), the uncertainties were lower due to the averaging of many samples, resulting in a reasonable comparison to simulated distributions over a wide range of temperatures. 
We expect the same to apply for our LMC study, while individual $T_s$ measurements carry significant uncertainty, the statistical distributions and population-level properties (e.g., histograms, medians) are more robust indicators of the underlying physical conditions. Because the primary results of this LMC investigation rely on aggregated distributions of cold gas properties rather than on the precision of single component measurements, we expect our conclusions on population trends to be reliable.

\section{Individual component properties}\label{sec:comp_result}

We decompose a total of 155 sightlines across the LMC. Based on the HI absorption fits, we classify CNM components as those with spin temperatures $T_s < 250$ K, and UNM components as those with spin temperatures larger than 250 K. 
This temperature classification is based on Milky Way conditions. 
LMC's lower metallicity and a stronger UV radiation field 
shift the CNM--WNM phase balance in the same general direction, making the survival of cold gas more difficult and moving the two-phase equilibrium toward lower temperatures \citep{Dickey2000,Stelea2025} and higher densities/pressures \citep[e.g.,][]{Wolfire1995,Wolfire2003}. 
Therefore, our UNM classification is likely slightly conservative.
Using this criterion, we identify 330 CNM components and 2 UNM components.
From the HI emission fits, all Gaussian components that are not associated with detected HI absorption exhibit maximum kinetic temperatures $T_{k,\mathrm{max}} > 10^3$ K. We therefore classify these components as WNM and identify a total of 310 WNM components.


In this section, we analyze the physical properties of the CNM components derived from the HI absorption spectra. The two UNM components are discussed separately in Section~\ref{sec:unm}.
For comparison, we include two Galactic HI absorption studies, the BIGHICAT meta-catalog \citep{McClure-Griffiths2023} and the GASKAP-HI Magellanic Clouds Foreground Survey \citep[hereafter GASKAP-foreground][]{Nguyen2024}. At the time of publication, BIGHICAT combined publicly available Galactic HI absorption data from the most sensitive surveys to date ($\sigma_{\tau_{\mathrm{HI}}}\lesssim0.01$). Most of its sightlines lie away from the Galactic plane, since HI spectra in the plane are substantially more complex.
Following the same CNM selection criterion of $T_s < 250$ K adopted by \cite{Murray2018}, we select 766 CNM components from the BIGHICAT sample for comparison with our LMC data.
As for the GASKAP-foreground, only sightlines in the direction of the Magellanic Clouds ($b = -45^\circ$ to $-27^\circ$, $l = 270^\circ$–$305^\circ$) were targeted. This survey identified 691 CNM components. Although it covers only a limited spatial area at high Galactic latitude, its sensitivity and spectral resolution are the same as those of our LMC dataset.

\begin{figure*}
    \centering
    \includegraphics[width=1\textwidth]{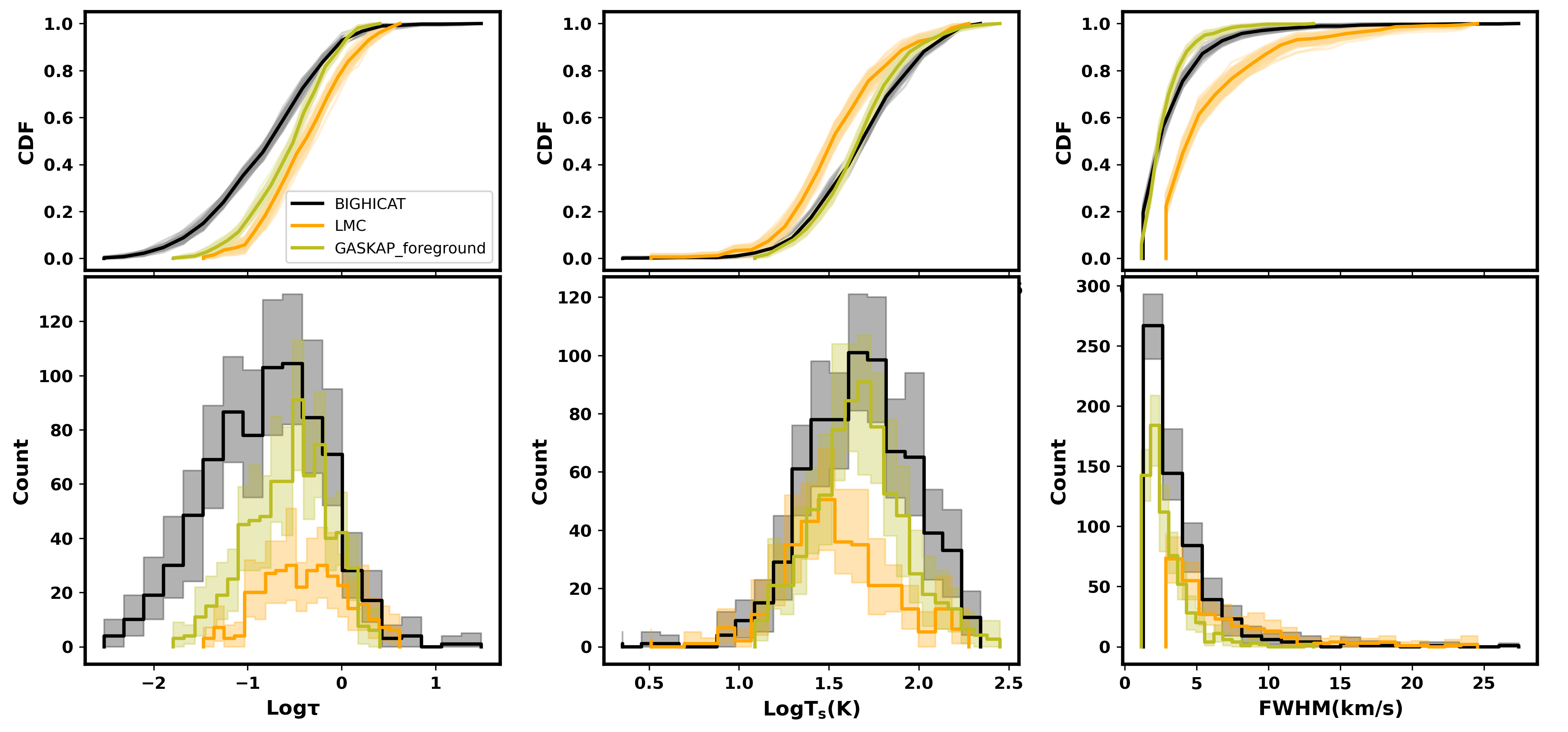}
    \caption{Cumulative distribution functions (CDFs, upper row) and number distribution (Count,bottom row) of individual CNM component properties are shown from left to right: peak optical depth, spin temperature, and line width. The orange lines represent our LMC results from the GASKAP-HI survey, while the Milky Way results from the BIGHICAT catalog \citep{McClure-Griffiths2023} (black) and GASKAP-HI  Magellanic Clouds foreground survey \citep{Nguyen2024} (olive) are included for comparison. We perform 100 bootstrap trials, the shaded regions indicate all trails, and center solid curves represent the median distribution for each bin. }
    \label{fig:cdf_comp}
\end{figure*}

\begin{figure*}
    \centering
    \includegraphics[width=\textwidth]{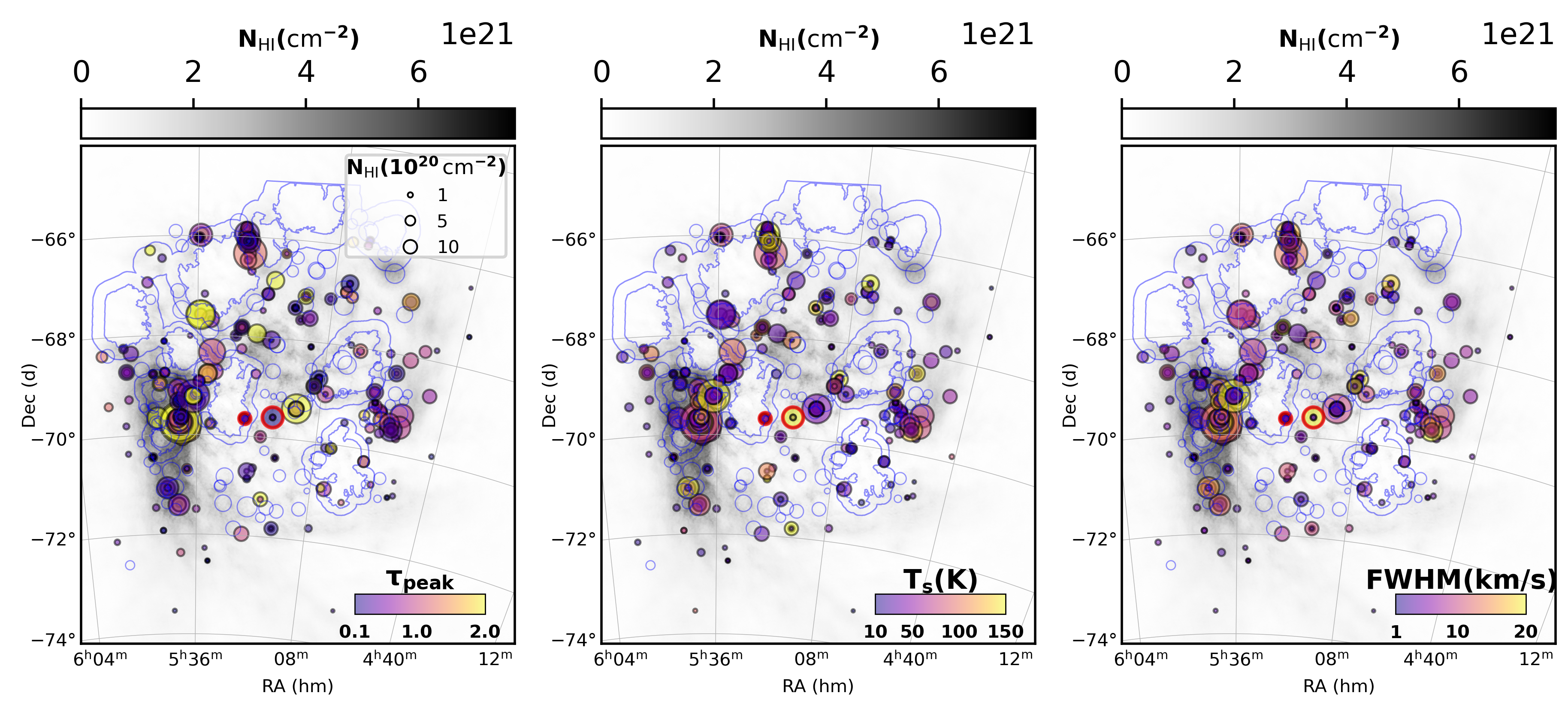}
    \caption{Properties of all components detected in absorption overplotted on the HI column density map of the LMC. From left to right, the color of each circle represents the optical depth, spin temperature ($T_s$), and line width of individual components. The size of each circle indicates the corresponding HI column density. The close two red dots in the center of the LMC highlight the position of the two UNM components. Large contours mark the locations of supergiant shells identified by \cite{Dawson2013}, while small open circles denote giant shells from \cite{Kim1999}. 
    }
    \label{fig:spatial_comp}
\end{figure*}

\subsection{Optical depth}

In the left panel of Figure~\ref{fig:cdf_comp}, we show the cumulative distribution functions (CDFs; top) and number distribution (bottom) of the peak optical depth for all CNM/UNM components in the LMC, with BIGHICAT and the GASKAP-foreground survey included for comparison.
The peak optical depth in the LMC ranges from 0.03 to 4.72, with a median of 0.46$\pm$0.04 and a mean of 0.76$\pm$0.04. The uncertainties in the mean and median are obtained by 100 bootstrap resampling.

Relative to the BIGHICAT sample, cold clouds in the LMC generally exhibit higher optical depths. This difference is partly a selection bias, as BIGHICAT incorporates several Galactic surveys with higher optical depth sensitivity, enabling the detection of lower optical depth absorption features that are largely inaccessible to the GASKAP-HI survey.
When compared with the GASKAP-foreground survey, which has the same sensitivity to our LMC data, the LMC still shows systematically higher optical depths. This is likely because the GASKAP-foreground focuses on high-latitude Galactic regions that are intrinsically more diffuse, whereas for the LMC we sample many cold clouds  embedded in denser environments where stronger absorption is expected.

Figure~\ref{fig:spatial_comp} presents the spatial distribution of CNM peak optical depth across the LMC. To further examine possible large-scale trends, Figure~\ref{fig:radius_comp} in Appendix~\ref{sec:radial} shows the radial variation of optical depth as a function of projected distance from the LMC kinematic center \citep[adopting the center from][]{Kim1998}. Although the scatter is substantial, we observe a mild decline in optical depth with increasing distance from the 30 Doradus region, particularly toward the outermost radii. 

\subsection{Spin temperature}\label{sec:temperature}
The spin temperature of the CNM is typically comparable to its kinetic temperature because the higher densities in this phase allow collisions to efficiently thermalize the HI hyperfine levels. 
In contrast, in the low-density WNM collisional coupling is weak. Radiative coupling via Ly$\alpha$ scattering (the Wouthuysen–Field effect) can in principle couple the spin temperature to the kinetic temperature, but its effectiveness depends on the local Ly$\alpha$ photon density. In regions where the radiation field is modest, the combined collisional and radiative coupling is insufficient to fully thermalize the hyperfine levels, and the spin temperature can be much lower than the kinetic temperature \citep[e.g.][]{Liszt2001, Wolfire2003}.
Given the sensitivity of our data, the majority of absorption-detected components correspond to the CNM. By contrast, Milky Way studies such as 21-SPONGE \citep{Murray2015, Murray2018} achieved much higher sensitivity and were able to detect not only the CNM but also the UNM and even WNM components in absorption.

The middle panel of Figure~\ref{fig:cdf_comp} shows the CDF and number distributions of the spin temperature for cold clouds in the LMC, compared with the Milky Way studies.
The CNM components in the LMC span spin temperatures of $\sim$10-200 K, with a median of $37 \pm 2$ K and a mean of $50 \pm 2$ K. The uncertainties are estimated via bootstrap resampling. Our results are in agreement with previous studies, e.g. \cite{Marx-Zimmer2000,Liu2021}.

Overall, CNM in the LMC exhibits lower spin temperatures compared to the two Milky Way samples. Even though the GASKAP-foreground survey has the same optical depth sensitivity to the LMC sample, the spin temperature distribution of the LMC is still shifted toward smaller values. This difference persists even when the comparison is restricted to a common HI column density range of $10^{19}$ to $6\times10^{20},\mathrm{cm^{-2}}$. Under this selection, both the BIGHICAT and GASKAP foreground samples, which primarily probe relatively diffuse regions in the outskirts of the Milky Way, still show higher spin temperatures than those observed in the LMC, despite our survey covering the entire galaxy.

This suggests that CNM clouds in the LMC are colder, consistent with expectations for lower metallicity and higher UV radiation environments. In such environments, the reduced dust abundance tends to lower the photoelectric heating efficiency, while the stronger interstellar radiation field can compensate for this effect, resulting in a photoelectric heating rate comparable to, or even higher than, that in the Milky Way. On the cooling side, the reduced abundance of metals, particularly carbon, leads to less efficient fine-structure cooling (e.g., [C II]). At the same time, the CNM–WNM phase equilibrium is shifted toward higher pressures and densities \citep[e.g.,][]{Wolfire2003,Bialy2019}. Under these conditions, CNM in the LMC is expected to exist at higher densities than in the Milky Way and cooling becomes more effective overall, which can lead to systematically lower spin temperatures despite the comparable or enhanced heating rates.

The second panel of Figure \ref{fig:radius_comp} in Appendix~\ref{sec:radial} presents spin temperature as a function of galactocentric distance. We do not find a clear radial trend in $T_s$. A similar conclusion is reached when using the column-density–weighted mean spin temperature, $\langle T_s \rangle$, rather than the individually fitted $T_s$ values (see Appendix~\ref{sec:mean_Ts}). The absence of a strong radial dependence is consistent with results in the Milky Way, where \citet{Dickey2009} found that $\langle T_s \rangle$ remains approximately constant ($\sim$250–400 K) out to $\sim$40 kpc \citep[see also][]{Dickey2022}. Likewise, Galactic simulations by \citet{Smith2023} show that the mass-weighted mean CNM temperature remains relatively flat with galactocentric radius.  

However, the spatial distribution shown in the middle panel of Figure~\ref{fig:spatial_comp} reveals localized increases in spin temperature, particularly in the supergiant shell regions surrounding 30 Doradus. This is consistent with molecular gas studies, which find enhanced CO excitation and higher inferred gas temperatures near 30 Doradus, attributed to strong stellar feedback \citep[e.g.,][]{Lee2019}. At the same time, the overall spin temperature distribution shows large scatter and no clear systematic trend with the ambient UV radiation field (not shown here). In addition, the LMC does not exhibit a significant global radial metallicity gradient in either the ISM \citep{Pagel1978,Cipriano2017} or young stellar populations \citep{Grocholski2006,Cioni2009}. These temperature enhancements are unlikely to be driven by large-scale metallicity or UV radiation variations. Instead, we attribute them to local environmental effects like shells, as discussed in Section~\ref{sec:shell}. Sources located near shells exhibit systematically higher spin temperatures compared to those in more quiescent environments. This indicates that the spatial variation of the CNM temperature in the LMC is governed primarily by local environmental processes, such as stellar feedback and shell compression, rather than by global radial structure. 
\subsection{Linewidths}
\begin{figure*}
    \centering
    \begin{subfigure}[b]{0.35\textwidth}
        \includegraphics[width=\linewidth]{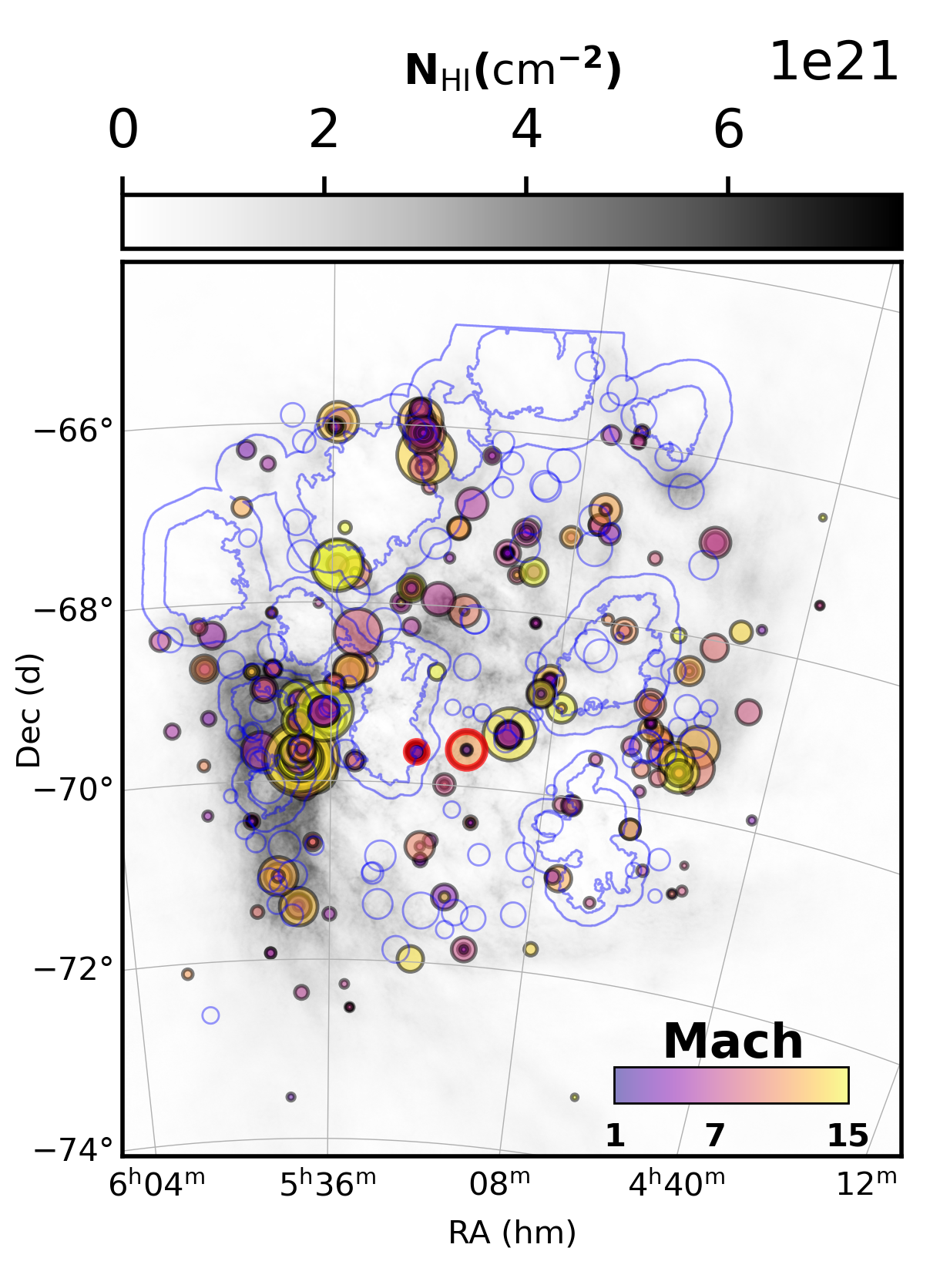}
    \end{subfigure}
    \hspace{3mm}
    \begin{subfigure}[b]{0.3\textwidth}
        \includegraphics[width=\linewidth]{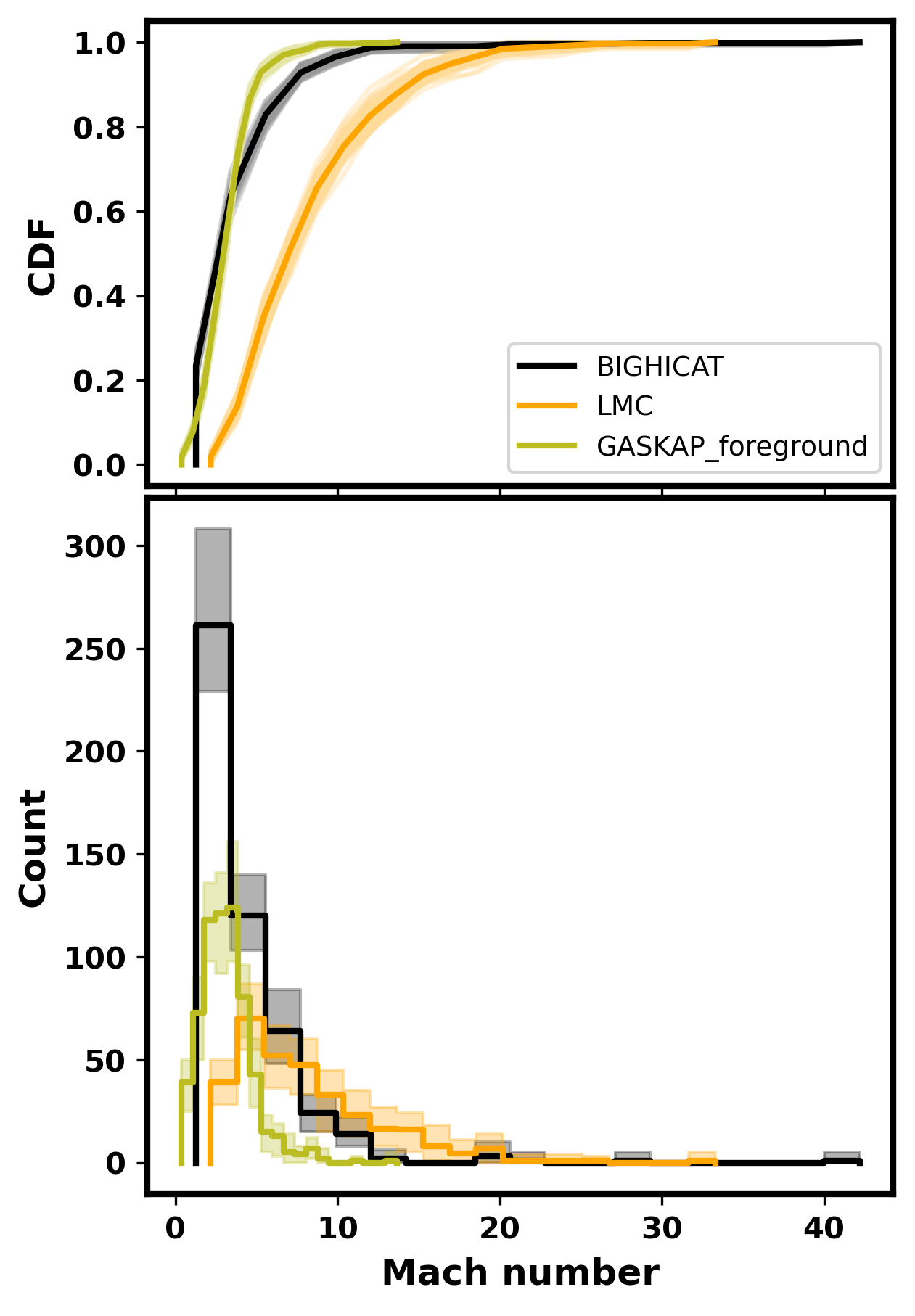}
    \end{subfigure}
    \hfill
    
    \caption{\textit{Left:} Mach number of the CNM overplotted on the HI column density map of the LMC. 
    \textit{Right:} Cumulative distribution functions (CDFs) of mach number for each individual CNM component. The orange lines represent our LMC results from the GASKAP-HI survey, while the Milky Way results from the BIGHICAT catalog \citep{McClure-Griffiths2023} and the GASKAP-foreground \citep{Nguyen2024} are included for comparison. 
    For CDF/Count, we perform 100 bootstrap trials and the shaded region include all trails.  The solid lines show the median values of the distributions.}
    \label{fig:cdf_mach}
\end{figure*}

The right panel in Figure~\ref{fig:cdf_comp} presents the distribution of the FWHM for the cold HI components.
The linewidth ranges from $\sim$2 $\mathrm{km~s^{-1}}$ to 25 $\mathrm{km~s^{-1}}$ with a median of 4.9$\pm$0.2 $\mathrm{km~s^{-1}}$ and a mean of 6.3$\pm$0.2 $\mathrm{km~s^{-1}}$. The uncertainties here are calculated via 100 bootstrap resampling.

The cold HI clouds in the LMC generally exhibit larger linewidths than those typically found in the Milky Way. The total linewidth of a CNM component reflects both thermal broadening -- which depends on the spin temperature ($T_s$) -- and non-thermal (turbulent) motions. As shown by \citet{Heiles2003b}, the non-thermal motions of the CNM can be quantified through the turbulent Mach number. By combining the observed linewidth (which sets the maximum kinetic temperature, $T_{k,\max}$) with the spin temperature $T_s$ (which reflects the thermal contribution), we can separate thermal and non-thermal broadening and derive the sonic Mach number \citep{Heiles2003b}:
\begin{equation}
M^2 = 4.2\left(\frac{T_{k,\max}}{T_s} - 1\right).
\end{equation}

The left panel of Figure~\ref{fig:cdf_mach} shows the spatial distribution of Mach number and the right panel of Figure~\ref{fig:cdf_mach} shows the CDF/Count of turbulent Mach numbers in the Milky Way and the LMC. In the LMC, all CNM clouds are supersonic ($M > 1$), with significantly higher Mach numbers than those in the Milky Way. This large Mach number suggests a more turbulent environment in the LMC.

The enhanced turbulence is consistent with expectations for low-metallicity galaxies, which tend to exhibit higher star formation rates and more pervasive stellar feedback -- both of which can inject turbulent energy into the ISM \citep[e.g.,][]{McKee2007, Tamburro2009,Stilp2013}. Supporting this, we find that $\sim$ 20\% of the CNM components in our sample have FWHM $>$ 10 $\mathrm{km~s^{-1}}$, with some of these broad components located near HII regions or supergiant shells (see Figure~\ref{fig:spatial_comp}). A notable example is the region around 30 Doradus, where many cold components exhibit significantly broader linewidths and larger Mach numbers (see the left panel of Figure~\ref{fig:cdf_mach}). 30 Doradus is a well-known massive star-forming region, and the observed HI line broadening is consistent with stellar feedback driven turbulence.

A similar result has been reported in IC 10, another low-metallicity \citep[$Z/Z_{\odot} \sim 0.27$,][]{Karachentsev2004,margini2009}, starburst dwarf galaxy. Using HI absorption measurements, \citet{Stelea2025} identified a broad CNM component with FWHM exceeding 10 km s$^{-1}$ located near shell structures, which they attributed to turbulence induced by stellar feedback and shell expansion. However, as noted by \citet{Stelea2025}, not all broad components can be uniquely associated with known shells or star forming regions, and narrower components are also observed toward HII regions. A similar situation is seen in our LMC samples and we further explore the impact of shells on the cold gas properties in Section~\ref{sec:shell}. 

Consequently, while stellar feedback and shell structures likely contribute to the increased linewidths observed in some regions, other factors may also play a role. In particular, limited spatial and spectral resolution can lead to blending of multiple unresolved CNM components along the same line of sight (as mentioned in Section~\ref{sec:resolution}). Under the current GASKAP-HI resolution in the LMC, distinct cold HI clouds may not be separable and can instead appear as a single broad spectral component, artificially increasing the measured FWHM and Mach number. This scenario is particularly plausible given that the cold HI structures in the Milky Way are known to be clumpy and filamentary  \citep{Clark2019, Murray2020, Lei2023}. 
In such cases, the measured linewidth/Mach number reflects not only internal turbulence within each cloud but also bulk motions between unresolved structures, and should therefore be interpreted as an upper limit on the true turbulent linewidth/Mach number.

The spatial and radial distribution of FWHM (right panel of Figure~\ref{fig:spatial_comp} and the third panel of Figure~\ref{fig:radius_comp}) and Mach number (left panel of Figure~\ref{fig:cdf_mach} and the fourth panel of Figure~\ref{fig:radius_comp}) does not show a clear monotonic trend with galactocentric distance. The median values remain broadly similar across most radii, with substantial scatter at all distances. There is a small enhancement appearing around $\sim$1.7 kpc, corresponding to the 30 Doradus region as well as several active star-forming regions and shells. CNM components in these regions show  broader linewidths and higher Mach numbers, suggesting that intense local stellar feedback enhances turbulent motions in the cold gas. This interpretation is consistent with molecular gas studies in the LMC. \cite{Wong2019} found that CO linewidths increase toward regions with stronger $8\mu$m emission, likely reflecting both gravity and local star formation driven energy injection. \cite{Green2024} further showed that CO linewidths correlate strongly with recent star formation. Our CNM results suggest that a similar connection may extend to the cold atomic gas.

\subsection{Unstable neutral medium} \label{sec:unm}
As discussed in Section~\ref{sec:comp_result}, we classify UNM components using a temperature threshold of $T_s > 250$ K, following the convention adopted for the Milky Way. However, this criterion may not be directly applicable to the LMC. As shown in Figure~\ref{fig:cdf_comp}, the spin temperatures in the LMC are systematically lower than those in the Milky Way, suggesting that the boundary between CNM and UNM may also shift to lower temperatures in a lower metallicity and higher UV radiation environment. Quantifying this shift is challenging. In our sample, we identify 34 components with $T_s > 100$ K, including 4 with $T_s \sim 200$ K and 2 with $T_s \sim 600$ K. While the classification of components with moderate temperatures ($T_s \sim 100$–200 K) remains uncertain, the two components with $T_s \sim 600$ K are well above any plausible CNM–UNM boundary and can be robustly identified as UNM.
These two components (J052456$-$693855 and J051832$-$693521) have spin temperatures of $559 \pm 42$ K and $654 \pm 39$ K, respectively.
These detections provide direct observational evidence for thermally unstable atomic gas outside the Milky Way using radiative transfer calculations. The corresponding optical depths are $0.047 \pm 0.008$ and $0.029 \pm 0.003$, with FWHM values of $7 \pm 2$ km s$^{-1}$ and $31 \pm 3$ km s$^{-1}$, and column densities of $(3.5 \pm 1.0)\times10^{20}$ cm$^{-2}$ and $(1.2 \pm 0.2)\times10^{21}$ cm$^{-2}$. We also derive UNM fractions of $(22 \pm 5)\%$ and $(50 \pm 4)\%$, compared to CNM fractions of $(17 \pm 3)\%$ and $(6 \pm 1)\%$, respectively.

The locations of these two UNM components are marked by red dots in Figure~\ref{fig:spatial_comp}. Both are located in the central regions of the LMC but exhibit notably different linewidths, suggesting different local physical conditions. The source J052456–693855 lies near the supergiant shell SGS~3, as named by \citet{Dawson2013} based on earlier studies by \citet{Kim1999} and \citet{Meaburn1980}, suggesting that shell-driven compression or shocks may promote the formation of thermally unstable HI. In contrast, J051832–693521 is not associated with any known shell or H II region, indicating that UNM conditions can also arise in more quiescent regions of the LMC. 


Although the detection of UNM in these very different environments suggests that thermally unstable gas can exist in the various ISM of the LMC, identifying it in absorption remains challenging. Such components are expected to be broad and shallow, detecting them requires much high sensitivity compared to the CNM, this is especially harder in external galaxy. In our sample, UNM absorption is detected in only two out of 155 sightlines. The optical depth sensitivities for these two sources are $\sigma_\tau = 0.01$ (J052456$-$693855) and $\sigma_\tau = 0.007$ (J051832$-$693521). 18 sightlines in the sample reach $\sigma_\tau \leq 0.01$, and UNM is detected in just 2 of them. By comparison, the Galactic 21-SPONGE survey, which detects both UNM (25 sources out of 57) and WNM (4 out of 57 sources) in absorption, reaches $\sigma_\tau < 10^{-3}$. The limited number of UNM detections in our sample is therefore likely influenced by sensitivity constraints.

The fact that UNM absorption is detected even in the current, relatively shallow GASKAP-HI observations underscores the capability of the GASKAP survey. Planned deeper integrations as part of the full GASKAP program, including future $\sim$200 hr observations towards the Magellanic Clouds, will substantially improve sensitivity and are expected to enable more detections of thermally unstable gas and potentially direct absorption measurements of the WNM.


\section{Line of sight properties}\label{sec:los_result}
In this section, we examine the line of sight properties of HI in the LMC and compare them with those in the Milky Way. Our LMC sample includes 155 sightlines with reliable HI absorption detections. For comparison, we select 190 Milky Way sightlines from the BIGHICAT catalog \citep{McClure-Griffiths2023} that have non-zero CNM fractions.
While we used the GASKAP-foreground absorption sample in the previous section, this sample is not only biased toward high Galactic latitudes, but also probes a very limited set of HI structures, with many sightlines intersecting the same or closely related clouds. As a result, while the GASKAP-foreground survey serves as a useful control sample for assessing sensitivity effects, it does not sample a wide diversity of Galactic environments. In contrast, both the BIGHICAT catalog and the LMC sample probe a broad range of independent HI clouds. Therefore, in the following analysis of the column density and the CNM fraction, we do not include GASKAP-foreground samples for comparison.

\subsection{Column density}\label{sec:column}
\begin{figure*}
    \centering
    \includegraphics[width=0.9\textwidth]{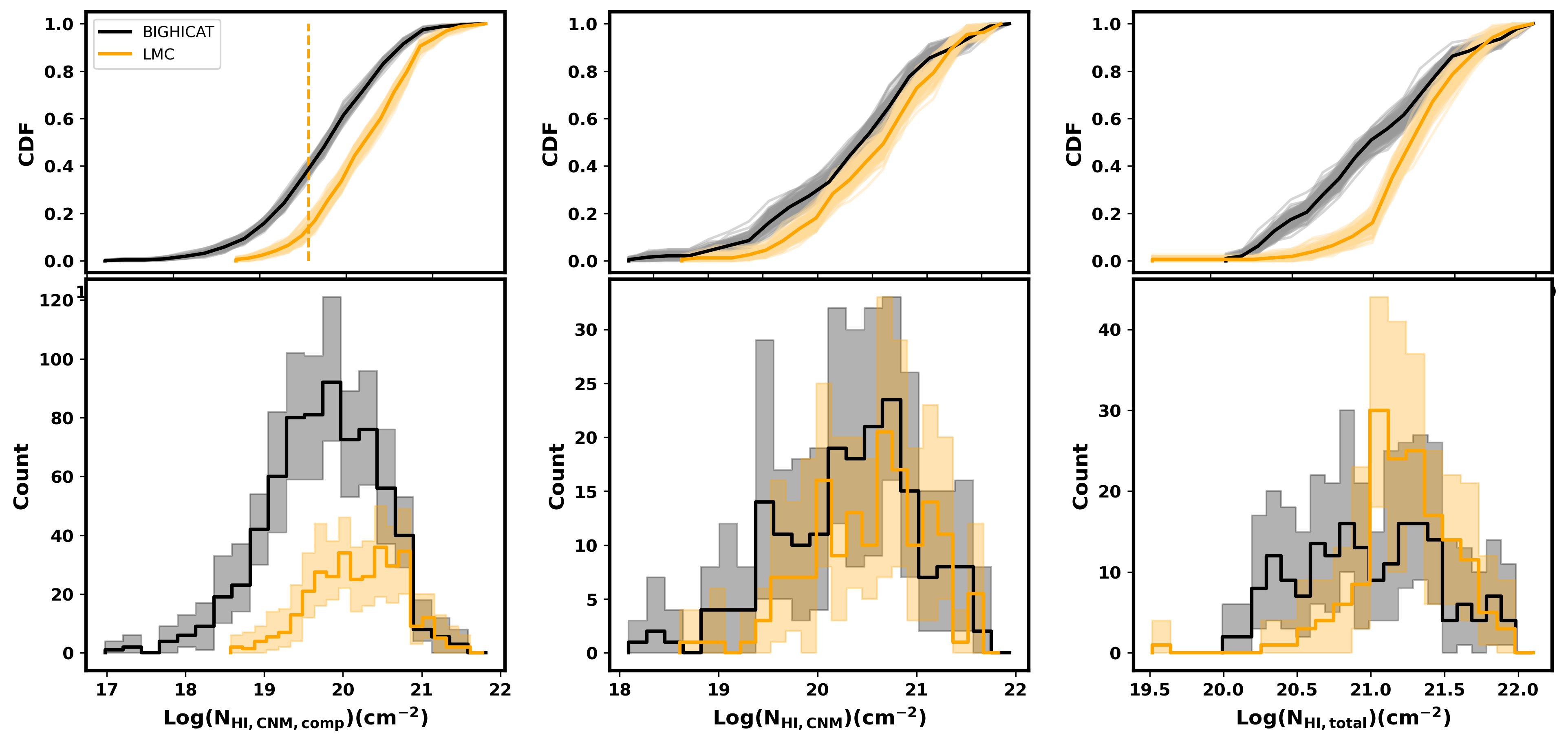}
    \caption{CDF (top) and number (bottom) distribution of HI column density for individual CNM components (left), total CNM column density summed along each line of sight (middle), and total HI column density including both CNM and WNM along each line of sight (right). The orange lines represent our LMC results from the GASKAP-HI survey, while the Milky Way results are from the BIGHICAT catalog \citep{McClure-Griffiths2023}. The orange dashed vertical line in the left panel indicates the CNM column density sensitivity limit ($\sim 4\times10^{19}~\mathrm{cm^{-2}}$), assuming a $3\sigma_\tau$ detection threshold.
    For each CDF/Count, we perform 100 bootstrap trials. The shaded regions indicate all trails and the solid lines show the median values of the distributions. }
    \label{fig:cdf_los}
\end{figure*}

The left panel of Figure~\ref{fig:cdf_los} presents the CDF and number distribution of column density for individual CNM components in the LMC and the Milky Way. In the LMC, individual CNM column densities span from $4.4\times10^{18}$ to $5.0\times10^{21}$ cm$^{-2}$, with a median of $(1.8 \pm 0.2)\times10^{20}$ cm$^{-2}$ and a mean of $(4.0 \pm 0.4)\times10^{20}$ cm$^{-2}$. Overall, individual CNM components in the LMC tend to have higher column densities than those observed in the Milky Way.

To assess whether this offset could arise from different sensitivity limits, we estimate the typical $3\sigma$ detection threshold for CNM component column density ($\sim 4\times10^{19}~\mathrm{cm^{-2}}$) in the LMC, adopting the typical median spin temperature and FWHM derived in Section~\ref{sec:comp_result}, together with the optical depth noise of the GASKAP HI absorption data. About 40\% of the BIGHICAT CNM components fall below this threshold. To ensure a fair comparison, we perform an additional test by applying a common lower limit of $N_{\rm CNM} > 4\times10^{19}~\mathrm{cm^{-2}}$ on both samples (not shown here). Even under this selection, the LMC distribution remains systematically shifted toward higher CNM component column densities relative to the Milky Way. This suggests that the observed offset is unlikely to be driven solely by different sensitivity in both datasets.

Two additional effects may contribute to this difference. First, individual CNM components in the LMC may be less well resolved than those in the Milky Way, so that multiple nearby structures are blended into a single fitted component, artificially increasing the inferred column density per component. Second, the higher CNM column densities in the LMC are consistent with expectations for lower metallicity and higher UV radiation environments. As also discussed in Section~\ref{sec:temperature}, reduced metal and dust abundances modify both heating and cooling processes in the neutral ISM, shifting the CNM–WNM coexistence regime toward higher thermal pressures \citep{Wolfire2003, Bialy2019}. Under these conditions, cold HI can remain thermally stable only at higher densities, which in turn leads to larger column densities for individual CNM structures.

In contrast, the middle panel of Figure~\ref{fig:cdf_los} shows that the total CNM column density summed along each line of sight in the LMC is only slightly higher than that in the Milky Way. For the LMC, the total CNM column density per sightline ranges from $4.5\times10^{18}$ to $5.0\times10^{21}$ cm$^{-2}$, with a median of $(4.4 \pm 0.8)\times10^{20}$ cm$^{-2}$ and a mean of $(8.4 \pm 0.9)\times10^{20}$ cm$^{-2}$. On average, LMC sightlines contain $\sim$2 CNM components, compared to $\sim$3 components per sightline in the Milky Way. This difference in component number may suggest that Milky Way sightlines intersect more CNM clouds with lower individual column densities, whereas LMC sightlines are characterized by fewer but systematically higher column density CNM clouds.
Such a trend is broadly consistent with expectations for a lower-metallicity and higher UV radiation environment, in which thermally stable CNM may require higher density and pressure thresholds \citep{Wolfire1995,Wolfire2003,Bialy2019,Kim2023}, potentially favoring fewer and denser cold clouds.
However, this difference may also arise from the lower resolution of the LMC data, which limits our ability to separate nearby narrow components as effectively as in the Milky Way and may blend multiple CNM structures into a single fitted component. 

By combining the decomposed CNM and WNM components, we compute the total (true) HI column density by summing column densities from both phases along each line of sight.
The right panel of Figure~\ref{fig:cdf_los} shows the CDF/Count of total HI column density for full sightlines through the LMC, compared with those from the Milky Way (from the BIGHICAT survey). In the LMC, total HI column densities range from 3.8$\times10^{19}$ to 1.1$\times10^{22}$ cm$^{-2}$, with a median of (2.0$\pm$0.1)$\times10^{21}$ cm$^{-2}$ and a mean of (2.6$\pm$0.2)$\times10^{21}$ cm$^{-2}$. 
Given that the total CNM column density is similar between the two datasets (middle panel of Figure~\ref{fig:cdf_los}), while the total HI column density is systematically higher in the LMC, the excess HI in the LMC must primarily arise from the WNM. This indicates that LMC sightlines contain a larger WNM contribution per line of sight, especially in regions of low total HI column density.

The higher total HI column density in the LMC reflects the different sampled path lengths between the Milky Way and the LMC. The LMC is viewed nearly face-on, so a typical sightline probes the full depth of the disk, sampling a long path length through gas-rich regions that can accumulate sufficient WNM. In contrast, BIGHICAT sightlines in the Milky Way often probe intermediate or high Galactic latitudes. These sightlines preferentially sample gas away from the Milky Way's main HI disk and have a shorter path length for the line of sight, which naturally leads to systematically lower total WNM column densities and thus lower total HI column densities. 
In addition, the lower metallicity and higher UV radiation of the LMC may intrinsically bias the atomic ISM toward warmer phases. Simulations show that reduced metal and dust abundances shift the thermal phase balance of HI, favoring warmer gas at a given pressure \citep{Wolfire2003, Bialy2019, Kim2023}. 

Future application of the same radiative transfer analysis by Gaussian decomposition to the SMC \citep{Dempsey2022} (and other external galaxies, e.g. LGLBS \citep{Koch2025}) will potentially help disentangle these two effects. As an external galaxy, the SMC is observed with sightlines going through the entire galaxy similar to those in the LMC, while its much lower metallicity of $\sim$0.2 $Z_\odot$ \citep{Dufour1984} provides a critical test of how metallicity influences the HI phase balance across different galactic environments.

\subsubsection{Column density correction factor}
\begin{figure}
    \centering
    \includegraphics[width=0.48\textwidth]{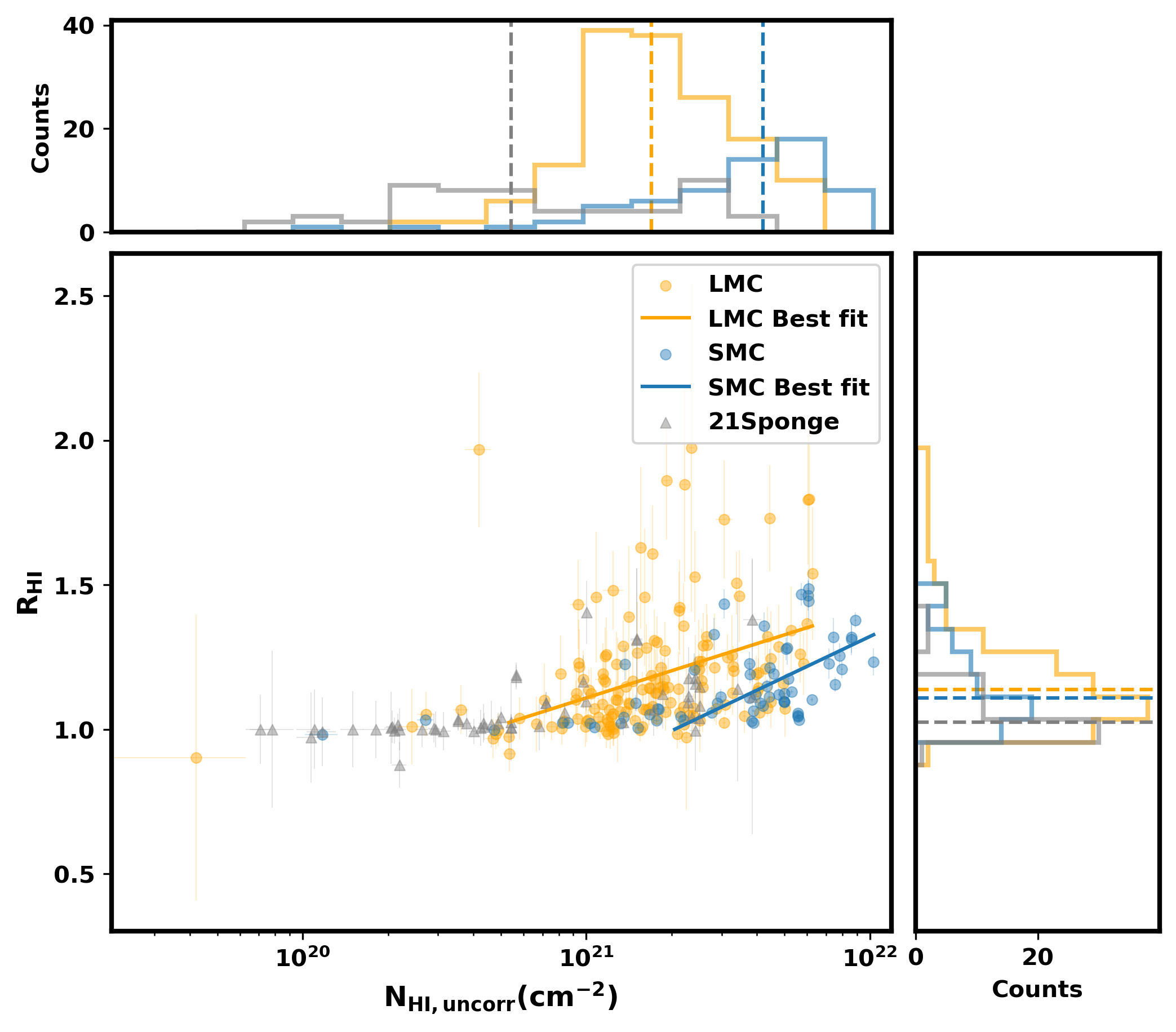}
    \caption{Optical depth correction factor $R_{HI}$ as a function of the original uncorrected HI column density. The orange points represent LMC sightlines and the blue points show the SMC sightlines from \cite{Dempsey2022}. The solid lines represent a linear fit towards the LMC and SMC. The gray triangles correspond to measurements from the 21-SPONGE survey of the Milky Way \citep{Murray2018}.
    The marginal histograms along the top and right axes display the number distribution of uncorrected HI column density and $R_{\rm HI}$, for all samples. Dashed lines in the histograms indicate median values. }
    
    \label{fig:RHI_NHI}
\end{figure}
In the absence of absorption-line information, the 
HI column density is calculated under the assumption of low optical depth ($\tau\ll1$). However, this assumption breaks down in regions where the gas is optically thick. Based on radiative transfer calculations, we have information about individual CNM and WNM components, which allow us to estimate the true HI column density by summing the column densities of all components.
To quantify the impact of optical depth, we define the HI column density correction factor as in \cite{Heiles2003}:
\begin{equation}
R_{\rm HI} = \frac{N_{\rm HI,CNM,all} + N_{\rm HI,WNM,all}}{N_{\rm HI,uncorr}},
\end{equation}
where $N_{\rm HI,CNM,all}$ and $N_{\rm HI,WNM,all}$ are the total CNM and WNM column densities along the line of sight, derived from our radiative transfer solutions. $N_{\rm HI,uncorr}$ is the uncorrected HI column density obtained directly from the HI emission spectrum, assuming optically thin conditions, and is given by \citep[Equation 3 in][]{Dickey1990}:
\begin{equation}
N_{\rm HI,uncorr} = 1.823 \times 10^{18} \int T_{\rm B}(v) dv \quad \text{[cm}^{-2}\text{]}.
\end{equation}

In Figure~\ref{fig:RHI_NHI}, we show the optical depth correction factor ($R_{\rm HI}$) as a function of the uncorrected HI column density for the LMC, the SMC \citep{Dempsey2022}, and the Milky Way (21-SPONGE survey; \citealt{Murray2018}).
For the LMC and the Milky Way, $R_{\rm HI}$ values are derived from Gaussian decomposition of individual CNM and WNM components using the radiative transfer method described above. In contrast, for the SMC, $R_{\rm HI}$ is obtained from integrated HI emission and absorption spectra (e.g. Eq. 5 in \citealt{Dickey1982} and Eq. 4 in \citealt{Dempsey2022}). 

At low uncorrected HI column densities ($N_{\mathrm{HI}} \lesssim 3\times10^{20}~\mathrm{cm^{-2}}$), $R_{\rm HI}$ is close to 1, indicating that the emission is optically thin and the optical depth correction is negligible. At higher column densities, however, $R_{\rm HI}$ exhibits a large scatter. Most sightlines in the Milky Way, SMC, and LMC have $1 \lesssim R_{\rm HI} \lesssim 1.5$. In contrast, several LMC sightlines show $R_{\rm HI}$ values up to $\sim$2, and some of these high values even exist in regions with relatively low uncorrected column densities. These high $R_{\rm HI}$ values are primarily found near shells and active star-forming regions, where dense and clumpy CNM structures are common. This suggests that standard optically thin assumptions can substantially underestimate the true HI column density in these environments, highlighting the need for caution when interpreting the uncorrected HI column densities in such environments without accounting for optical depth effects.

Despite large scatter, we perform linear fits for LMC and SMC samples and find correction factor relations with column density of $R_{\rm HI} \simeq 1 + 0.31(\log_{10} N - 20.67)$ for the LMC and $R_{\rm HI} \simeq 1 + 0.47(\log_{10} N - 21.3)$ for the SMC. These fits are restricted to the regime where $R_{\rm HI} > 1$, where optical depth corrections are significant. For the 21-SPONGE sample, only a small number of sightlines exhibit $R_{\rm HI} > 1$, and we therefore do not perform a linear fit for that dataset.

The SMC results show a steeper dependence of the correction factor on the uncorrected column density compared to the LMC. However, because the SMC $R_{\rm HI}$ values are derived using a different methodology, the comparison should be treated with caution. Although \citet{Lee2015} demonstrated that the two approaches yield consistent correction factors, their analysis was limited to the regime of $\tau \ll 1$ and $T_{\rm sky} \ll T_s$ where the two approaches converge. In contrast, both the LMC and SMC samples include denser clouds with $\tau > 1$, where the two methods may diverge. A more direct and robust comparison will be possible once Gaussian decomposition–based analyses are also applied to the SMC in future work.


\subsection{ CNM fraction}\label{sec:cnm fraction}

\begin{figure*}
    \centering
    \begin{subfigure}[b]{0.35\textwidth}
        \includegraphics[width=\linewidth]{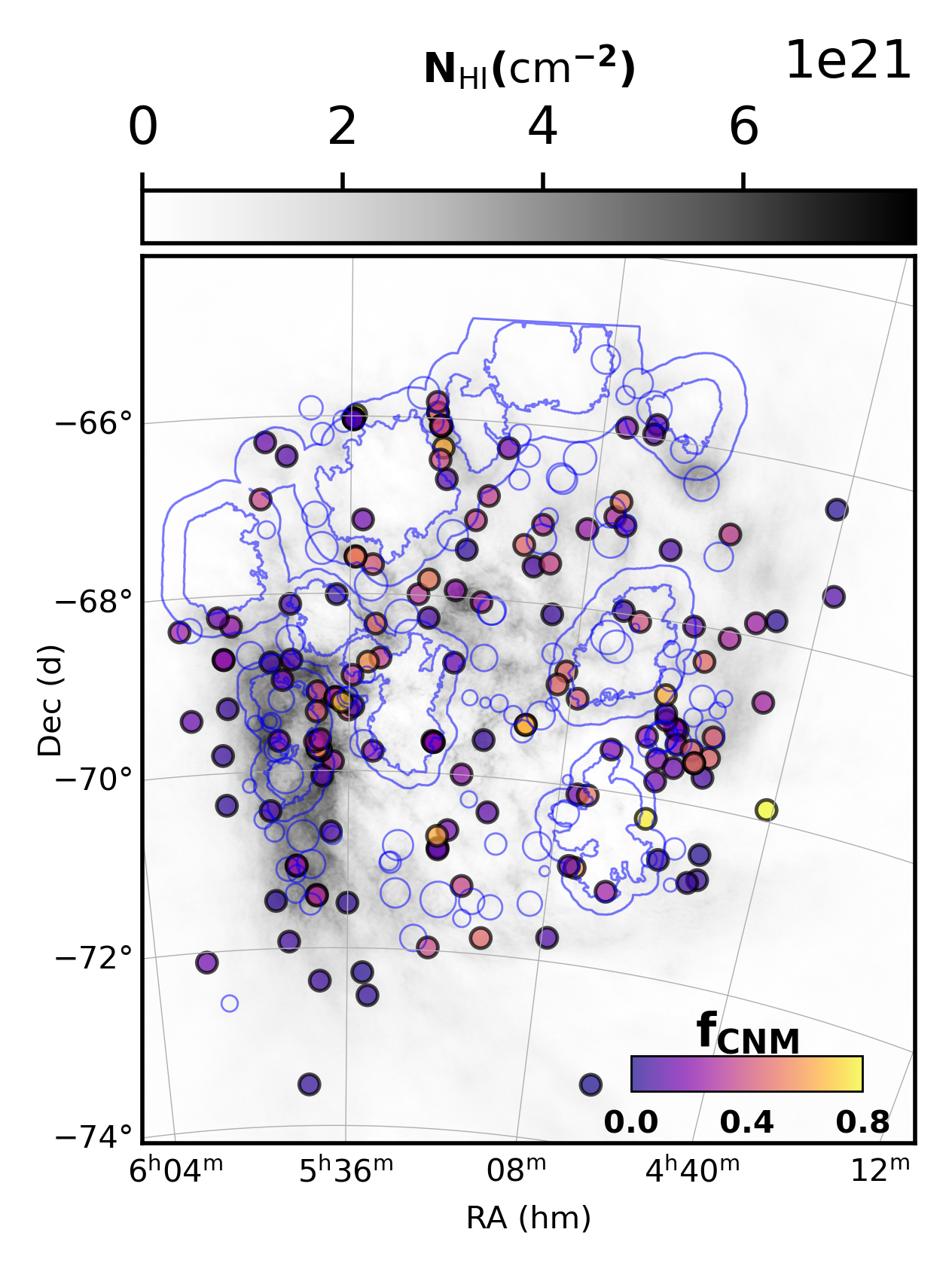}
    \end{subfigure}
    \hfill
    \begin{subfigure}[b]{0.24\textwidth}
        \includegraphics[width=\linewidth]{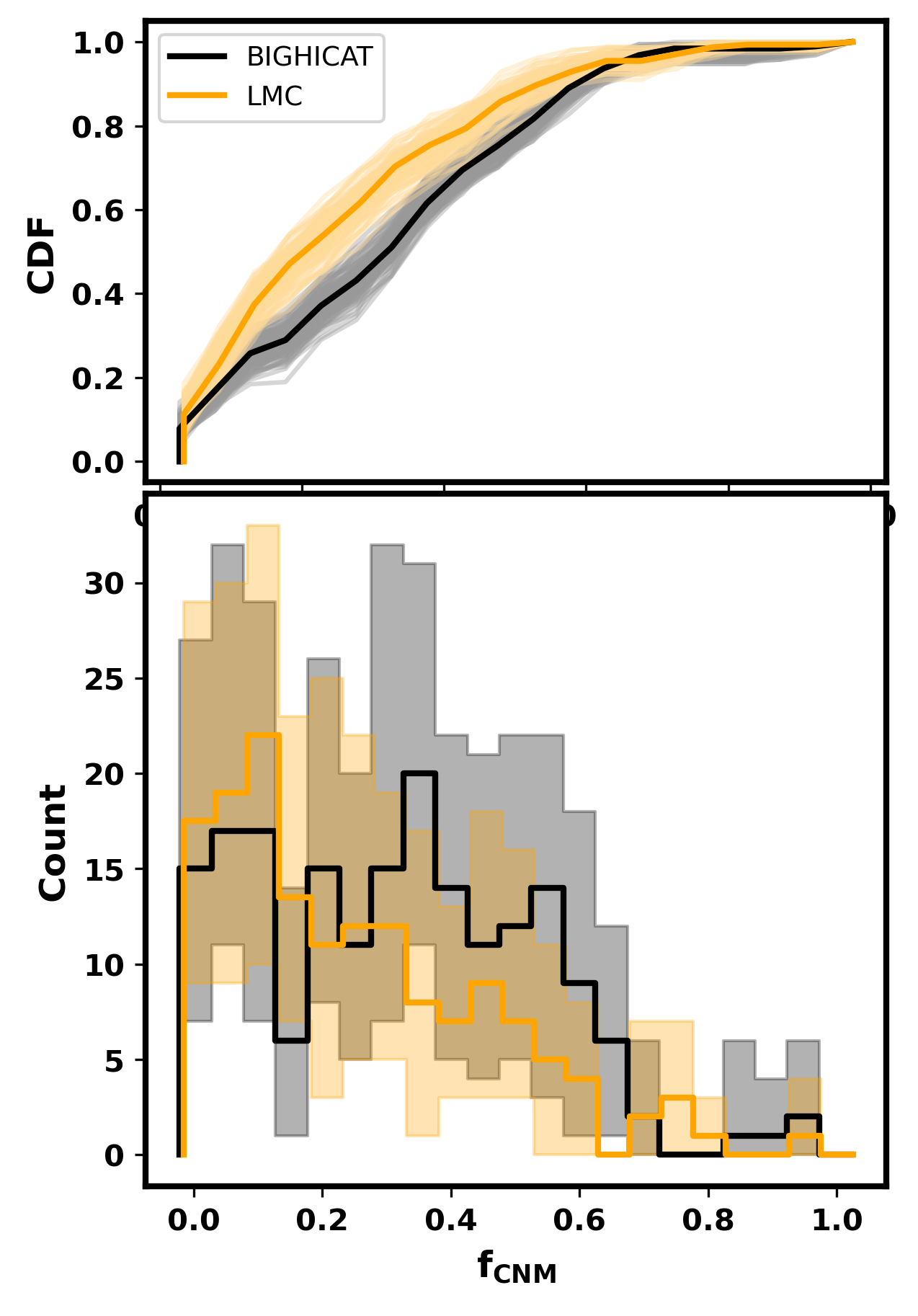}
    \end{subfigure}
    \hfill
    \begin{subfigure}[b]{0.37\textwidth}
        \includegraphics[width=\linewidth]{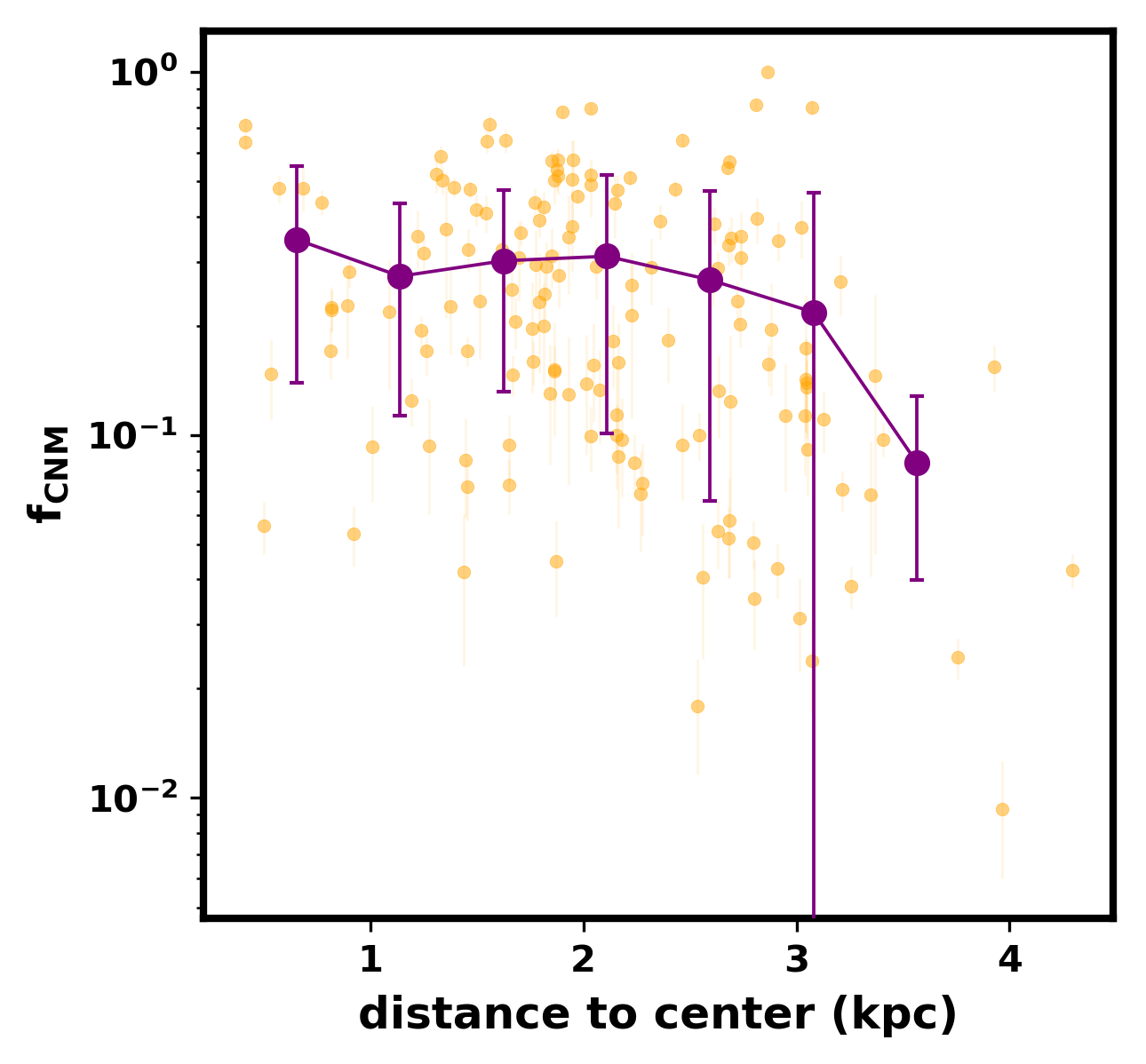}
    \end{subfigure}
    
    \caption{\textit{Left:} CNM fraction (colored circles) overplotted on the HI column density map of the LMC. Large contours mark the locations of supergiant shells identified by \cite{Dawson2013}, while small open circles denote giant shells from \cite{Kim1999}.\textit{Middle:} CDF and number distribution of CNM fraction for each line of sight. The orange lines represent our LMC results from the GASKAP-HI survey, while the Milky Way results from the BIGHICAT catalog \citep{McClure-Griffiths2023} 
    is included for comparison. We perform 100 bootstrap trials. The shaded regions indicate all trails and the solid lines show the median values of the distributions.\textit{Right:} CNM fraction as a function of the source distance to the LMC's kinematic Center. Purple filled circles with error bars indicate the mean and standard deviation of $f_{\rm CNM}$ within different radial bins.
    }
    \label{fig:fcnm}
\end{figure*}



\begin{figure}
    \centering
    \includegraphics[width=0.45\textwidth]{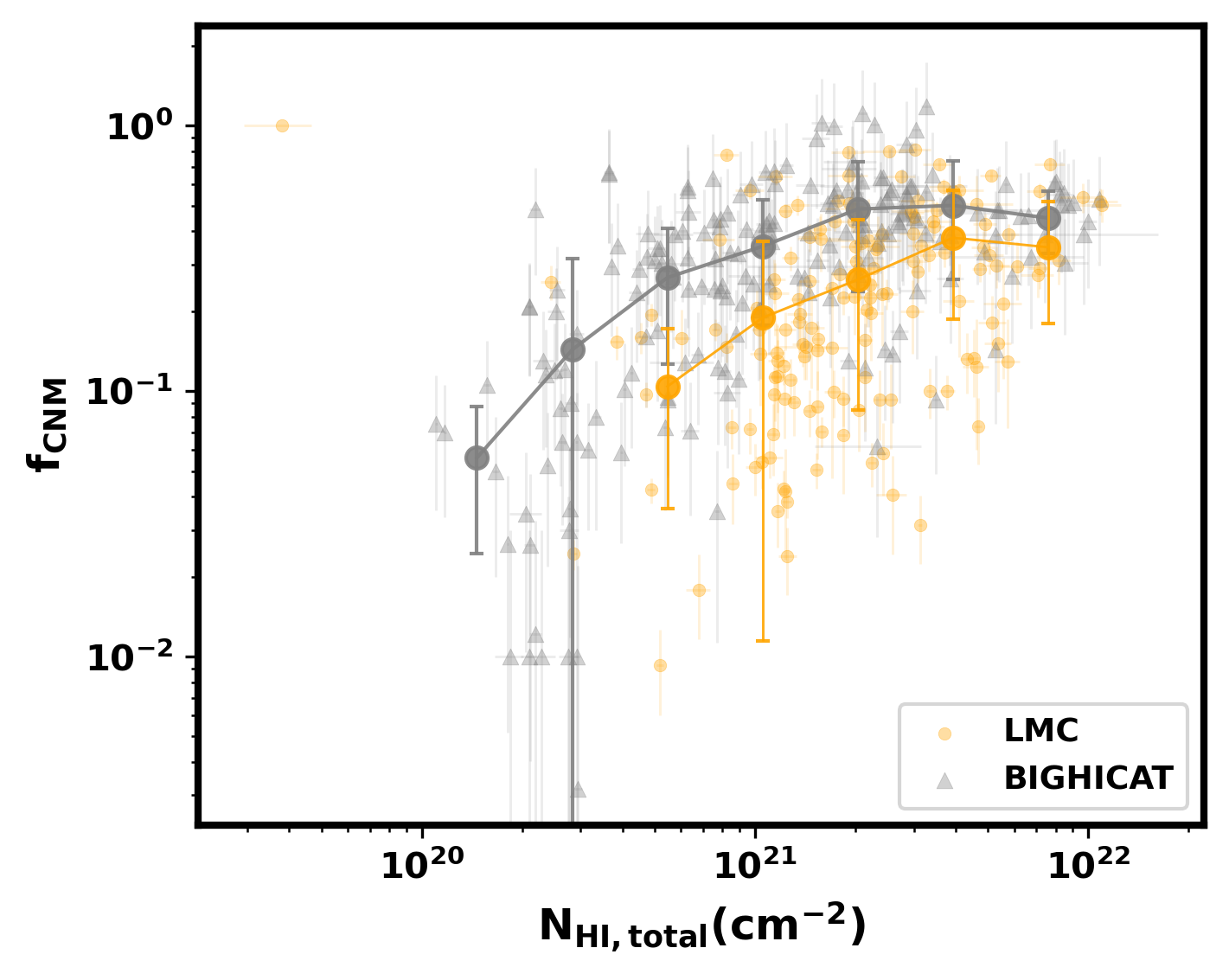}
    \caption{CNM fraction ($f_{\rm CNM}$) as a function of total HI column density ($N_{\rm HI,CNM,all} + N_{\rm HI,WNM,all}$). Orange circles show individual LMC sightlines, while orange filled circles with error bars indicate the mean and standard deviation of $f_{\rm CNM}$ within bins of total HI column density. Gray triangles show individual Milky Way sightlines from the BIGHICAT sample, with gray filled circles and error bars representing the corresponding binned mean and standard deviation. 
    }
    \label{fig:fcnm_NHI}
\end{figure}

The left panel of Figure~\ref{fig:fcnm} shows the spatial distribution of CNM fraction overplotted on the HI column density. Enhanced CNM fractions are found in the vicinity of major supergiant shells. The influence of shells on the CNM fraction is investigated in detail in Section~\ref{sec:shell}.

The middle panel of Figure~\ref{fig:fcnm} shows the CDF and number distribution of the CNM fraction ($f_{\mathrm{CNM}}$) for the LMC and the Milky Way. In the LMC, $f_{\mathrm{CNM}}$ spans a broad range from nearly 0.9\% to 100\%, with a median of (23$\pm$3)\% and a mean of (28$\pm$2)\%. 
A Kolmogorov–Smirnov test yields a statistic of 0.2 with a p value of 0.001, indicating that two CNM fraction CDF distributions are statistically distinct. Compared to the Milky Way BIGHICAT sample, the LMC exhibits a larger fraction of sightlines with low CNM fractions ($f_{\mathrm{CNM}} < 20$ percent) and fewer sightlines with high CNM fractions ($f_{\mathrm{CNM}} > 60$ percent), suggesting that cold gas contributes less to the total neutral medium in the LMC.
The slightly higher CNM fractions observed in the Milky Way can be interpreted in terms of the factors discussed in Section~\ref{sec:column}. Differences in sampled path length and physical conditions (metallicity and radiation field) act to increase the total WNM column density relative to the CNM in the LMC. Given that the total CNM column density is comparable between the two samples, the enhanced WNM contribution in the LMC naturally leads to a lower inferred CNM fraction.

Figure~\ref{fig:fcnm_NHI} presents the CNM fraction as a function of the total HI column density for both the LMC and the Milky Way. 
Although there is significant scatter, both datasets exhibit a general trend of increasing $f_{\rm CNM}$ with increasing column density, consistent with expectations that cold gas formation is favored in denser environments where cooling and shielding are more effective \citep{Wolfire2003, Kanekar2011}.

In the BIGHICAT sample, $f_{\rm CNM}$ tends to level off beyond $N_{\rm HI} \sim 2\times10^{21}~\mathrm{cm}^{-2}$, suggesting that further increases in column density no longer boost the CNM fraction. This transition corresponds to the HI column density threshold required for sufficient  shielding for H$_2$, implying that atomic gas is increasingly converted to H$_2$ beyond this point \citep{Heiles2003b, Lee2012, Stanimirovic2014}.

In contrast, the LMC exhibits a systematic shift toward higher $N_{\rm HI}$ values at a given CNM fraction, with $f_{\rm CNM}$ leveling off at $N_{\rm HI} \sim 4\times10^{21},\mathrm{cm}^{-2}$. This indicates that the transition from WNM to CNM -- and ultimately to H$_2$ -- requires higher column densities in the LMC,  consistent with models showing that reduced dust abundance and weaker shielding in low-metallicity environments shift the HI-to-H$_2$ transition to higher column density \cite{Wolfire2010,Richings2014}.
Notably, one special cloud in the LMC has the lowest total column density but an $f_{\rm CNM}$ of 1, indicating a purely CNM structure with no surrounding WNM (see top left corner of Figure 11). As discussed in \citet{Chen2025}, this cloud likely originated within the LMC and subsequently lost its WNM gas, possibly due to stripping processes as it moved outward.

The right panel of Figure~\ref{fig:fcnm} shows the radial distribution of the CNM fraction across the LMC. Overall, the distribution is broadly flat and exhibits substantial scatter at all radii. There is, however, a weak tendency for lower $f_{\rm CNM}$ values at larger galactocentric distance, particularly beyond $\sim 3$ kpc. A Spearman rank test gives $\rho = -0.28$ with $p = 3.4 \times 10^{-4}$, indicating that this trend is statistically significant but relatively weak. We therefore interpret the radial behavior of $f_{\rm CNM}$ as showing, at most, a mild decline with distance. This tendency may suggest that the outer LMC is somewhat less favorable for the formation or survival of cold gas, possibly because of lower pressure, weaker shielding, or differences in the local ISM environment. However, given the large scatter, local environmental variations (e.g. shells as shown in Section~\ref{sec:shell}) appear to be at least as important as any global radial effect.

This result differs from \citet{Marx-Zimmer2000}, who found no clear radial decrease in the cold gas fraction in the LMC. Their analysis, however, relied on integrating the emission and absorption spectra and assuming a constant cold gas temperature, which may oversimplify the thermal temperature of the CNM in the LMC. 
As discussed in Section~\ref{sec:temperature}, spin temperature in the LMC spans a wide range, meaning that a single-temperature assumption can bias the CNM fraction estimates. In addition, their study is based on only 9 detections concentrated near 30 Doradus and a LMC shell out of 20 sightlines, whereas the GASKAP-HI sample covers many more sightlines over a much larger area. 
Our larger and higher-sensitivity dataset therefore provides a more complete view, although the radial trend itself remains weak.

\section{The environment surrounding the cold HI gas}\label{sec:environment}
In this section, we examine how cold HI interacts with its surrounding environment in the LMC. Cold atomic gas is thought to represent a transitional stage prior to molecular gas formation and can therefore indirectly influence star formation. We first investigate the connections between cold HI, molecular gas, and star formation in order to trace how cold gas evolves toward the next stage. We then explore the relation between cold HI and dust, which provides insight into how cold gas survives in the lower metallicity environment compared to the Milky Way. Finally, we compare the properties of cold gas near shells with those in more quiescent regions, aiming to isolate the local environmental processes that affect cold gas properties. 

\subsection{Molecular gas}\label{sec:mole}
\begin{figure}
    \centering
    \begin{subfigure}{\linewidth}
        \centering
        \includegraphics[width=0.95\textwidth]{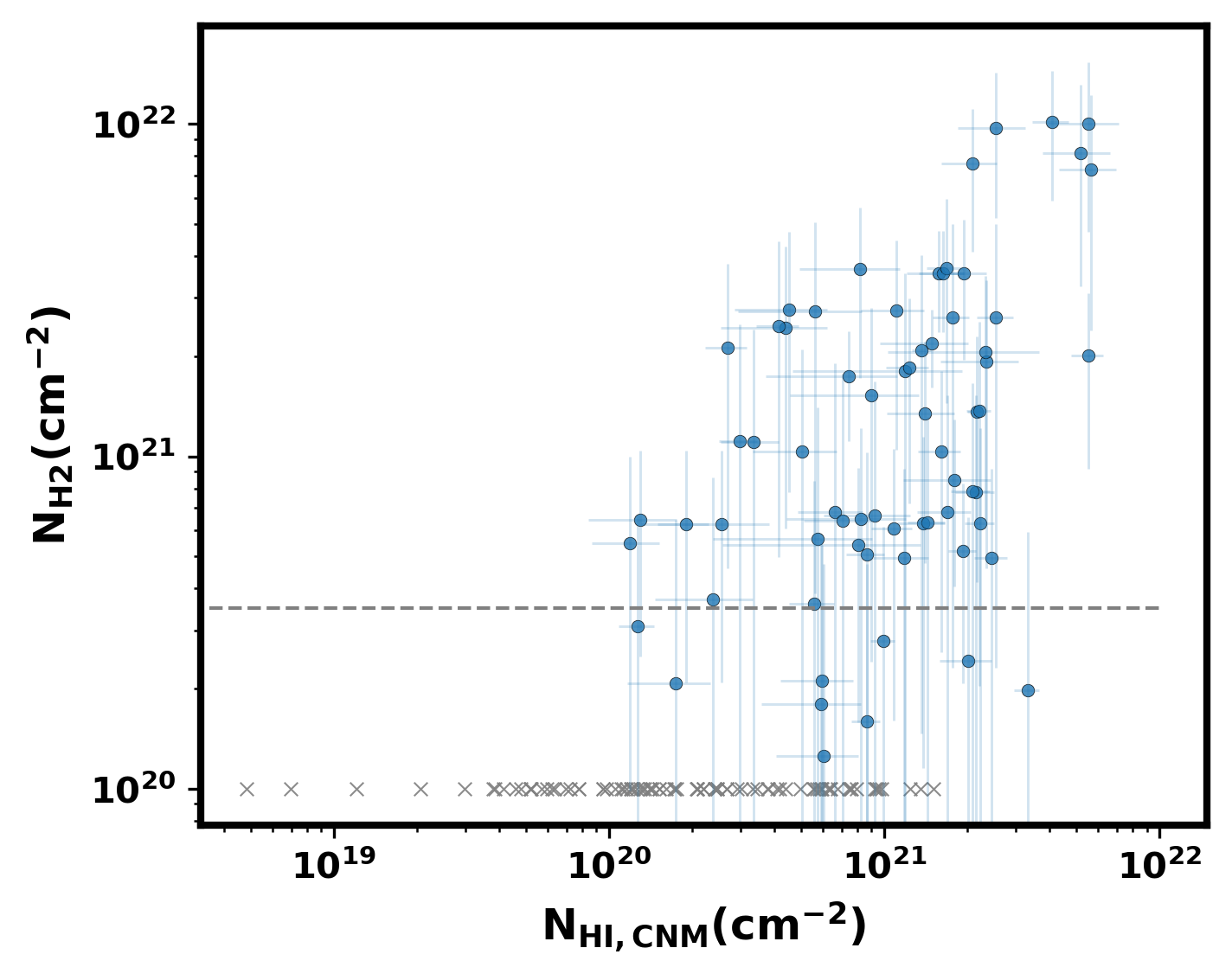}
    \end{subfigure}

    \vspace{0.2cm}

    \begin{subfigure}{\linewidth}
        \centering
        \includegraphics[width=0.95\textwidth]{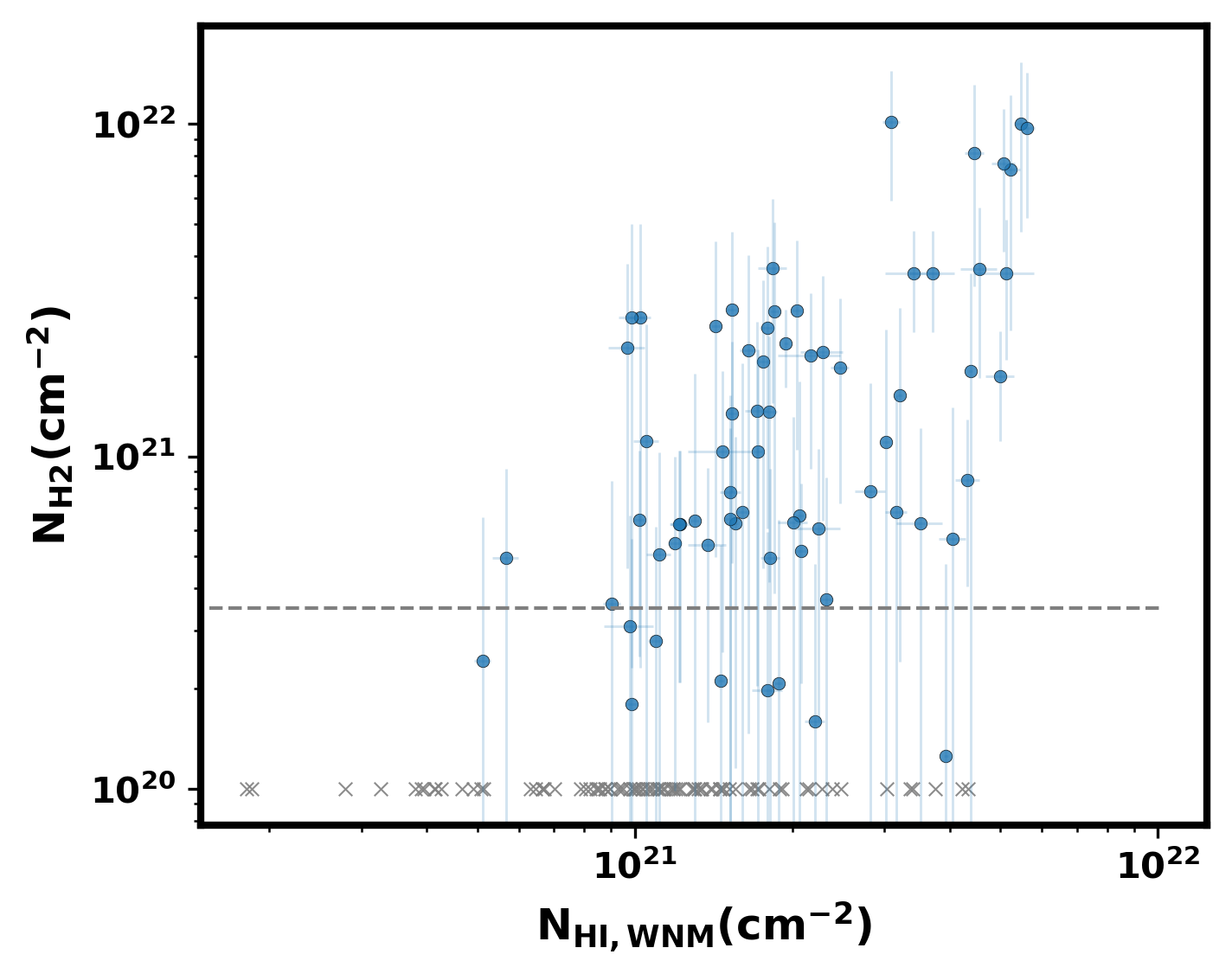}
    \end{subfigure}

    \caption{The H$_2$ column density as a function of the total CNM HI column density (top) and total WNM column density (bottom) along each line of sight. Blue circles denote sightlines where molecular gas is detected ($N_{\rm H_2} > 0$), while gray crosses indicate those without associated H$_2$. The H$_2$ column densities are taken from \citet{Jameson2016}. The gray dashed horizontal line marks the $1\sigma$ sensitivity limit ($N_{\rm H_2} \sim 3.5\times10^{20}~\mathrm{cm^{-2}}$). For clarity, only sources with $N_{\rm H_2} > 10^{20}~\mathrm{cm^{-2}}$ are shown as blue circles to avoid including highly diffuse values below the reliable detection threshold.}
    \label{fig:NHI_NH2}
\end{figure}


Molecular gas represents the next evolutionary stage of cold HI and serves as the primary reservoir for star formation. To examine how the CNM relates to molecular gas within the LMC,
we use the H$_2$ column density ($N_{\rm H_2}$) map from \citet{Jameson2016}, who derived molecular gas distributions in the Magellanic Clouds using dust emission from the Herschel HERITAGE survey and HI emission from \cite{Kim2003}. Though CO is traditionally used to trace H$_2$, low metallicity galaxies have a significant amount of H$_2$ gas that is not traced by CO (the CO-dark gas) as H$_2$ can better self-shield against dissociating UV photons compared to CO. Therefore, dust based estimates of H$_2$ are more reliable than CO-based estimates in low metallicity environments \citep{Bolatto2011,Bolatto2013}. The $N_{\rm H_2}$ map from \citet{Jameson2016} has a spatial resolution of $20''$ and a 1$\sigma$ sensitivity of $N_{\rm H_2}\sim 3.5\times10^{20} cm ^{-2}$. For each of our sightlines, we calculate the local $N_{\rm H_2}$ by averaging values within a $120''$ radius around its position, and we adopt the corresponding standard deviation within this region as the uncertainty. This radius is chosen to match the scale over which the HI emission spectra are averaged in our analysis, ensuring that the CNM and molecular gas measurements probe comparable spatial regions. Although this averaging scale is relatively large ($\sim$ 30 pc), it provides a consistent and fair comparison for the two datasets. The same averaging radius is adopted in the subsequent sections for all comparisons.

We identify 75 sightlines with detected $N_{\rm H_2}$. The upper panel of Figure~\ref{fig:NHI_NH2} shows the H$_2$ column density as a function of the total CNM HI column density along each line of sight. Although there is large scatter, $N_{\rm H_2}$ generally increases with increasing CNM column density.
The formation of H$_2$ requires sufficient shielding against dissociating far-ultraviolet radiation, which is provided by total neutral gas and dust \citep[e.g.,][]{Krumholz2009, Sternberg2014}. Both the CNM and WNM therefore contribute to the attenuation of the ambient radiation field. Consistent with this, we also find a weak positive correlation between $N_{\rm H_2}$ and the total WNM column density (lower panel of Figure~\ref{fig:NHI_NH2}), albeit with significantly larger scatter. The tighter correlation with CNM column density suggests that, while the WNM also contributes to shielding, the CNM more directly traces the reservoir of cold, dense atomic gas that can efficiently convert into H$_2$. The observed trends therefore reflect both the role of total HI gas in providing shielding and the role of the CNM in regulating the local H$_2$ formation rate.

The lowest CNM column density at which molecular gas is detected in the LMC above the noise level is  $\sim 10^{20}\mathrm{cm^{-2}}$, while for the WNM is $\sim10^{21}\mathrm{cm^{-2}}$. These values suggest the typical CNM column density thresholds for the onset of H$_2$ formation in the LMC. \citet{Wong2009} found that molecular gas detected by CO in the LMC starts to appear at total HI column densities of $\sim10^{21}\mathrm{cm^{-2}}$, which aligns with what we find here as the WNM is the major component of the total HI column density.  A similar analysis in the solar neighborhood by \citet{Park2023} found lower thresholds in the Milky Way, with minimum CNM and WNM column densities associated with CO detections of $\sim2\times10^{19}\mathrm{cm^{-2}}$ and $\sim2\times10^{20}\mathrm{cm^{-2}}$, respectively. The lower thresholds reported for the Milky Way may partly reflect the higher sensitivity of Galactic HI absorption measurements. In addition, the higher CNM and WNM column density thresholds observed in the LMC are consistent with expectations for lower metallicity environments, where reduced dust abundance and weaker shielding require larger atomic gas columns before molecular gas can form and survive efficiently \citep{Krumholz2009, Wolfire2010, Sternberg2014}.

However, some regions with $N_{\rm HI}$ $\sim10^{20-21}~\mathrm{cm^{-2}}$ still lack detectable H$_2$, indicating that dynamical processes could be important for H$_2$ formation too. For example, \cite{Dawson2013} showed that 
the presence of supergiant shells has a positive effect on the molecular gas fraction in the LMC. 
Beyond $N_{\rm HI}$ $\sim10^{21}~\mathrm{cm^{-2}}$, we find that molecular gas is almost always present, implying that the gas becomes fully self-shielded and efficiently transitions into the molecular phase.

\subsection{Star formation}\label{sec:sfr}
\begin{figure*}
    \centering
    \begin{subfigure}[b]{0.38\textwidth}
        \includegraphics[width=\linewidth]{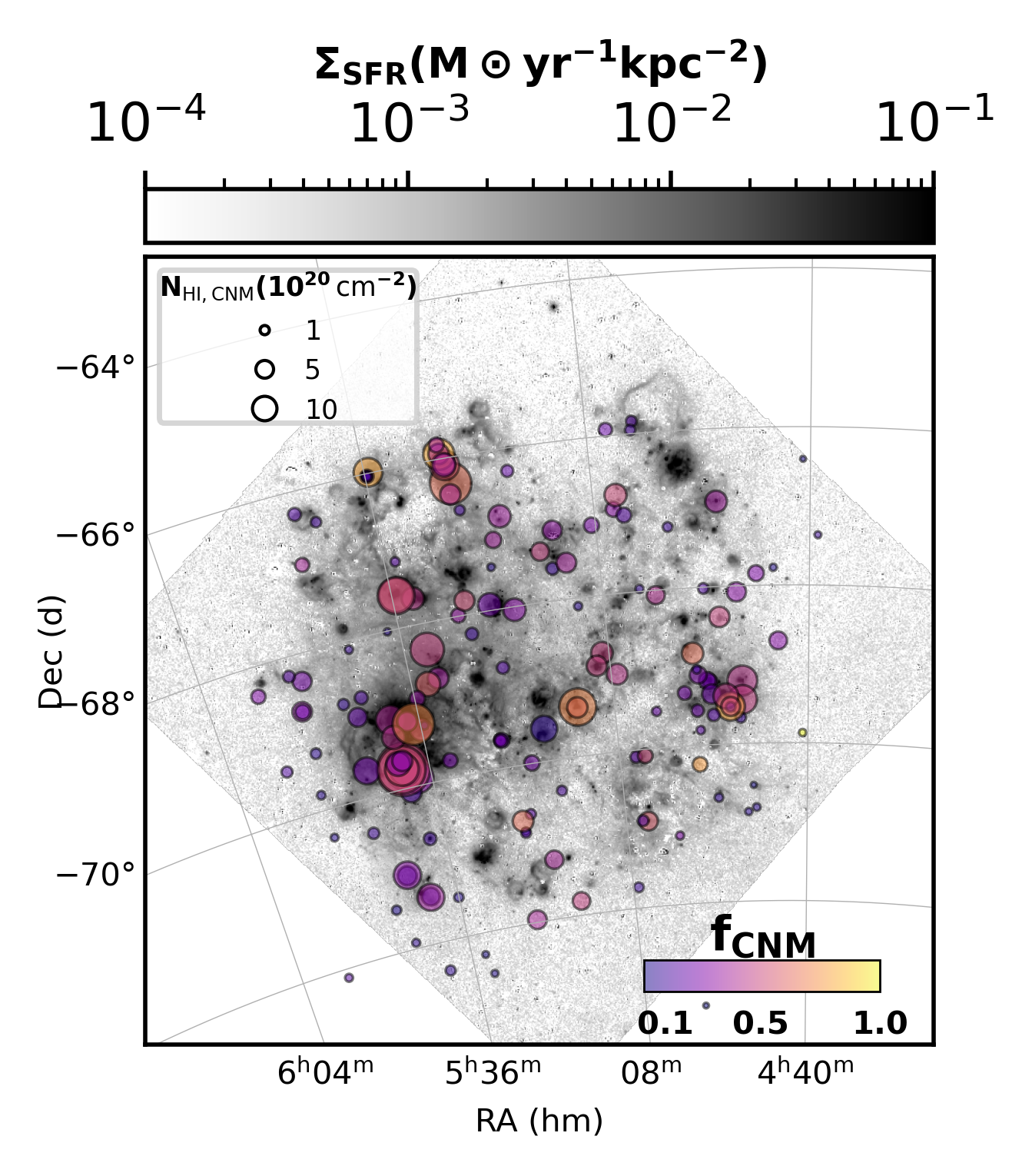}
    \end{subfigure}
    \hfill
    \begin{subfigure}[b]{0.61\textwidth}
        \includegraphics[width=\linewidth]{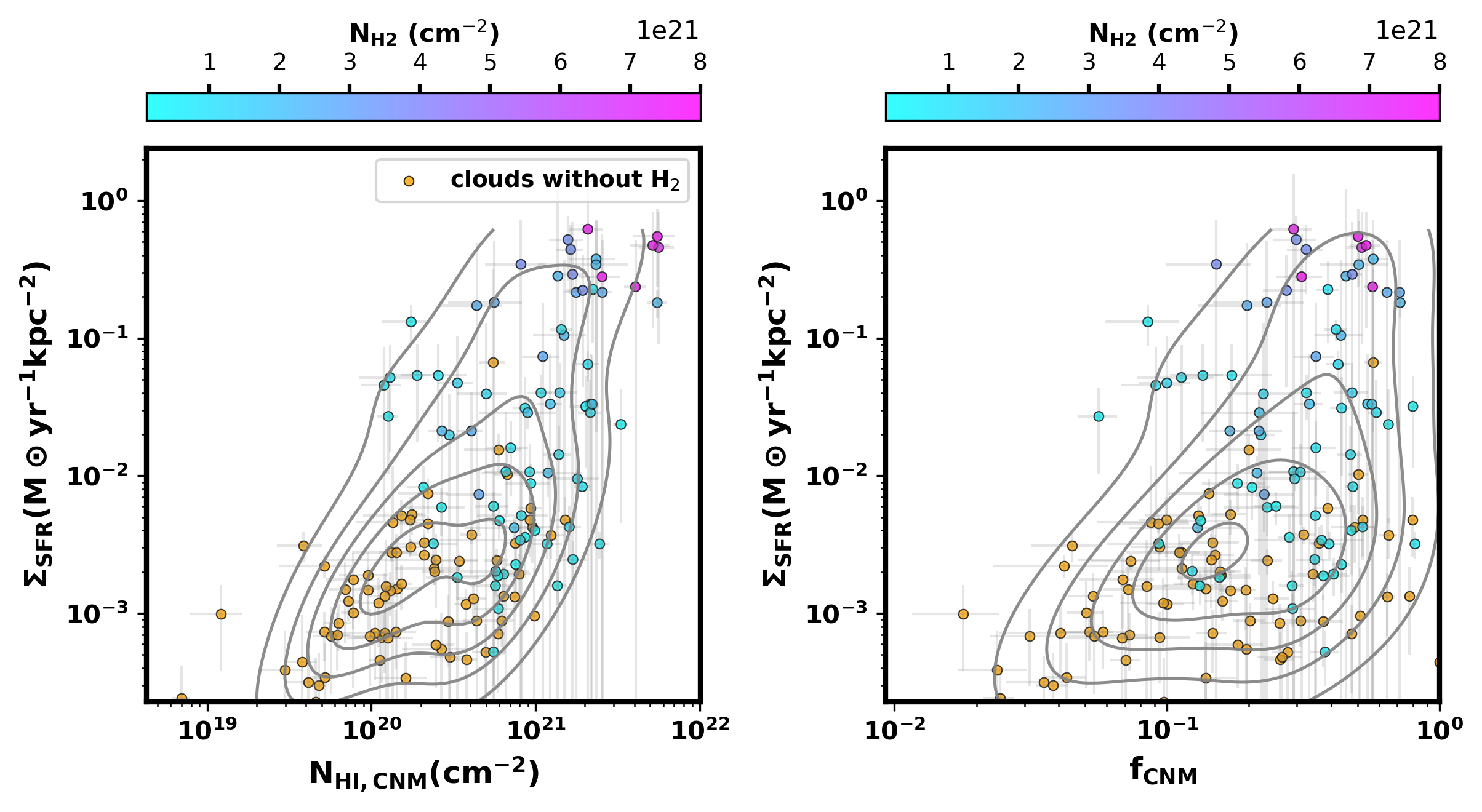}
    \end{subfigure}
    \caption{Positions of our sources overplotted on the SFR surface density map of the LMC (left). The color of the circle shows the CNM fraction at that location and the size of each circle indicates the corresponding total CNM HI column density along the line of sight. The middle and right panel show star formation rate surface density as a function of total CNM HI column density along the line of sight and CNM fraction, respectively. 
    The yellow circles show the cold clouds without detection of H$_2$, while the color of other dots presents the H$_2$ column density.
    Contours in the middle and right panels show kernel density estimates of the data distribution, computed in logarithmic space and plotted at five equally spaced density levels.}
    \label{fig:sfr}
\end{figure*}

As numerical simulations have suggested a positive correlation between star formation rate (SFR) and CNM fraction \citep[e.g.,][]{Smith2023}, we investigate here whether such a correlation exists in the LMC.

We use H$\alpha$, locally with small correction for extinction using 24 $\mu m$ dust emission, to trace recent star formation. We use the calibrated, continuum-subtracted H$\alpha$ map from the Southern H$\alpha$ Sky Survey Atlas \citep[SHASSA;][]{Gaustad2001} at $0.8^\prime$ resolution. As for the 24 $\mu m$ dust emission,  we use the Multiband Imaging Photometer (MIPS) 24 $\mu m$ map from the Spitzer Survey “Surveying the Agents of Galaxy Evolution” \citep[SAGE;][]{Meixner2006}. The luminosities of both H$\alpha$ and 24 $\mu m$ are derived from $L=4\pi D^2\theta^2 I$, where $D=50$ kpc \citep{Freedman2001}, $\theta$ is the angular size of the region of interest in the unit of arcsec (here we use the resolution of the two maps), and $I$ is the intensity of H$\alpha$ or 24 $\mu m$ map.
We then use the  star formation rate (SFR) calibration by \cite{Calzetti2007} to convert H$\alpha$ and 24 $\mu m$ luminosities into SFR:
\begin{equation}
\begin{aligned}
\mathrm{SFR}\left(M_{\odot} \mathrm{yr}^{-1}\right) & =5.3 \times 10^{-42}[L(H \alpha) \\
& +(0.031 \pm 0.006) L(24 \mu m)].
\end{aligned}
\end{equation}
The star formation rate surface density ($\Sigma_{\rm SFR}$) is finally obtained by dividing the SFR in each pixel by its area in kpc$^2$, yielding units of $M_\odot~\mathrm{yr}^{-1}~\mathrm{kpc}^{-2}$. The resulting $\Sigma_{\rm SFR}$ map has a sensitivity of $1 \times 10^{-4} M_{\odot} \mathrm{yr}^{-1} \mathrm{kpc}^{-2}$. To assign a representative $\Sigma_{\rm SFR}$ value to each of our sources, we compute the average $\Sigma_{\rm SFR}$ within a $2'$ radius around each position. The associated uncertainty is estimated using the standard deviation of values within that region. The left panel of Figure~\ref{fig:sfr} shows our sources overplotted on the SFR surface density map.

The middle and right panels of Figure~\ref{fig:sfr} show the relationship between the star formation rate surface density ($\Sigma_{\rm SFR}$) and cold gas properties. The middle panel reveals that $\Sigma_{\rm SFR}$ increases with CNM column density, though with substantial scatter compared to the tighter correlation between $\Sigma_{\rm SFR}$ and molecular gas \citep[e.g.,][]{Jameson2016}. In contrast, no clear relationship is found between $\Sigma_{\rm SFR}$ and the total WNM column density (see Appendix~\ref{app:sfr}). Although the WNM contributes a substantial fraction of the total atomic gas reservoir (e.g., Figure~\ref{fig:cdf_los}), its abundance alone does not appear to set the conditions necessary for molecular cloud formation. Instead, these results suggest that the presence of cold atomic gas, rather than the total atomic gas content, plays a more direct role in regulating star formation.

Nearly all CNM clouds associated with substantial star formation rates ($\Sigma_{\rm SFR} > 10^{-2}~M_{\odot}~\mathrm{yr^{-1}~kpc^{-2}}$) also contain molecular gas, and higher $\Sigma_{\rm SFR}$ values correspond to higher $N_{\rm H_2}$ and $N_{\rm HI,CNM}$. This indicates that active star formation in the LMC predominantly occurs in regions where cold atomic gas has already transitioned into the molecular phase, and molecular gas serves as a more closely connected reservoir for star formation than the cold atomic phase. The close connection between CNM column densities, molecular gas and star formation rate supports the view that the CNM provides the immediate reservoir for H$_2$ formation and subsequent star formation \citep[e.g.,][]{Heiles2003, Krumholz2009}.

The right panel of Figure~\ref{fig:sfr} shows the relationship between $\Sigma_{\rm SFR}$ and the CNM fraction. Although we do not find a clear, monotonic correlation between $\Sigma_{\rm SFR}$ and $f_{\rm CNM}$ -- as suggested in some simulations such as \citet{Smith2023} -- the data still indicate that active star-forming regions are biased toward higher CNM fractions. In particular, nearly all regions with ongoing star formation have $f_{\rm CNM} \gtrsim 0.1$, suggesting that a minimum CNM fraction of roughly 10\% is required to sustain star formation in the LMC. Below this threshold, the lack of sufficient cold gas likely prevents efficient molecular hydrogen formation and gravitational collapse. Beyond $f_{\rm CNM} \sim 0.1$, however, no clear trend is observed. This indicates that once a sufficient amount of CNM is achieved, further increases in $f_{\rm CNM}$ no longer enhance star formation activity. In addition, the CNM fraction measured along a given line of sight may include contributions from warm gas that are not physically associated with the local star-forming regions, which can further dilute any intrinsic correlation.

The star formation efficiency likely depends more on the local physical conditions of the dense gas, such as the absolute CNM column density (as shown in the middle panel of Figure~\ref{fig:sfr}), rather than on the relative CNM fraction compared to the surrounding WNM. While our results suggest that a minimum CNM fraction of 10\% is associated with ongoing star formation, this threshold should not be interpreted as a necessary but not sufficient condition. An alternative explanation is that the physical conditions necessary for star formation, such as high pressure and gas compression, naturally produce CNM fractions at or above this level. In this view, the observed $f_{\rm CNM} \gtrsim 0.1$ reflects the outcome of star formation favorable environments rather than a direct predictor of star formation activity.

\subsection{Dust}\label{sec:dust}
\begin{figure}
    \centering
    \includegraphics[width=0.45\textwidth]{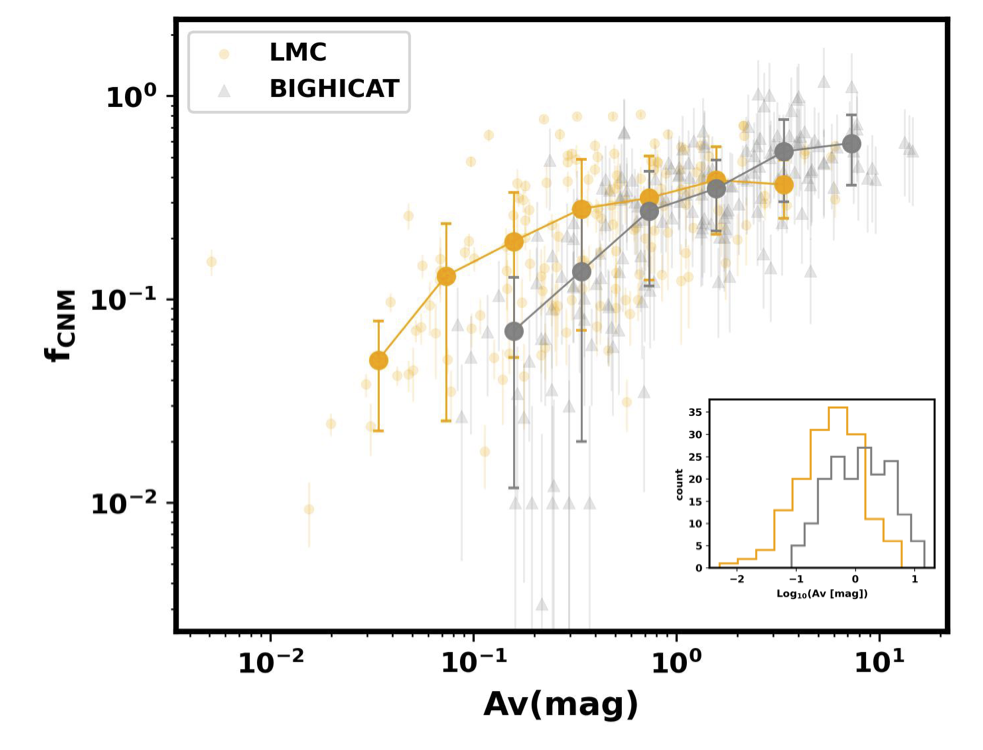}
    \caption{The cold neutral medium (CNM) fraction as a function of optical extinction ($A_V$). The orange points show the distribution of LMC, while the black points show the BIGHICAT distribution. The $A_V$ data for the LMC is derived from the dust surface density map from \citet{Clark2023}, see Section~\ref{sec:dust} for more details. The solid connected lines represent the binned mean CNM fraction in logarithmic $A_V$ bins, with vertical error bars indicating the standard deviation within each bin. The inner panel shows the histogram of $A_V$ in log space for these two galaxies.}
    \label{fig:fcnm_Av}
\end{figure}

Dust grains play a crucial role in regulating the physical conditions of the CNM by both shielding the gas from radiation and mediating heating and cooling processes. Dust extinction protects CNM clouds from far-ultraviolet (FUV) radiation, allowing the gas to cool efficiently and facilitating the formation of molecules such as H$_2$. At the same time, dust grains contribute to photoelectric heating, where UV photons eject electrons from grain surfaces, injecting thermal energy into the gas. 

Observational studies in the Milky Way have shown that regions with high optical extinction ($A_V$) tend to exhibit higher CNM fractions \citep{McClure-Griffiths2023}. High $A_V$ not only indicates enhanced dust shielding against far-ultraviolet radiation, which reduces heating, but also traces higher gas densities, which increase radiative cooling efficiency. Together, these effects promote the formation and survival of cold neutral gas. To investigate whether a similar relationship holds in the LMC, we present the CNM fraction as a function of $A_V$ in Figure~\ref{fig:fcnm_Av}.

We derive an $A_V$ map of the LMC using the dust surface density map ($\Sigma_{\rm dust}$) from \citet{Clark2023}, which was obtained by SED fitting of Herschel far-infrared emission in five bands spanning 100–500~$\mu$m. The $\Sigma_{\rm dust}$ map has an angular resolution of $\sim1'$. Following the approach of \citet{LeeC2015}, we convert dust optical depth to visual extinction using $A_V \sim 2200\tau_{160}$, where $\tau_{160}$ is the dust optical depth at 160~$\mu$m. The optical depth is computed as $\tau_{160} = \kappa_{160}\Sigma_{\rm dust}$, with $\kappa_{160}$ being the dust mass absorption coefficient at 160~$\mu$m. We adopt $\kappa_{160} = 1.24~\mathrm{m^2~kg^{-1}}$, same as the values used in \citet{Roman-Duval2017} and \citet{Clark2023}.  We note that the dust map is defined only in regions with hydrogen surface density $\Sigma_{\rm H} > 2.8~M_\odot~\mathrm{pc^{-2}}$, corresponding to $N_{\rm H} \gtrsim 3.5 \times 10^{20}~\mathrm{cm^{-2}}$. For each HI absorption sightline in our sample, we estimate the local $A_V$ by averaging the extinction values within a $2'$ radius centered on the source position, and we take the standard deviation within this region as the associated uncertainty. 

In Figure~\ref{fig:fcnm_Av}, the LMC follows a similar overall trend to the Milky Way, with the CNM fraction generally increasing with visual extinction ($A_V$).
The Galactic samples generally exhibit higher $A_V$ compared to the LMC, which explicitly indicates the reduced dust content in low metallicity galaxies.
In the low-extinction regime ($A_V<1$), the same CNM fraction in the LMC corresponds to significantly lower dust extinction but higher total HI column density compared to the Milky Way (Figure~\ref{fig:fcnm_NHI}). This suggests that similar CNM fractions can be maintained in the LMC despite weaker dust shielding. 
At low $A_V$ the LMC directions have higher HI column densities but much smaller H$_2$ column densities relative to the Milky Way. For example, at $A_V<0.5$, the HI column density in LMC is is approximately an order of magnitude higher than in the Milky Way. As shown in Figure~\ref{fig:cdf_los}, the reason for this is a higher HI column density for individual CNM components, as well as a longer line of sight depth in the LMC. This suggests that a likely higher local density in individual CNM components, as expected theoretically at low metallicity \citep{Bialy2019}, may result in enough cooling to form and sustain a CNM fraction of a few percent. Together with a longer line of sight, the enhanced local cooling may compensate the effect of weaker dust shielding and sustain a similar level of the CNM fraction as the Milky Way.

As $A_V$ increases, the difference in $A_V$ between the LMC and the Milky Way at fixed CNM fraction becomes smaller, suggesting that dust shielding plays an increasingly important role in regulating the CNM fraction in the LMC.  We note, however, that part of this trend may be influenced by beam-dilution effects in the dust measurements in the case of the LMC, which are expected to be more significant in low-$A_V$ regions where the dust distribution is more spatially inhomogeneous, and less important at higher $A_V$ where the gas and dust are more beam-filling.

At $A_V \gtrsim 1$,  the CNM fraction begins to flatten in both galaxies. This suggests that further increases in dust shielding no longer enhance the formation or survival of cold atomic gas at this level and  the atomic gas is transitioning to molecular gas in this region \citep{McClure-Griffiths2023}.

In addition, in the Milky Way, high CNM fractions ($>60\%$) are typically found in regions where $A_V > 1$ mag, yet in the LMC, we find cold clouds that reach similarly high CNM fractions even at $A_V \sim 0.1$ mag.
This unusually high CNM fraction at low extinction suggests that additional processes, such as pressure from expanding shells (see Section~\ref{sec:shell}) and stellar feedback, turbulent compression, or locally enhanced atomic line cooling, help maintain cold HI in regions with weak dust shielding.  

We note, however, that the comparison between the LMC and Milky Way is subject to several caveats. The $A_V$ conversion adopted here for the LMC is calibrated for diffuse Milky Way environments and may not fully capture variations in denser regions. In addition, differences in angular resolution between the LMC and Galactic datasets may affect the measured $A_V$ and CNM fraction on small spatial scales.

\subsection{Shells}\label{sec:shell}
\begin{figure*}
    \centering
    \begin{subfigure}[b]{0.5\textwidth}
        \includegraphics[width=\linewidth]{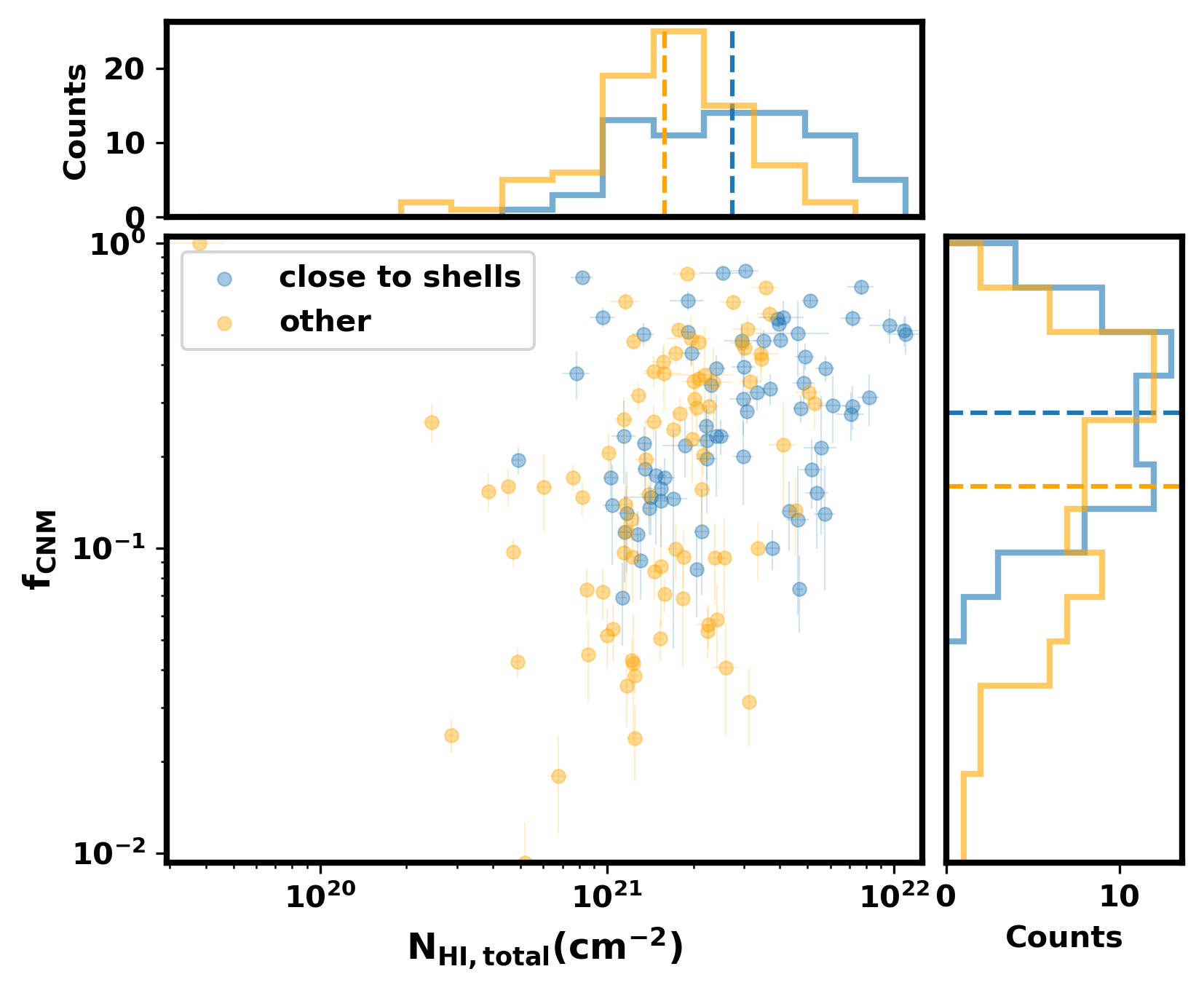}
    \end{subfigure}
    \hspace{3mm} 
    \begin{subfigure}[b]{0.3\textwidth}
        \includegraphics[width=\linewidth]{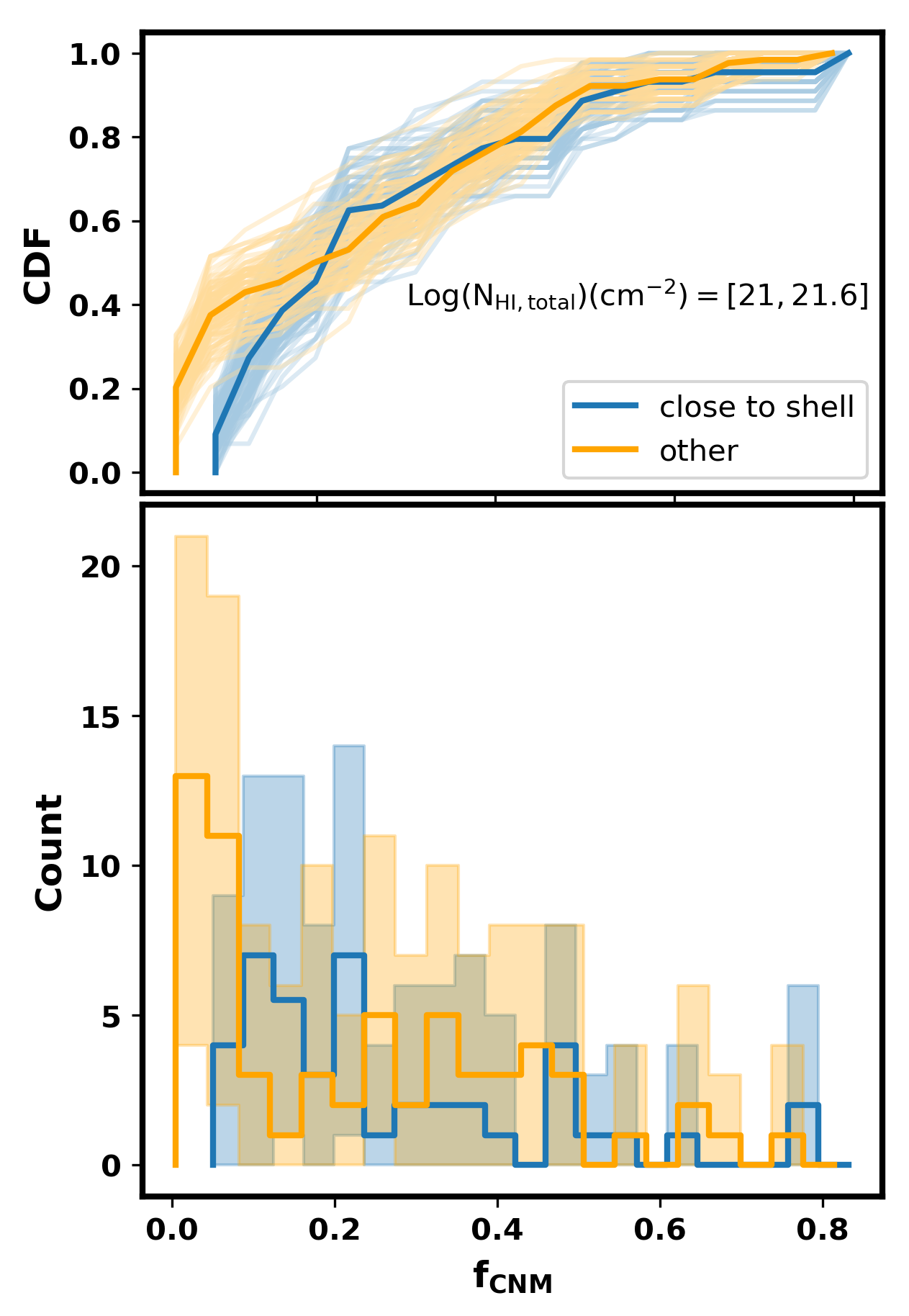}
    \end{subfigure}
    \caption{Left: CNM fraction ($f_{\rm CNM}$) as a function of total HI column density ($N_{\rm HI,CNM} + N_{\rm HI,WNM}$) for clouds located near shell structures (blue) and those outside shell regions (orange). The marginal histograms along the top and right axes display the distributions of total HI column density and $f_{\rm CNM}$, respectively, for both samples. Dashed lines in the histograms indicate median values.
    Right: Cumulative (CDF) and number distributions of $f_{\rm CNM}$ for the same two groups, restricted to clouds with total HI column densities from $10^{21}$ to $4\times10^{21}~\mathrm{cm^{-2}}$.}
    \label{fig:shell_fcnm_NHI}
\end{figure*}

\begin{figure*}
    \centering

    \begin{subfigure}{\textwidth}
        \centering
        \includegraphics[width=1\linewidth]{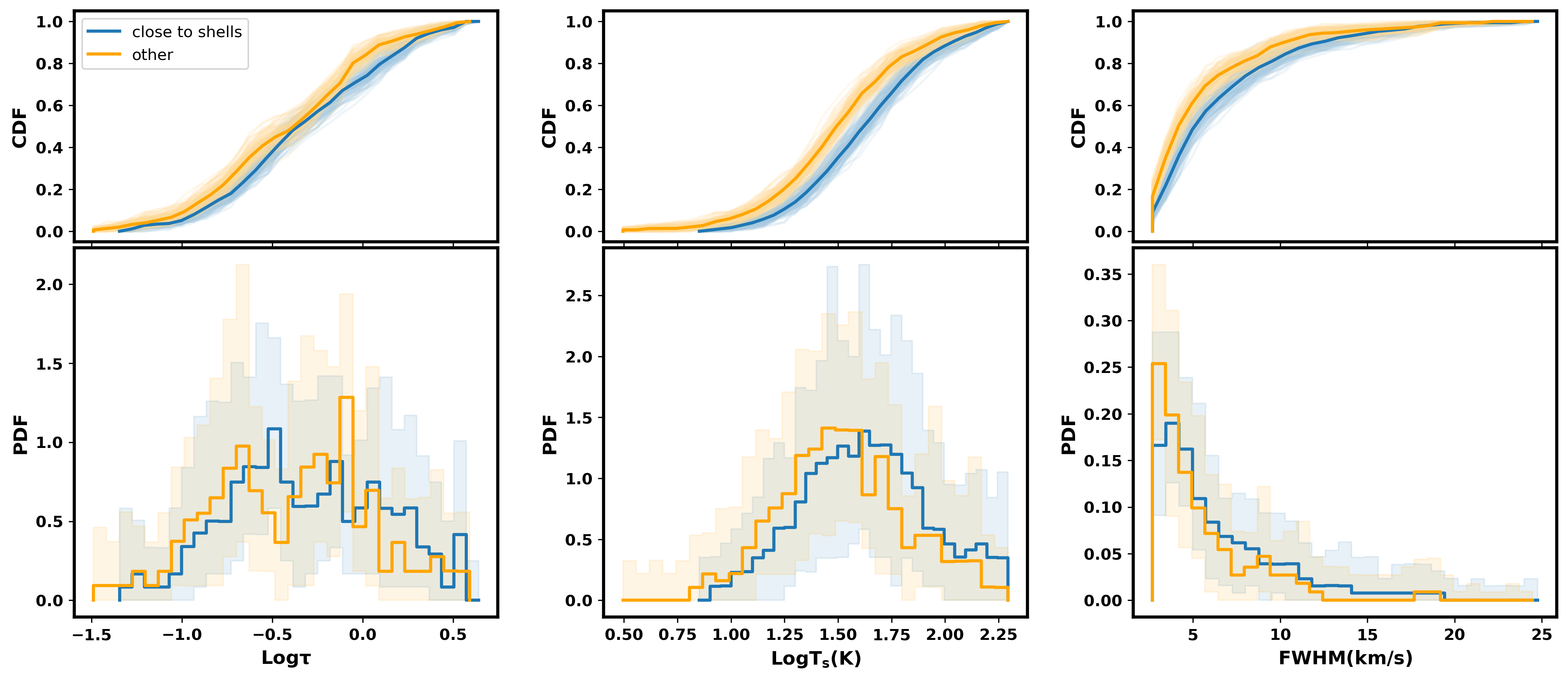}
    \end{subfigure}

    \vspace{1em}  

    \begin{subfigure}{\textwidth}
        \centering
        \includegraphics[width=0.7\linewidth]{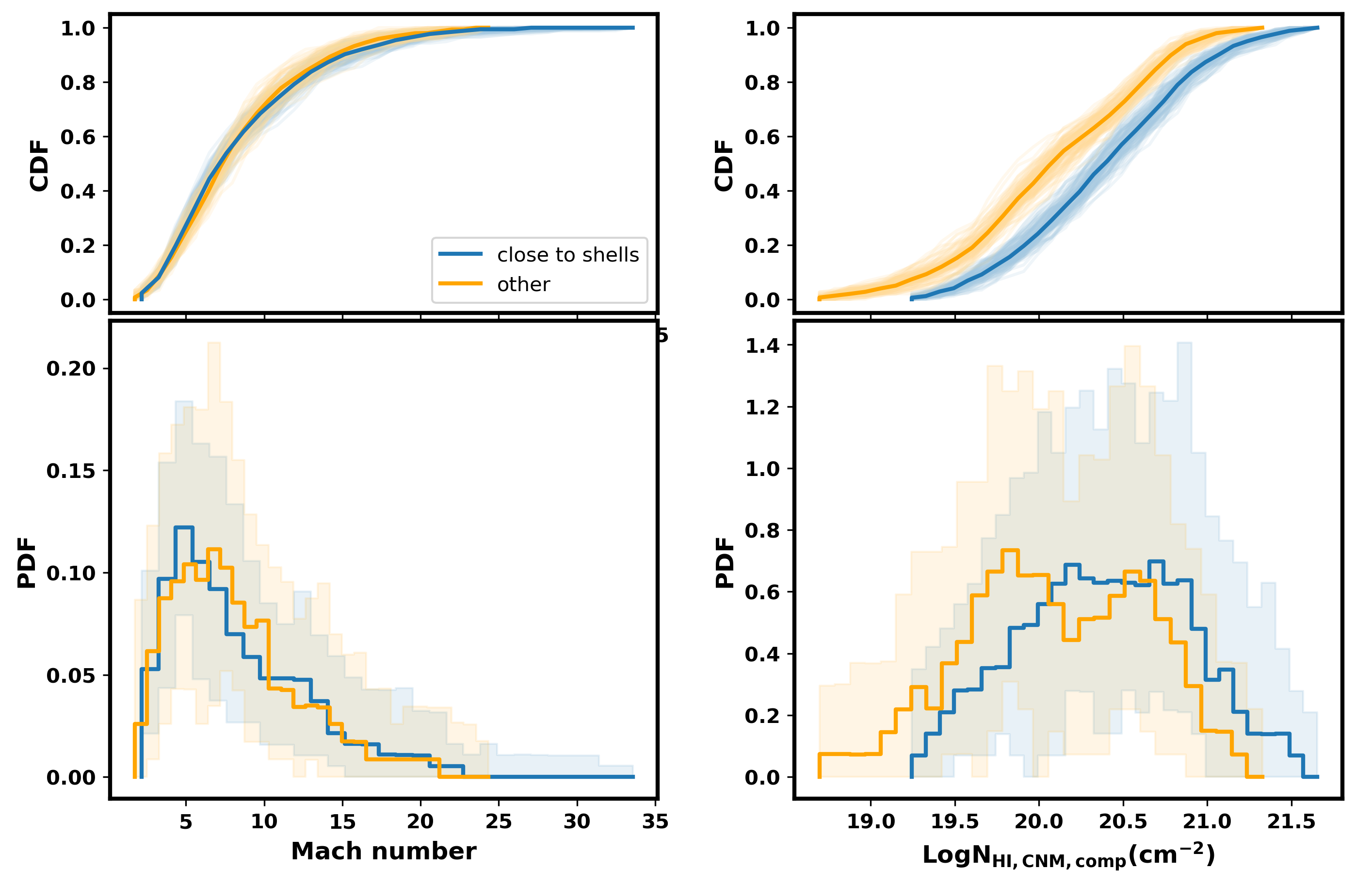}
    \end{subfigure}
    \caption{Property distributions of the CNM components close to shells (blue) and the other cold clouds (orange). The top row shows the peak optical depth (left), spin temperature (middle), and FWHM (right), while the bottom row presents the mach number and the HI column density for each CNM component. Each panel shows the CDF (top) and the corresponding histogram (bottom). }
    \label{fig:shell_cdf}
\end{figure*}

The LMC is a vigorously star-forming galaxy, characterized by widespread and ongoing massive stellar feedback. Its rich population of young, massive stars generates intense stellar winds, ionizing radiation, and frequent supernova explosions, all of which inject substantial energy into the ISM. These feedback processes drive the formation of expanding shells, superbubbles, and supergiant shells (SGSs) in the LMC \citep[e.g.,][]{Kim1999,Dawson2013}. As these expanding shells propagate through the ambient WNM gas, they compress and sweep up the surrounding gas. The post-shock regions behind these expanding fronts often reach sufficiently high densities and pressures that thermal instability is triggered, allowing the shocked gas to condense into the CNM. This process leads to the formation of dense, cold atomic clouds and clumps within the shell walls, and eventually to molecular clouds \citep{Guo2024}. 

The LMC hosts numerous shells and supergiant shells, providing an excellent environment to study the impact of shells on cold gas formation. In Figure~\ref{fig:spatial_comp} and the left panel of Figure~\ref{fig:fcnm}, we show the locations of these shells using blue contours and circles, where supergiant shells are adopted from \citet{Dawson2013} and giant shells from \citet{Kim1999}. In \citetalias{Chen2025}, we highlighted that shells can help maintain cold gas in regions of low column density within the LMC. With the expanded samples used in this work, we now quantify this effect statistically. Specifically, we separate clouds located near shells from those that are not to systematically assess how the presence of shells influences local CNM properties across the LMC.

To identify clouds associated with supergiant shells, we use the SGS mask from \citet{Dawson2013}, where shell regions are assigned a value of 1 and other regions 0. We calculate the mean mask value within a $2^{\prime}$ radius around each source, classifying a source as “close to shells” if this average exceeds 0.5. For giant shells identified by \citet{Kim1999}, we use their published central coordinates and radii. For each source, we calculate the projected distance to the nearest shell center. To account for the finite thickness of the shell, we define a shell rim annulus of width $2^{\prime}$ centered on the shell radius. Specifically, a source is classified as shell-associated if its projected distance satisfies $R_{\rm shell} - 1^{\prime} \le d \le R_{\rm shell} + 1^{\prime}$, where $R_{\rm shell}$ is the shell radius and $d$ is the distance between the source and the shell center. Combining both criteria, we identify 72 sources located near shells and 83 located outside shell regions. 

The distributions of these two populations in the CNM fraction versus total HI column density plane are shown in Figure~\ref{fig:shell_fcnm_NHI}. The CNM fractions of sources located near shells are systematically higher than those not. This is consistent with previous findings by \citet{Marx-Zimmer2000}, who reported enhanced cold gas fractions around the supergiant shell LMC 4 \citep{Meaburn1980} and the 30 Doradus region compared to the other regions in the LMC. They attributed this enhancement to a higher cooling rate in these regions compared to other parts of the LMC, probably due to an increased pressure near the shock fronts. However, Figure~\ref{fig:shell_fcnm_NHI} also shows that clouds near shells tend to have higher total HI column densities. In particular, beyond $N_{\mathrm{HI}} \sim 4\times10^{21}~\mathrm{cm^{-2}}$, nearly all clouds are located near shell regions. This suggests that the apparent increase in CNM fraction near shells may partly arise from the fact that these regions coincide with denser environments, where both shielding and cooling are more efficient (see Section~\ref{sec:cnm fraction}).

To mitigate this bias, we restrict the comparison to clouds with total HI column densities between $1\times10^{21}$ and $4\times10^{21}~\mathrm{cm^{-2}}$, ensuring both samples span a similar density range. As shown in the right panel of Figure~\ref{fig:shell_fcnm_NHI}, the median CNM fractions of the two populations are comparable, although sources near shells show fewer extremely low $f_{\rm CNM}$ ($<10\%$) cases. This suggests that, while shell environments do not substantially elevate the overall CNM fraction, they help sustain a baseline level of cold HI.


Using the same spatial selection criteria for supergiant shells (SGSs) and giant shells (GSs), we identify 152 CNM components associated with shells and 176 components not associated with any shell structures. Each shell in the catalog has a defined central velocity, however, we do not restrict our CNM cloud selection to components whose central velocities closely match those of the shells. This is because cold gas with large velocity offsets may still be physically related to the shells -- for instance, as part of expanding or outflowing motions triggered by shell feedback.

Figure~\ref{fig:shell_cdf} presents the property distributions of CNM components for clouds located near shells (blue) and those farther away (orange). CNM components in the vicinity of shells have systematically higher spin temperatures, linewidths and HI column density and have a similar distribution of optical depth and Mach number.  The higher spin temperatures near shells indicate slightly warmer CNM, possibly due to mild heating from stellar feedback or shock compression/dissipation. Although these clouds also exhibit broader FWHM values, their Mach numbers remain similar to those of CNM farther from shells. Since the Mach number primarily traces turbulent motions, while the FWHM reflects both turbulence and thermal broadening, this similarity implies that the turbulent energy does not significantly differ between the two samples. Therefore, the broader linewidths of CNM near shells are most likely driven by enhanced thermal motion rather than the increased turbulence. 
The comparable optical depths suggest similar local absorption conditions between the samples.
The corresponding higher CNM column densities for shell-associated components indicate that expanding shells not only heat and compress the surrounding gas but also promote CNM formation and increase its overall mass content in the LMC.

\section{Kinematics of the cold gas} \label{sec:kine}
\begin{figure*}
    \centering
    \includegraphics[width=0.9\linewidth]{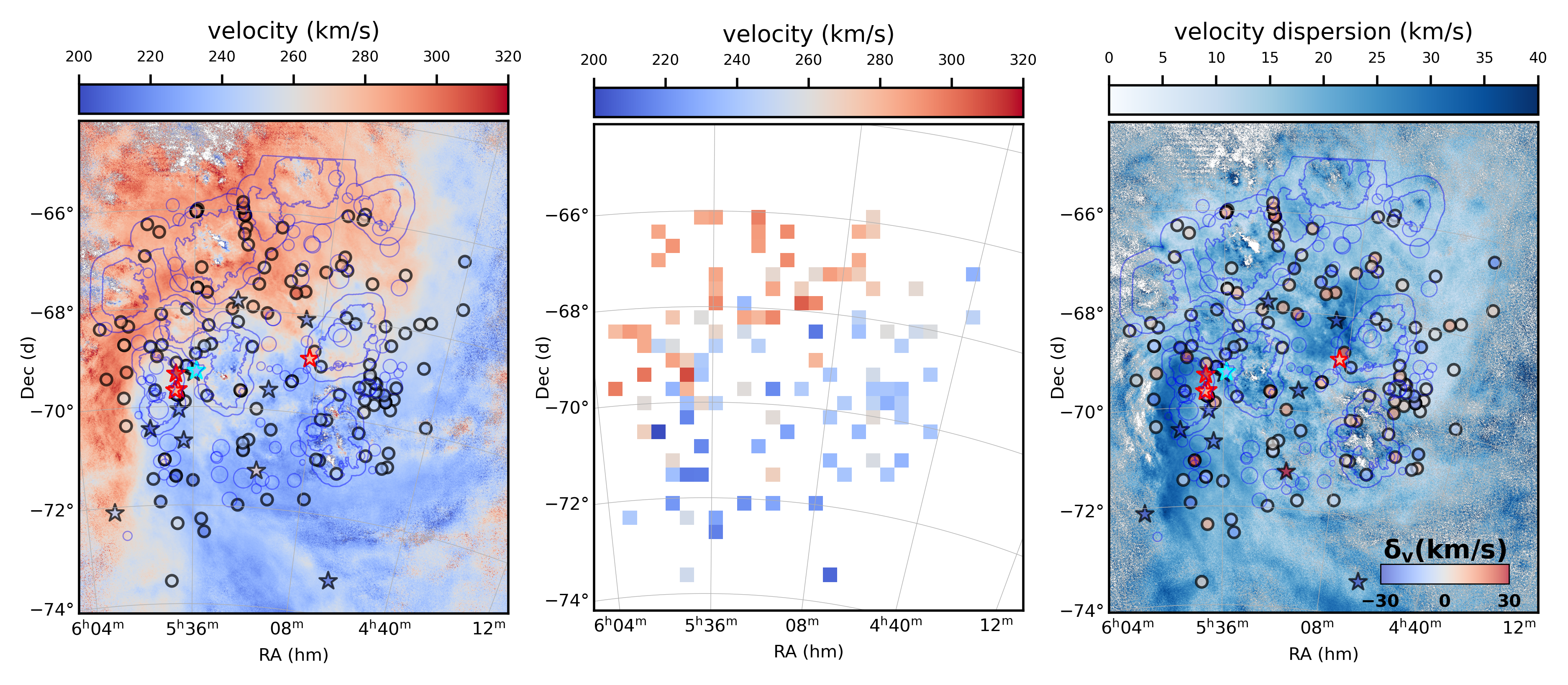}
    \caption{Left: The background shows the intensity-weighted velocity (moment 1) map of the LMC. Dots/stars mark the locations of all our selected sources.
    They are color-coded by the central velocities of CNM components with the largest velocity differences ($\delta_v$) between their central velocities and the local moment 1 values, using the same velocity scale as the background for consistency.
    Middle: Binned map of the velocity of CNM component with the largest velocity differences. Each pixel represents the mean velocity of CNM components within a spatial bin/resolution of 0.3 degree, with blank regions indicating no available data. The color has the same scale as the left panel.
    Right: The same CNM components are overlaid on the velocity dispersion (moment 2) map, where dot/star colors represent their velocity offsets ($\delta_v$). In both maps, supergiant shells (SGSs; contours) and giant shells (GSs; open circles) from \citet{Dawson2013} and \citet{Kim1999} are shown. The maps are restricted to regions with $N_{\rm HI} > 10^{19}~\mathrm{cm^{-2}}$. The stars in the left and right panels show the location of CNM components that have $|\delta_v|>25~\mathrm{km~s^{-1}}$ and the dots show all other clouds. Among the large-offset components, red and cyan stars denote sources located within the LMC, where red stars indicate inflowing CNM components and the cyan star indicates an outflowing CNM component.
    }
    \label{fig:spatial_velocity}
\end{figure*}

\begin{figure}
    \centering
    \centering
    \begin{subfigure}[b]{0.42\textwidth}
        \includegraphics[width=\linewidth]{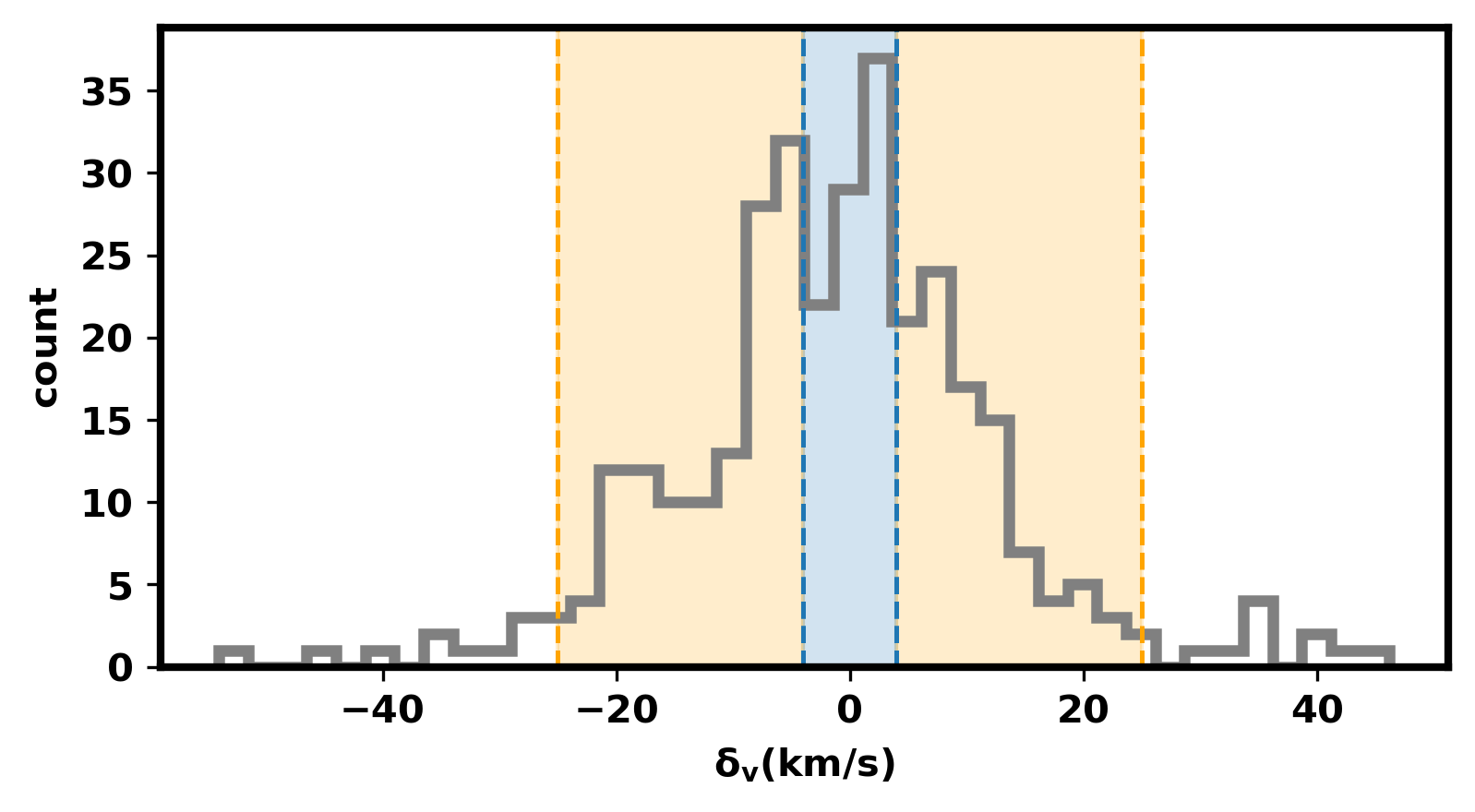}
    \end{subfigure}
    \vspace{0.3cm}
    \begin{subfigure}[b]{0.45\textwidth}
        \includegraphics[width=\linewidth]{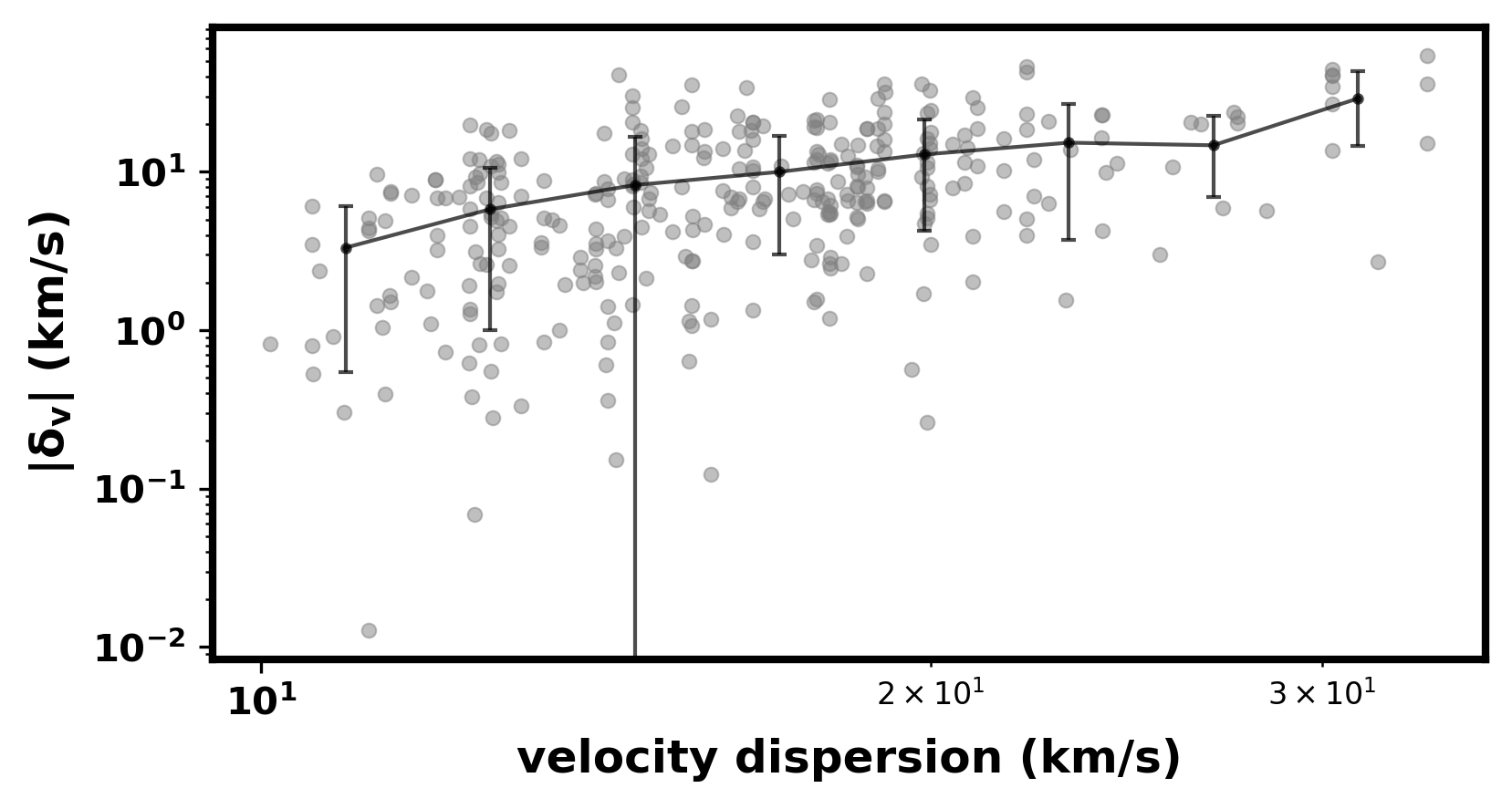}
    \end{subfigure}
    \caption{Top: Histogram of velocity offsets ($\delta_v$) for all the CNM components. The blue-shaded region marks components within $\pm4~\mathrm{km~s^{-1}}$, the orange region marks $4$–$25~\mathrm{km~s^{-1}}$, and components outside this range are classified as potentially tracing local inflows or outflows. 
    Bottom: The absolute velocity difference $|\delta_v|$ as a function of the surrounding velocity dispersion of the CNM components. The solid connected lines represent the binned mean $|\delta_v|$  in logarithmic velocity dispersion bins, with vertical error bars indicating the standard deviation within each bin.
    }
    \label{fig:delta_v_v_disp}
\end{figure}

The left panel of Figure~\ref{fig:spatial_velocity} displays the central velocities of the CNM components that have the largest deviations compared to their surroundings along each line of sight, overlaid on the LMC’s intensity-weighted velocity (moment 1) map. The middle panel presents the spatial distribution of these component velocities alone, revealing a large-scale rotational pattern consistent with that of the LMC disk. 

The color difference between individual CNM components and the underlying velocity field highlights the extent to which some clouds deviate from the global disk rotation.
To quantify these deviations, we define the velocity difference as $\delta_v = v_0 - v_{\rm disk}$, where $v_0$ is the central velocity of the CNM component and $v_{\rm disk}$ is the averaged local moment 1 within $2'$. Because the bulk of the HI gas in the LMC is known to co-rotate with its young stellar disk \citep{Olsen2007, Olsen2011, Zheng2024}, we choose the moment 1 velocity as a reliable proxy for the local disk velocity. 
By comparing these velocities with the local disk velocity field, we can determine whether each CNM cloud is kinematically consistent with the rotating LMC disk or exhibits significant velocity offsets, potentially indicating the inflow/outflow of gas. 

The top panel of Figure~\ref{fig:delta_v_v_disp} shows the histogram distribution of $\delta_v$ for all CNM components. Following the criterion adopted in \citet{Chen2025}, components with $|\delta_v| < 4~\mathrm{km~s^{-1}}$ likely represent small-scale turbulent fluctuations or localized systematic motions, such as gravitational collapse or mild stellar activity. We find 111 CNM components (34\%) within this range. Components with $4 \leq |\delta_v| \leq 25~\mathrm{km~s^{-1}}$ correspond to the typical shell expansion velocity range in the LMC and are likely influenced by shell-driven motions. We identify 196 components (59\%) within this range, indicating that the majority of the CNM components are likely affected by the shell motions. Although not all are spatially associated with known supergiant or giant shells from \citet{Dawson2013} and \citet{Kim1999}, some may correspond to previously unidentified or fragmented shells along the line of sight. 

Finally, 23 CNM components (7\%) exhibit $|\delta_v| > 25~\mathrm{km~s^{-1}}$, indicative of significant kinematic deviations from disk rotation. These components may trace inflows or outflows of cold gas. Using the source classifications listed in Table~\ref{tab:full}, which indicate whether a background source lies within the LMC, we find that seven of these components toward 5 sightlines 
correspond to sightlines located inside the LMC and are spatially close to HII regions or supernova remnants. The corresponding sightlines and $\delta_v$ values are J053528$-$691612 ($-29~\mathrm{km~s^{-1}}$), J054024$-$694014 ($+33~\mathrm{km~s^{-1}}$), J050953$-$685302 ($+25$ and $+30~\mathrm{km~s^{-1}}$), J054009$-$691951 ($+43$ and $+46~\mathrm{km~s^{-1}}$), and J053943$-$693847 ($+36~\mathrm{km~s^{-1}}$). Following the interpretation by \citet{Park2026}, and considering that the CNM absorption components must lie in front of the main HI emission along the line of sight, blueshifted components (negative $\delta_v$) can be interpreted as outflowing gas, while redshifted components (positive $\delta_v$) correspond to inflowing gas. Among these 7 components, only one in J053528$-$691612 (near SN1987A) exhibits a negative $\delta_v$ (the location is marked as a cyan star in Figure~\ref{fig:spatial_velocity}), while the remaining six have positive $\delta_v$ (the locations are marked as red stars in Figure~\ref{fig:spatial_velocity}). This suggests the presence of one outflowing and six inflowing CNM components within the LMC.
For the remaining sources, distinguishing between inflow and outflow requires precise three-dimensional spatial information that is not available with the current data. Therefore, we analyze them collectively here. 

The background of the right panel of Figure~\ref{fig:spatial_velocity} shows the velocity dispersion (moment 2) map of the LMC. Regions of large velocity dispersion trace areas where gas motions are enhanced, either through turbulence, star forming activities and supernova feedback \citep{Tamburro2009}. These processes can broaden spectral lines and introduce multiple overlapping velocity components along the line of sight, leading to large velocity dispersion. To examine how the velocity dispersion relates to the kinematics of the cold gas, Figure~\ref{fig:delta_v_v_disp} shows the relationship between the CNM velocity deviation ($\delta_v$) and the surrounding velocity dispersion, averaged within a $2'$ radius. 
Although there is a large scatter, we find a statistically positive correlation between these two quantities (the Spearman rank test yields $\rho = 0.47$ with $p = 9.3 \times 10^{-20}$). This indicates that CNM clouds located in regions of higher velocity dispersion tend to show larger velocity offsets from the local disk rotation.
This result implies that the same energetic processes that stir the surrounding ISM, such as shell expansion, stellar feedback, and supernova-driven turbulence, also influence the kinematics of the cold neutral gas.

The stars in the left and right panels of Figure~\ref{fig:spatial_velocity} mark the locations of CNM components identified as potential inflowing or outflowing clouds ($|\delta_v| > 25~\mathrm{km~s^{-1}}$).  We note that some high velocity absorbers projected toward the LMC may instead trace foreground Milky Way halo gas \citep{Richter2015,Poudel2025} and those velocities can reach up to 175 $\rm km ~ s^{-1}$. However, all CNM components we identified in the LMC have velocities $>200~\mathrm{km~s^{-1}}$, so they lie well beyond the foreground halo velocity range and are therefore unlikely to be associated with Milky Way gas. Several of these components are near 30 Doradus, suggesting that strong stellar feedback can accelerate or entrain cold atomic gas, driving both inflow or outflow. This interpretation is consistent with previous studies that identified large-scale outflows/inflows from the 30 Doradus region \citep[e.g.,][]{Poudel2025,Horton2025}, demonstrating that intense starburst activity can expel substantial amounts of gas from the disk. 

A few other high $|\delta_v|$ CNM clouds are found near the LMC’s central regions where star formation is active, close to shells and regions with high velocity dispersion. Interestingly, two of them lie in the very outskirts of the disk, far from major star-forming complexes. One of them is located in the outer disk of the LMC, another one is located toward the direction of the Magellanic Bridge that connects to the SMC. This suggests that the presence of CNM clouds with large velocity offsets depends on multiple factors. On one hand, strong stellar feedback or expanding shells can eject or entrain cold gas locally. On the other hand, large-scale environmental interactions, such as ram pressure from the Milky Way halo \citep{Indu2015} or the tidal interaction within the Magellanic system, may also accelerate CNM in the outer disk \citep[e.g.,][]{McClure-Griffiths2018,Dempsey2020,Chen2025}.

We note, however, that these interpretations may be affected by projection effects along the line of sight. In regions with significant depth, the moment 1 velocity could be dominated by one major HI component, while the detected absorption arises from a physically distinct structure at a different velocity. In such cases, a large $|\delta_v|$ would not necessarily indicate true non-circular motion relative to the local disk, but could instead reflect the presence of multiple kinematically distinct layers along the same sightline. Similar line-of-sight complexity has been reported in the SMC, where distinct velocity components trace physically separate structures within a single direction \citep[e.g.,][]{Murray2024}. Therefore, while $\delta_v$ can highlight kinematic deviations, it may in some cases trace superposed structures rather than inflow or outflow.

\section{Summary}\label{sec:conclu}
In this paper, we investigate the properties of the cold atomic gas within the LMC. We use data from the GASKAP-HI survey, which currently provides the largest sample of HI absorption detections in the LMC, also with the highest available spectral and spatial resolution. We apply the radiative transfer method to decompose the HI emission and absorption spectra of 155 sightlines. Using this method, we identify 330 CNM components, 2 UNM components from the HI absorption and 310 WNM components from the HI emission. Our main findings are summarized below.

\begin{itemize}
    \item For individual CNM components, we find that the LMC exhibits systematically higher HI optical depths compared to the Milky Way, both relative to the BIGHICAT catalog \citep{McClure-Griffiths2023} and the GASKAP-foreground Survey \citep{Nguyen2024}, despite the latter having comparable sensitivity. 
Most CNM components in the LMC exhibit spin temperatures between $\sim$10 and 200 K, with a median of $\sim$37 K and a mean of $\sim$50 K. These values are  lower than those observed in both Milky Way surveys, indicating that CNM clouds in the LMC are on average colder, which is consistent with expectations for a lower metallicity and higher UV radiation environment. 
The CNM linewidths range from 2 to 25 km s$^{-1}$, with a median of 4.9 km s$^{-1}$ and a mean of 6.3 km s$^{-1}$. All components have Mach numbers $\mathrm{M}>1$, indicating supersonic turbulence. Both linewidths and Mach numbers are larger than in the Milky Way, suggesting that the LMC’s CNM is more turbulent. Some broadening may also arise from unresolved multiple components within the GASKAP beam. 

    \item We identify two UNM components toward sources J052456–693855 and J051832–693521, both with spin temperatures of $\sim$600 K. This is the first direct detection of thermally unstable HI using radiative transfer method outside the Milky Way. Despite similar temperatures, their properties and surrounding environments differ.

    \item CNM components in the LMC generally exhibit higher individual column densities than those in the Milky Way, while the total CNM column density per line of sight shows only modest differences. This may reflect a combination of the less resolved ability of CNM components and intrinsic environmental effects (lower metallicity and stronger UV radiation fields) in the LMC. As a result of longer sampled path length and thoes intrinsic environmental effects, the total HI column density in the LMC is higher the CNM fraction in the LMC (median of 23\% and mean of 28\%) is slightly lower than in the Milky Way. The CNM fraction increases with total HI column density, and the LMC is shifted toward higher HI column densities relative to the Milky Way, suggesting that the HI-to-H$_2$ transition occurs at higher HI columns in the LMC. We also find a mild decrease in CNM fraction with increasing distance from the LMC center.

    \item We compare the CNM column density with its subsequent evolutionary stages of molecular gas and star formation in the LMC. We find that regions with high H$_2$ column density also show high CNM column densities. The CNM-to-H$_2$ transition begins near $N{\rm _{HI}} \sim 10^{20}~\mathrm{cm^{-2}}$, marking the onset of sufficient shielding for H$_2$ formation. Beyond $N{\rm _{HI}} \sim 10^{21}~\mathrm{cm^{-2}}$, molecular gas is always present, indicating that the gas is fully self-shielded and efficiently converts into the molecular phase. The star formation rate surface density ($\Sigma_{\rm SFR}$) increases with CNM column density, with nearly all regions showing molecular gas when $\Sigma_{\rm SFR} > 10^{-2}~M_{\odot}~\mathrm{yr^{-1}~kpc^{-2}}$. This correlation supports the view that CNM serves as the immediate reservoir for H$_2$ formation, while molecular gas is more directly linked to subsequent star formation. 

     \item We find a positive correlation between the CNM fraction and visual extinction ($A_V$), providing insight into the conditions required for CNM survival. In low-extinction regions ($A_V < 1$), the LMC can sustain higher CNM fractions compared to the Milky Way. This likely because of the higher HI columns associated with a given CNM fraction in the LMC, which may trace physical conditions that favor the cold phase, including higher local density or thermal pressure.
     Both distributions flatten at $A_V > 1$, indicating that dust shielding reaches a saturation level. In addition, clouds with high CNM fractions in the Milky Way are typically found at $A_V > 1$, whereas in the LMC similarly high CNM fractions can already occur at $A_V \sim 0.1$, implying that additional processes may also contribute.

    \item To probe the role of local environments, we compare CNM properties in regions near shells with those not. We find that clouds near shells generally exhibit higher CNM fractions, though this trend partly reflects their higher HI column densities. When restricting the comparison to a fixed column density range, the two populations show similar CNM fractions, except that clouds near shells rarely fall below $f_{\rm CNM}\sim10\%$. Individual CNM components show that those near shells tend to have higher spin temperatures, broader linewidths, and larger column densities, while their optical depth and Mach number distributions remain comparable to those farther from shells.

    \item We investigate the kinematics of the CNM by comparing the central velocity of each component to the surrounding bulk HI motion. About 34\% of the CNM components have velocity differences smaller than 4 km s$^{-1}$, likely reflecting small-scale turbulent or systematic fluctuations. The majority (59\%) show $4 < |\delta_v| < 25$ km s$^{-1}$, consistent with typical shell expansion velocities, suggesting that much of the CNM is influenced by shell-driven motions. The remaining 7\% exhibit $|\delta_v| > 25$ km s$^{-1}$, indicative of potential inflow or outflow. These large velocity CNM components are mainly concentrated in the central regions of the LMC, such as around 30 Doradus, though a few are also found in the outskirts. This distribution suggests that both strong stellar feedback in the LMC’s active star-forming regions and large-scale environmental interactions, such as with the Milky Way or the Magellanic system, may accelerate CNM formation.
\end{itemize}

Our study provides the most comprehensive view to date of the cold atomic gas in the LMC, based on the unprecedented sensitivity and resolution of the GASKAP HI absorption survey. By decomposing the HI spectra into CNM, UNM, and WNM components, we present the first detailed comparison of CNM properties between the LMC and the Milky Way. The observed differences are broadly consistent with the lower metallicity and higher UV radiation field in the LMC, while also reflecting the different observing path length of the two galaxies. By linking these results to molecular gas and star formation, we find that CNM is more closely associated with molecular gas, and molecular gas correlates more strongly with star formation, highlighting the next stage of cold gas evolution. 
The LMC also maintains relatively high CNM fractions at low dust extinction compared to the Milky Way, likely because of higher HI column density for individual CNM components that 
provides effective cooling due to higher local density and thermal pressure.
suggesting a different cold gas survival picture compared to the Milky Way. We also examine the local impact of shells and find systematic variations in cold gas properties. Overall, this work provides the first spatially-resolved comparison between the cold atomic phase and its surrounding environment in the LMC, establishing an observational foundation for understanding how metallicity/UV radiation field and local feedback shape the multiphase ISM.

In future work, we will extend this analysis to the SMC. Although the SMC has been explored in previous GASKAP HI absorption studies \citep[e.g.,][]{Dempsey2022}, no detailed spectral decomposition has yet been performed. A comparable analysis of the SMC would enable a more systematic comparison of cold HI properties across three different metallicity environments (SMC $\sim 0.2~Z_\odot$, LMC $\sim 0.5~Z_\odot$, and Milky Way $\sim 1~Z_\odot$). While these galaxies differ in viewing geometry and line of sight depth, intrinsic cloud scale properties such as optical depth, spin temperature, and CNM component column density can still provide key insight into how individual CNM clouds vary with metallicity. Combined with the surrounding environment (e.g. dust, molecular gas), this will help clarify the physical conditions that regulate the formation and survival of the CNM beyond the Milky Way.





\section*{acknowledgments}
We thank the referee for suggestions that improved the clarity of this paper.
We thank  H. Hassani for providing the synchrotron emission maps for the LMC. We thank Dr. Change-Goo Kim for helpful discussions and comments that improved the manuscript.
This work made use of the SIMBAD database and the Aladin sky atlas operated by CDS, Strasbourg, France. 
GASKAP-HI is partially funded by the Australian Government through an Australian Research Council Australian Laureate Fellowship (project number FL210100039 awarded to NMc-G). The GASKAP HI emission data were imaged using
CHTC services—Center for High Throughput Computing (2006): doi:10.21231/GNT1-HW21.
SS acknowledges support provided by the University of Wisconsin–Madison 
through the Vilas Distinguished Achievement Professorship, as well as funding provided by the NSF Award AST-2108370 and NASA award 80NSSC21K0991. This scientific work uses data obtained from Inyarrimanha Ilgari Bundara, the CSIRO Murchison Radio-astronomy Observatory. We acknowledge the Wajarri Yamaji People as the Traditional Owners and native title holders of the Observatory site. CSIRO’s ASKAP radio telescope is part of the Australia Telescope National Facility (https://ror.org/05qajvd42). Operation of ASKAP is funded by the Australian Government with support from the National Collaborative Research Infrastructure Strategy. ASKAP uses the resources of the Pawsey Supercomputing Research Centre. Establishment of ASKAP, Inyarrimanha Ilgari Bundara, the CSIRO Murchison Radio-astronomy Observatory and the Pawsey Supercomputing Research Centre are initiatives of the Australian Government, with support from the Government of Western Australia and the Science and Industry Endowment Fund.

\section*{Data Availability}

This work makes use of archived data obtained through the CSIRO ASKAP Science Data Archive (CASDA; \citealt{Chapman2017, Huynh2020}), available at \url{https://research.csiro.au/casda}. 
The radiative transfer code used in this work is publicly available at \url{https://github.com/retarchen/specRT}. All spectral fitting results including the tables and figures can be accessed at \url{https://retarchen.github.io/fitting_results_LMC/index.html}.

\appendix

\section{Spectral Decomposition Method}\label{sec:method_app}
As shown in \citetalias{Chen2025}, we decompose the HI emission and absorption spectra into individual CNM or WNM structures along the line-of-sight using the radiative transfer method \citep{Heiles2003}, assuming that the HI absorption spectrum (1$-e^{-\tau}$) mainly originates from the CNM, while the HI emission spectrum (brightness temperature) is a product of both CNM and the WNM.
Combining with the contribution from the diffuse radio continuum emission $T_{sky}$ , the total expected brightness temperature can be expressed as (equations (2) to (6) in \citetalias{Chen2025}):
\begin{equation}\label{eq:Texp}
T_{\exp }(v)=T_{B, \mathrm{CNM}}(v)+T_{B, \mathrm{WNM}}(v)+T_{\text {sky }}(e^{-\tau(v)}-1),
\end{equation}
with
\begin{align}
\tau(v)&=\sum_0^{N-1} \tau_{0, n} e^{-4 \ln 2 \left[\left(v-v_{0, n}\right) / \delta v_n\right]^2},\\
T_{B, \mathrm{CNM}}(v)&=\sum_0^{N-1} T_{s, n}\left(1-e^{-\tau_n(v)}\right) e^{-\sum_0^{M_n-1} \tau_m(v)},\\
T_{B, \mathrm{WNM}}(v) &= \sum_0^{K-1}\bigg[F_k+\left(1-F_k\right) e^{-\tau(v)}\bigg] \times \nonumber \\
&\quad T_{0, k} e^{-4 \ln 2 \left[\left(v-v_{0, k}\right) / \Delta v_k\right]^2},
\end{align}
where the optical depth $\tau(v)$ is a combination of $N$ Gaussian functions, corresponding to $N$ CNM components.
$\tau_{0, n}$ is the peak optical depth, $v_{0, n}$ is the central velocity, and $\Delta v_n$ is the FWHM of component $n$. 
$\tau_m(v)$ represents each of the $M_n$ CNM clouds that lie in front of cloud $n$ and we will consider all possible orders of the CNM clouds. $T_{s,n}$ represents the derived spin temperature for each order.
$T_{0, k}$, $v_{0, k}$ and $\Delta v_k$ represent the Gaussian fitting parameters (peak in units of brightness temperature, central velocity, FWHM) of original unabsorbed $k$-th WNM cloud. $F_k$ is the fraction of the $k$-th WNM component located in front of all CNM components, while $(1 - F_k)$ represents the fraction lying behind the CNM and therefore subject to absorption. We examine three specific values for the fraction $F_k: (0, 0.5, 1)$. The two extremes where $F_k=0$ and $F_k=1$ correspond to all WNM lying behind the CNM with total absorbed WNM and all WNM being in front of CNM with unabsorbed WNM, respectively. We set the minimum brightness temperature for WNM $T_{0, k}$ to be the average of the 1$\sigma$ noise level of the emission spectra.

The diffuse radio continuum emission  $T_{\text{sky}}$ includes the Cosmic Microwave Background (CMB) and the synchrotron emission, which are always present but typically weak. 
We use the synchrotron emission maps $T_{syn}$ of the LMC created by \cite{Hassani2022} combining multiband observations in the LMC. The CMB temperature is $T_{\text{CMB}}=2.73$ K \citep{Fixsen2009}. Thus, $T_{\text{sky}}=T_{syn} +T_{\text{CMB}}$. 
Note that some of the CNM components identified here may, based on their spin temperatures, actually belong to the thermally unstable neutral medium (UNM). We discuss these components in detail in Section~\ref{sec:unm}.

As described in \citetalias{Chen2025}, we determine the optimal number of Gaussian components and the best fit using the Bayesian Information Criterion (BIC), which balances goodness of fit against model complexity by penalizing additional free parameters.
We first fit the optical depth spectrum with $N$ Gaussian components, starting from one component and iteratively adding more until the BIC no longer decreases by more than 10. With the resulting CNM components fixed, we then fit Equation~\ref{eq:Texp} to determine the CNM spin temperatures and the number of WNM components. Beginning with zero WNM components, we incrementally add components and compute the BIC for all possible configurations. For each iteration, there are $N!$ orderings of CNM components and $3^K$ combinations of WNM components (due to three possible values of $F_k$), yielding $N! \times 3^K$ total combinations. The configuration with the minimum BIC is selected. This process continues until the representative BIC decreases by less than $\Delta \mathrm{BIC} = 1$, at which point the previous iteration is adopted as the best model. During fitting, we allow a velocity shift of up to $\pm 4~\mathrm{km~s^{-1}}$ (four channels) between emission and absorption CNM components to account for small velocity offsets.


For each fitting model, we set the minimum spin temperature for each CNM component the same as $T_{\text{sky}}$. For the clouds with very low density, in the absence of collisions, the spin temperature would converge to $T_{\text{sky}}$ \citep{Draine2011}.
The final spin temperatures $T_s$ and the corresponding errors for each CNM component are calculated by a weighted average over all  $N!$ $3^k$ trials of $N$ absorption components and $K$ WNM components that yield the least BIC (cf. Equations (21a) and (21b) of \cite{Heiles2003}).

\section{Source location and size comparison}\label{sec:extend_point}

\begin{figure*}
    \centering

    \begin{subfigure}{\textwidth}
        \centering
        \includegraphics[width=1\linewidth]{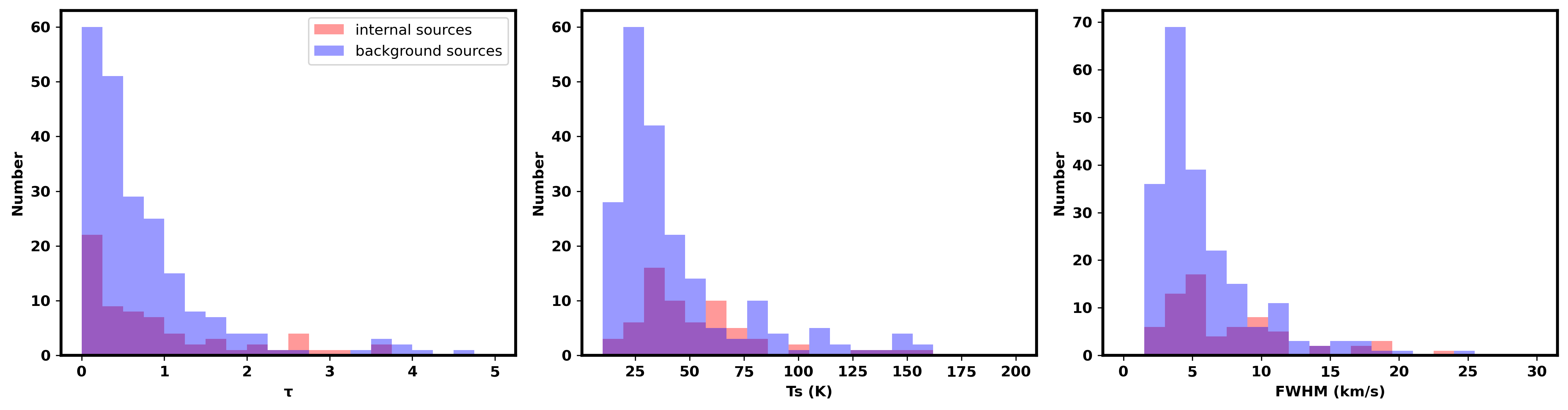}
    \end{subfigure}

    \vspace{1em}  

    \begin{subfigure}{\textwidth}
        \centering
        \includegraphics[width=1\linewidth]{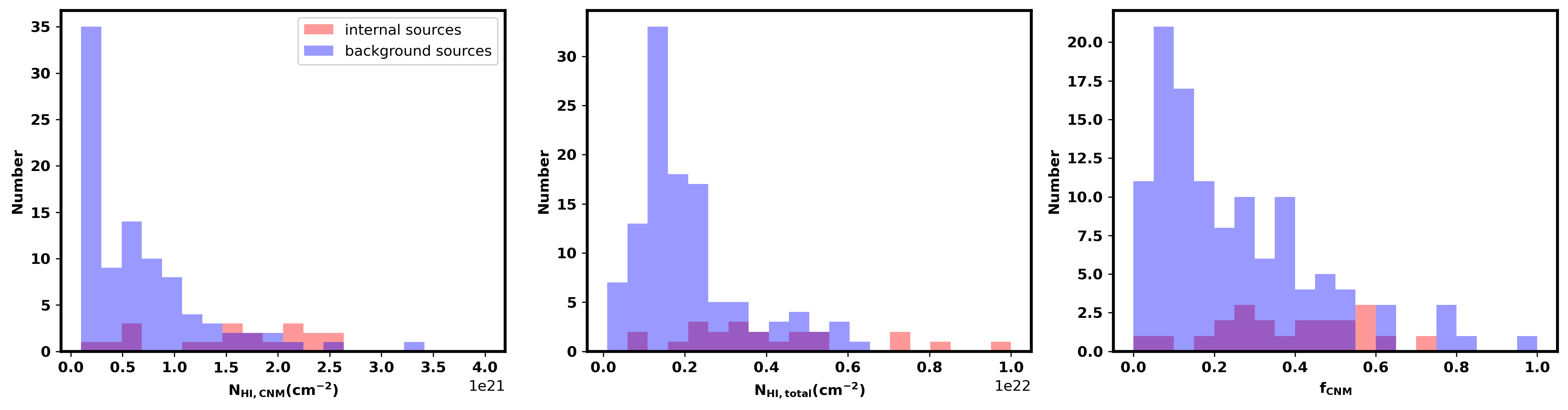}
    \end{subfigure}
    \caption{Distributions of cold HI cloud properties for internal and background sources. Panels show the optical depth (upper left), spin temperature (upper middle), FWHM (upper right), total CNM column density for LOS (lower left), total HI column density (lower middle), and CNM fraction (lower right).}
    \label{fig:internal_vs_external}
\end{figure*}

\begin{figure*}
    \centering

    \begin{subfigure}{\textwidth}
        \centering
        \includegraphics[width=1\linewidth]{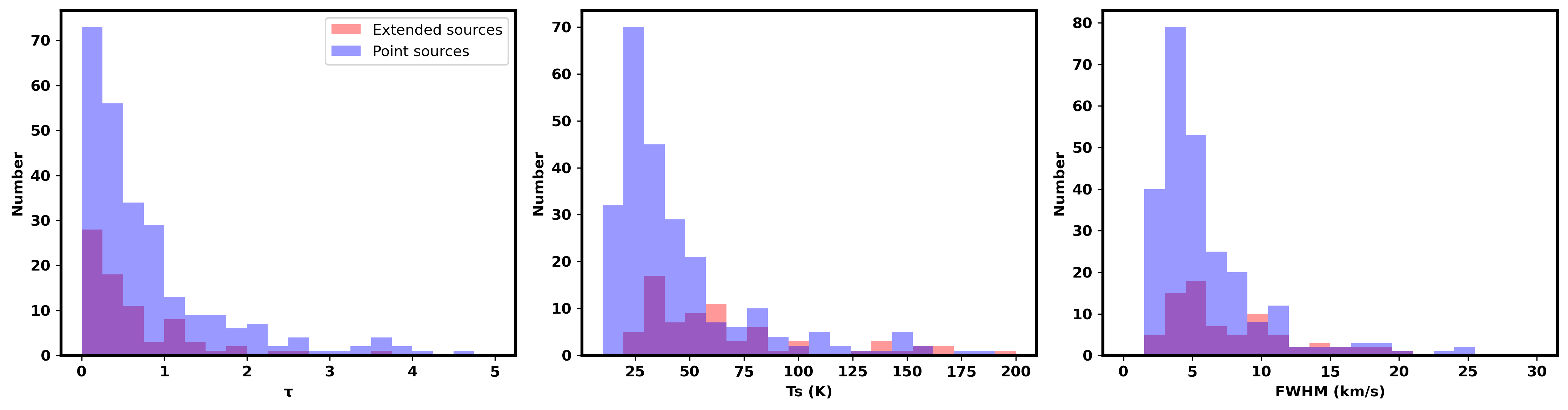}
    \end{subfigure}

    \vspace{1em}  

    \begin{subfigure}{\textwidth}
        \centering
        \includegraphics[width=1\linewidth]{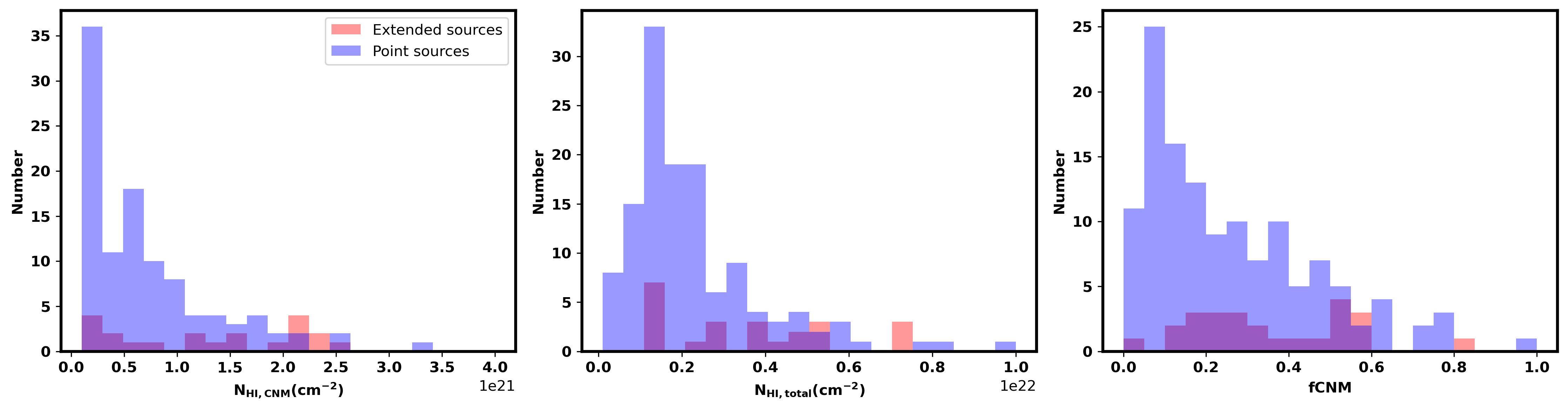}
    \end{subfigure}
    \caption{Distributions of cold HI cloud properties for extended and point sources. Panels show the optical depth (upper left), spin temperature (upper middle), FWHM (upper right), total CNM column density for LOS (lower left), total HI column density (lower middle), and CNM fraction (lower right).}
    \label{fig:extend_vs_point}
\end{figure*}
By visually inspecting the environments of the background continuum sources using SIMBAD and Aladin, we identify possible counterparts or associated structures for each source (see Table~\ref{tab:full}). Based on this classification, 22 sources (with 67 CNM components) are identified as internal to the LMC, 115 sources (with 215 CNM components) are classified as background objects, and 18 sources  (with 48 CNM components) remain unclassified. The average CNM components number for internal sources is $\sim$ 4 and  $\sim$ 2.5 for external sources, indicating the more complex cold gas structure for internal sources. Figure~\ref{fig:internal_vs_external} compares the distributions of optical depth, spin temperature, linewidth, total CNM column density, total HI column density, and CNM fraction for internal and background sources. Overall, the two populations span similar parameter ranges. The internal sources tend to show comparable optical depths, slightly higher spin temperatures and broader linewidths, as well as higher total column densities and CNM fractions. This tendency is expected as internal sources are typically associated with dense, compact star-forming regions such as YSOs or ultra-compact H II regions. A similar tendency is also seen for background sightlines intersecting star-forming regions (e.g. Section~\ref{sec:sfr} shows that CNM column densities are enhanced in such environments).

In addition, 25 sources (with 76 CNM components) in our sample are classified as extended and 130 sources (with 254 CNM components) as point-like. Among the 25 extended sources, 9 are also classified as internal sources. Figure~\ref{fig:extend_vs_point} compares the cold HI properties between these two groups, including optical depth, spin temperature, linewidth, total CNM column density, total HI column density and CNM fraction. Overall, the two populations have similar ranges, but the extended sources tend to show slightly higher spin temperatures and broader linewidths, and are skewed toward higher column densities and CNM fractions. Part of this difference is likely due to the fact that some extended sources are internal LMC sources, while many of the remaining ones are located near dense environments such as shells. This trend is therefore expected, since point-like sources probing similar environments show the similar tendency. For example, Section~\ref{sec:shell} shows that sources near shells tend to have higher temperatures, linewidth and CNM column densities.

We therefore conclude that whether a source is internal or background, or extended or point-like, does not appear to strongly affect the overall CNM properties by itself. Instead, the modest differences between these subsamples are more likely driven by the environments they preferentially probe, and the overall CNM parameter distributions remain broadly similar.

\section{Radial distribution}\label{sec:radial}
\begin{figure}
    \centering
    \includegraphics[width=1\textwidth]{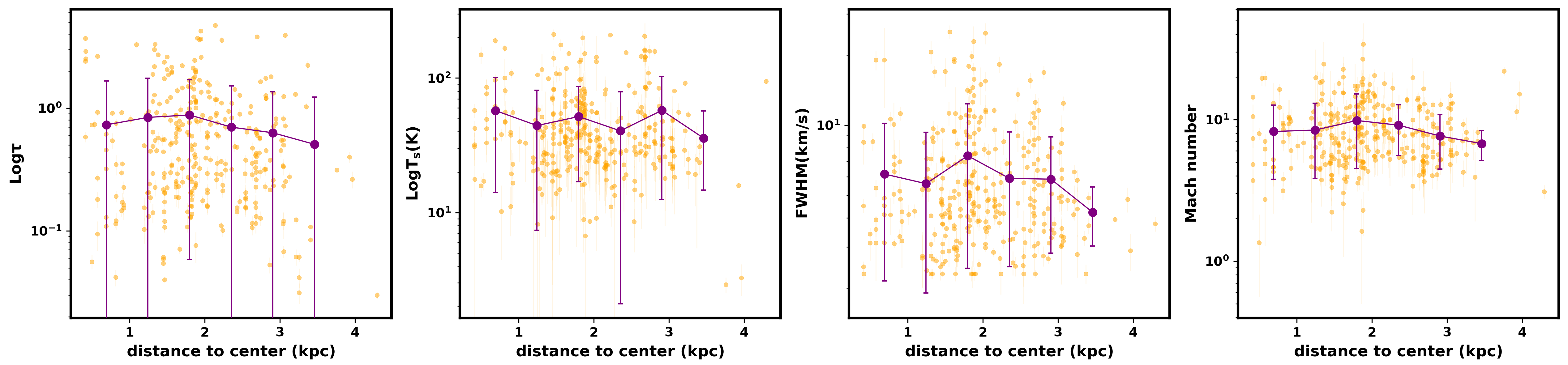}
    \caption{CNM properties as a function of source distance to the LMC's kinematic center. From left to Right, it shows the optical depth, spin temperature, FWHM and Mach number. Purple filled circles with error bars indicate the mean and standard deviation of $f_{\rm CNM}$ within different radial bins.}
    \label{fig:radius_comp}
\end{figure}

In this section, we examine the radial distribution of CNM component properties as a function of projected galactocentric distance in the LMC. Figure~\ref{fig:radius_comp} shows the optical depth ($\tau$), spin temperature ($T_s$), linewidth (FWHM), and Mach number of individual CNM components as a function of distance from the LMC kinematic center. Though  almost face-on to us \citep[with inclination degree of $23.4^\circ\pm0.5^\circ$, ][]{Choi2022}, the LMC is not a well-ordered disk galaxy like the Milky Way. The highest density region in the LMC -- 30 Doradus -- lies at $\sim$ 1.7 kpc from the kinematic center. Therefore, galactocentric distance in the LMC does not strictly trace global structural gradients in the same way as in massive spiral galaxies. The radial trends shown here should thus be interpreted as a reference framework rather than as indicators of classical inside-out disk evolution.

Compared to \citet{Marx-Zimmer2000}, who also examined the radial distribution of cold gas properties in the LMC, our sample includes a significantly larger number of sources and provides broader spatial coverage across the LMC. Overall, we do not find strong monotonic radial gradients in CNM properties. Instead, the highest values are concentrated around $\sim$1.7 kpc, corresponding to the location of 30 Doradus, and generally decline outward from this region. The outermost radial bin consistently exhibits the lowest values, suggesting that CNM properties weaken toward the extreme outskirts of the LMC. However, the substantial scatter at all radii indicates that any radial trends are secondary to localized variations. In particular, the enhancement of optical depth, spin temperature, linewidth, and Mach number near 30 Doradus highlights the strong influence of intense star formation and feedback in shaping the properties of the cold gas component.

\section{Weighted mean spin temperature}\label{sec:mean_Ts}
\begin{figure}
    \centering
    \includegraphics[width=0.45\textwidth]{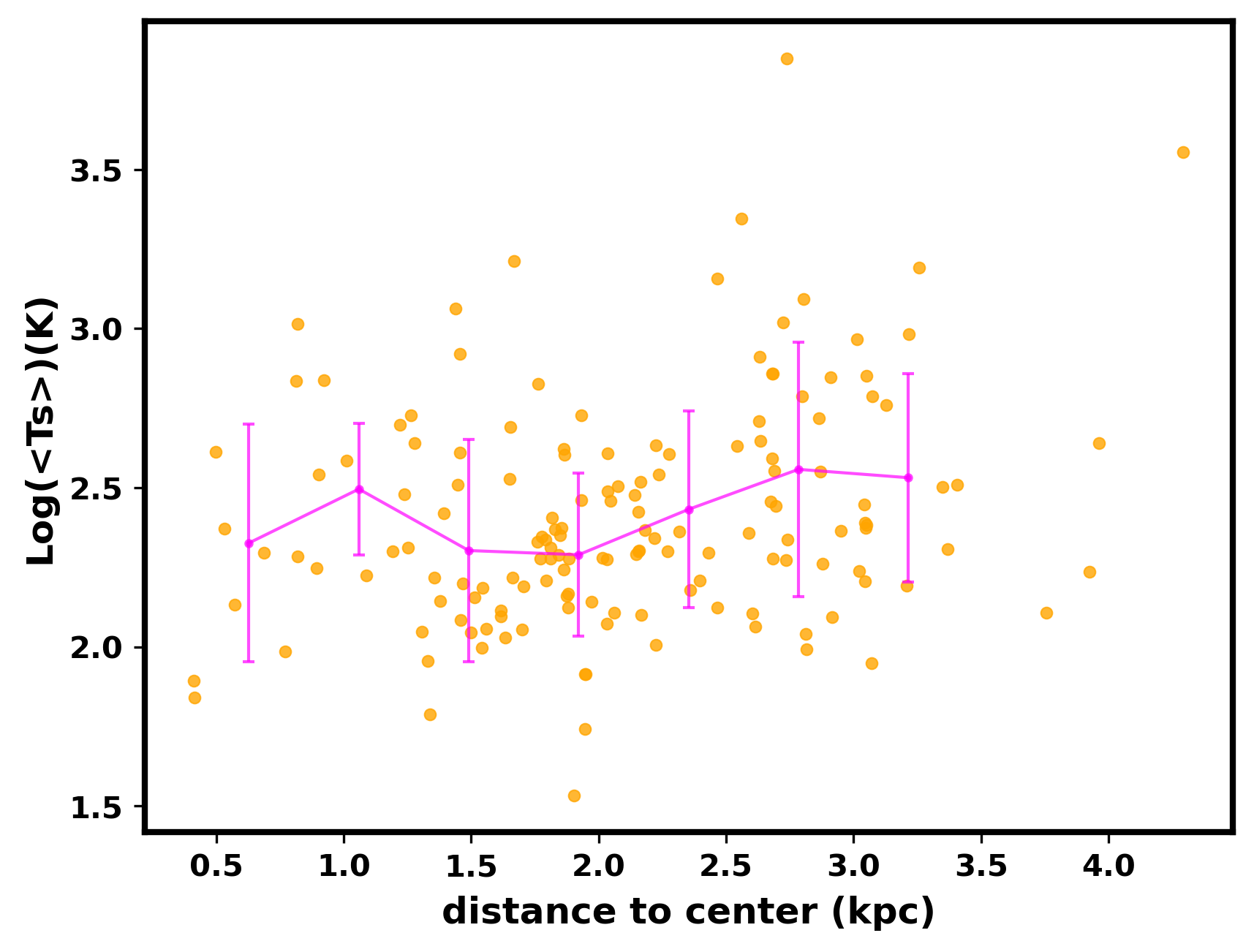}
    \caption{Weighted mean spin temperature as a function of the sources distance to the LMC's kinematic center.}
    \label{fig:mean_Ts_radius}
\end{figure}

In this section, we calculate the density-weighted mean spin temperature using the integrals of brightness temperature and optical depth, following \citet[Eq.~4 and Eq.~7]{Dickey2000}:
\begin{equation}
\langle T_{\mathrm{S}} \rangle = \frac{\int T_{\mathrm{B}}(v) dv}{\int \left[1-e^{-\tau(v)}\right] dv} \simeq \frac{T_s}{ f_{CNM} }.
\end{equation}

Figure~\ref{fig:mean_Ts_radius} shows $\langle T_{\mathrm{S}} \rangle$ as a function of galactocentric distance. The distribution reveals no clear radial trend. Since the CNM spin temperature also shows no significant radial variation in Figure~\ref{fig:radius_comp}, the approximately constant $\langle T_{\mathrm{s}} \rangle$ suggests that $f_{\rm CNM}$ remains broadly constant across most of the LMC. This is consistent with the direct radial distribution of $f_{\rm CNM}$ shown in Figure~\ref{fig:fcnm}, except for the outermost bin beyond $\sim3$ kpc, where $f_{\rm CNM}$ decreases. Because $\langle T_{\mathrm{s}} \rangle$ is less robust than the values derived from our full radiative transfer decomposition, individual bins, particularly the outermost one, may be more sensitive to measurement uncertainties.

\section{$\Sigma_{\rm SFR}$ vs. WNM column density}\label{app:sfr}

\begin{figure}
    \centering
    \includegraphics[width=0.4\linewidth]{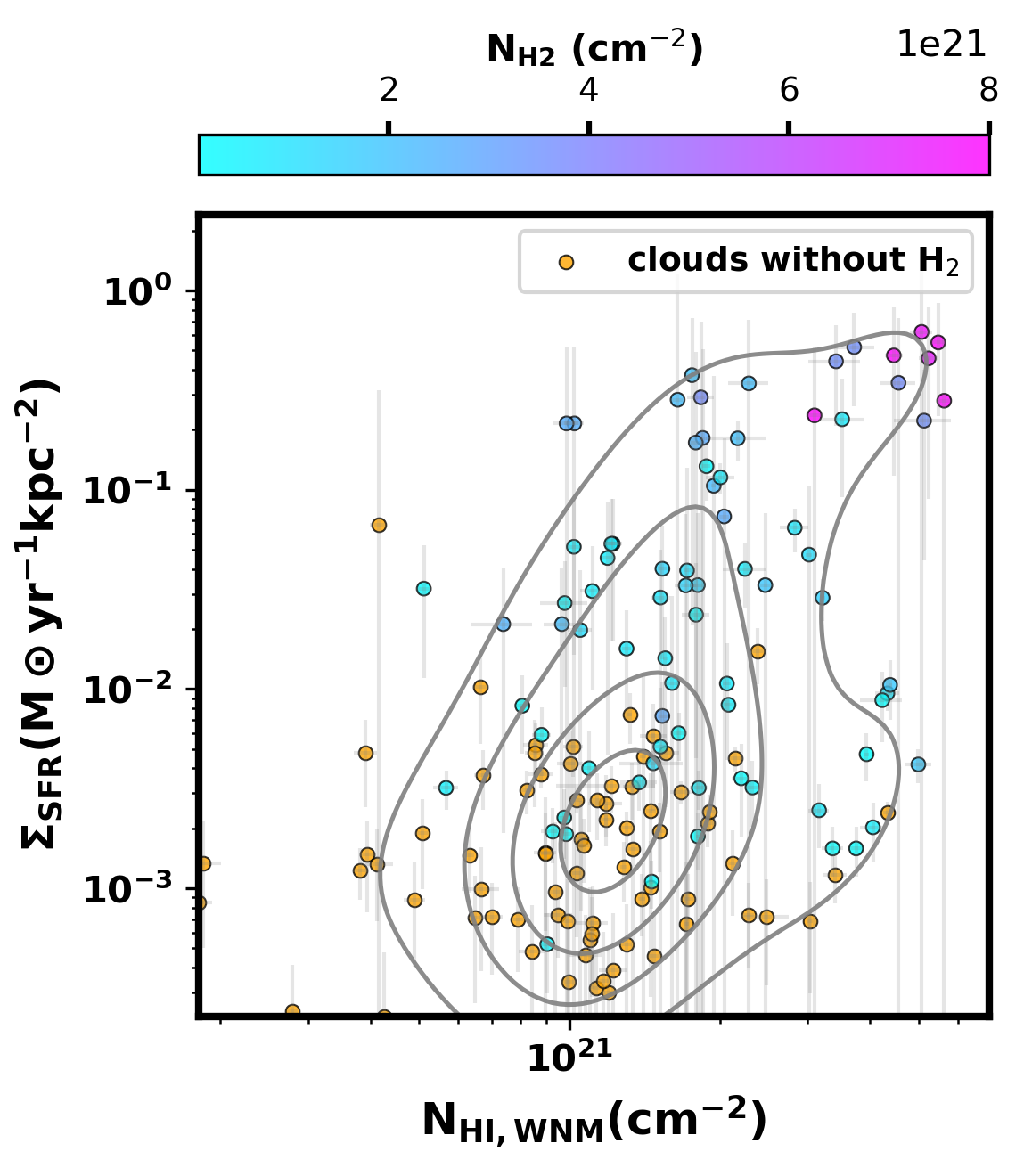}
    \caption{Star formation rate surface density as a function of total WNM HI column density along the line of sight.
    The yellow circles show the cold clouds without detection of H$_2$, while the color of other dots presents the H$_2$ column density.
    Contours show kernel density estimates of the data distribution, computed in logarithmic space and plotted at five equally spaced density levels.}
    \label{fig:sfr_wnm}
\end{figure}

In this section, we plot the star formation rate surface density as a function of the total WNM column density along each sightline, similar to Figure~\ref{fig:sfr}. It shows no clear correlation between these two quantities. In particular, even at high WNM column densities, many sightlines exhibit low $\Sigma_{\rm SFR}$, indicating that a large reservoir of warm atomic gas alone is not sufficient to trigger star formation. 



\bibliography{sample7}{}
\bibliographystyle{aasjournalv7}



\end{document}